\documentclass[twocolumn]{aa}  
\usepackage{natbib}
\usepackage{graphicx}
\usepackage{txfonts}
\usepackage{ulem}
\usepackage{threeparttable}
\usepackage{mathrsfs}
\usepackage{amsmath}
\usepackage{comment}
\usepackage{mathtools}
\usepackage{hyperref}
\usepackage{placeins}
\hypersetup{
    colorlinks=true,
    linkcolor=blue,
    citecolor=blue,
    urlcolor=green,
    }
\begin{document}

\title{A three-component giant radio halo: the puzzling case of the galaxy cluster Abell 2142}

   \author{L. Bruno 
          \inst{1,2},
        A. Botteon
         \inst{2},
        T. Shimwell
         \inst{3,4},
        V. Cuciti
       \inst{1,2,5},  
        F. de Gasperin
        \inst{2,5},  
           G. Brunetti
         \inst{2},
        D. Dallacasa
         \inst{1,2},
         F. Gastaldello
        \inst{6},
        M. Rossetti
         \inst{6},
        R. J. van Weeren
         \inst{4},
         T. Venturi 
        \inst{2}, 
        S. A. Russo 
        \inst{7},  
        G. Taffoni 
        \inst{7}, 
         R. Cassano 
        \inst{2},
        N. Biava 
        \inst{1,2},
        G. Lusetti
        \inst{5},
        A. Bonafede
         \inst{1,2},
        S. Ghizzardi
        \inst{6},
        S. De Grandi
        \inst{6} 
          }

   \institute{
    Dipartimento di Fisica e Astronomia (DIFA), Universit\`a di Bologna, via Gobetti 93/2, 40129 Bologna, Italy
    \and
    Istituto Nazionale di Astrofisica (INAF) - Istituto di Radioastronomia (IRA), via Gobetti 101, 40129 Bologna, Italy
    \and
    ASTRON, Netherlands Institute for Radio Astronomy, Oude Hoogeveensedijk 4, 7991 PD, Dwingeloo, The Netherlands         
    \and
    Leiden Observatory, Leiden University, PO Box 9513, 2300 RA Leiden, The Netherlands
    \and
    Hamburger Sternwarte, Universit\"at Hamburg, Gojenbergsweg 112, D-21029 Hamburg, Germany
    \and
    INAF - IASF Milano, Via A. Corti 12, I-20133, Milano, Italy
    \and
    Istituto Nazionale di Astrofisica (INAF) - Astronomical Observatory of Trieste, Trieste, Italy
%             
%             \and
%             INAF - Astronomical Observatory of Padova, Vicolo dell'Osservatorio 5, IT-35122 Padova, Italy
%             \and
%             
             \\
\email{luca.bruno4@unibo.it}
}

%   \date{Received September 15, 1996; accepted March 16, 1997}

% \abstract{}{}{}{}{} 
% 5 {} token are mandatory
 
  \abstract
   {Turbulence introduced into the intra-cluster medium (ICM) through cluster merger events transfers energy to non-thermal components, and can trigger the formation of diffuse synchrotron radio sources. Typical diffuse sources in the forms of giant radio halos and mini-halos are found in merging and relaxed cool core galaxy clusters, respectively. On the other hand, recent observations have revealed an increasing complexity of the non-thermal phenomenology.}
   {Abell 2142 (A2142) is a mildly disturbed cluster that exhibits uncommon thermal and non-thermal properties. It is known to host a hybrid halo consisting of two components (H1 and H2), namely a mini-halo-like and an enigmatic elongated radio halo-like structure. We aim to investigate the properties, origin, and connections of each component.}
   {We present deep LOFAR observations of A2142 in the frequency ranges $30-78$ MHz and $120-168$ MHz. With complementary multi-frequency radio and X-ray data, we analyse the radio spectral properties of the halo and assess the connection between the non-thermal and thermal components of the ICM. } 
   {We detected a third radio component (H3), which extends over the cluster volume on scales $\sim 2$ Mpc, embeds H1 and H2, and has a morphology that roughly follows the thermal ICM distribution. The radio spectral index is moderately steep in H1 ($\alpha=1.09\pm 0.02$) and H2 ($\alpha=1.15\pm 0.02$), but is steeper ($\alpha=1.57\pm 0.20$) in H3. The analysis of the thermal and non-thermal properties allowed us to discuss possible formation scenarios for each radio component. Turbulence from sloshing motions of low-entropy gas on different scales may be responsible for the origin of H1 and H2. We classified H3 as a giant ultra-steep spectrum radio halo, which could trace the residual activity from an old energetic merger and/or inefficient turbulent re-acceleration induced by ongoing minor mergers. }
   {}

   \keywords{Radiation mechanisms: thermal -- Radiation mechanisms: non-thermal -- Acceleration of particles -- Cosmology: large-scale structure of Universe -- Galaxies: clusters: intracluster medium -- Galaxies: clusters: individual: Abell 2142}
   
\titlerunning{The three-component radio halo in A2142}
\authorrunning{Bruno et al. 2023}
   \maketitle
%
%-------------------------------------------------------------------

\section{Introduction}

Galaxy clusters are the largest gravitationally bound structures in the Universe and accrete mass through mergers and infalling matter from the filaments of the Cosmic Web \citep[e.g.][]{kravtsov&borgani12}. The baryonic content is dominated by the intracluster medium (ICM), which is the hot ($T\sim 10^7-10^8$ K) and rarefied ($n_{\rm e}\sim10^{-2}-10^{-4}$ cm$^{-3}$) plasma that fills the space in-between the galaxy cluster members, and emits through thermal bremsstrahlung in the X-ray band \citep{sarazin}. Magnetic fields and relativistic particles are present in the ICM. These can emit synchrotron radiation that generates a variety of diffuse radio sources, such as giant radio halos and mini-halos that exhibit typically steep spectra\footnote{We define the spectral index from the flux density as $S_{\nu}\propto \nu^{-\alpha}$.} in the range $\alpha\sim 1-1.3$ (see \citealt{vanweeren19} for a review). 

Statistical studies showed that the mass and the dynamical state of the host cluster are key to understanding the processes that trigger the formation of a diffuse radio source \citep[e.g.][]{cassano10A,cassano13,kale13,cuciti15,cuciti21b,cassano23}. Giant radio halos extending on megaparsec-scales are preferentially found in massive and disturbed systems, whereas mini-halos are found to be confined within the cool core (on scales $\sim 100-500$ kpc) of relaxed clusters. According to re-acceleration models, a cluster merger dissipates a fraction of its energy in the ICM through turbulence, which can re-accelerate electrons and amplify magnetic fields, thus generating a radio halo \citep[e.g.][]{brunetti01,petrosian01,brunetti&lazarian07,beresnyak13,miniati15,brunetti&lazarian16}. These models predict a wide range of spectral properties, including a large number of halos with very steep $\alpha \gtrsim 1.5$ (also referred to as ultra-steep spectrum radio halos), which are rare at high ($\sim 1$ GHz) frequencies, but should be common at lower (hundreds of MHz) frequencies \citep[e.g.][]{cassano10,cassano23}. These can be old radio halos in which the bulk of turbulence has already been dissipated and/or trace inefficient turbulent re-acceleration associated with a minor merger and/or low mass clusters. In line with the key predictions of the turbulent re-acceleration scenario, low frequency observations have revealed very steep spectrum radio halos in a number of clusters \citep{brunetti08,dallacasa09,macario10,bonafede12,venturi17,bruno21,digennaro21b,duchesne21,edler22,pasini22,rajpurohit23}.

Current data support turbulent re-acceleration models, but the complex energy transfer mechanisms from megaparsec-scales down to smaller scales are still poorly understood \citep[e.g.][for a review]{brunetti&jones14}. Furthermore, the origin of the seed electrons that are re-accelerated by turbulence is unclear. Possibilities are that these are old, mildly-relativistic, populations of primary cosmic ray electrons (CRe) injected by AGN \citep[e.g.][]{fujita07,bonafede14,vanweeren17}, or secondary CRe produced by hadronic collisions between protons in the ICM \citep[e.g.][]{brunetti&blasi05,brunetti&lazarian11,brunetti17,pinzke17}. A similar scenario to that of giant halos was invoked for mini-halos as well, but in this case turbulence may be induced by sloshing of cold gas in the cluster core, after this was perturbed by a minor merger \citep[e.g.][]{mazzotta&giacintucci08,fujita&ohira13,zuhone13}. In the context of hadronic models \citep[e.g.][]{dennison80,blasi&colafrancesco99,dolag&ensslin00,pfrommer08}, mini-halos may also trace emission from fresh populations of secondary CRe, without requiring any re-acceleration process \citep{pfrommer&ensslin04}.

Sensitive radio observations are revealing an increasing complexity of the phenomenology of diffuse sources. The distinction between halos and mini-halos is being challenged by the discovery of giant halos in relaxed systems \citep{bonafede14b,kale19,raja20} and `hybrid' halos with an inner mini-halo-like component and an outer radio halo-like component \citep{farnsworth13,venturi17,savini18,savini19,biava21}. Furthermore, diffuse radio emission has been recently detected up to the canonical cluster outskirts \citep[e.g.][]{shweta20,rajpurohit21b,botteon22b,cuciti22} and between pairs of clusters in pre-merger phases \citep{botteon18a,govoni19,botteon20c,dejong22}, thus confirming the presence of non-thermal components and re-acceleration processes taking place on such large scales. In particular, by means of LOw Frequency ARray (LOFAR) observations below $\sim 200$ MHz, \cite{cuciti22} reported on the discovery of a new type of diffuse source in 4 merging clusters (ZwCl 0634.1+4750, A665, A697, and A2218) that surrounds classical giant radio halos, and has been called megahalo. Megahalos extend up to scales $\sim R_{\rm 500}$\footnote{$R_{500}$ is the radius enclosing $500\rho_{\rm c}(z)$, where $\rho_{\rm c}(z)$ is the critical density of the Universe at a given redshift; the corresponding mass is $M_{500}$.}, exhibit a shallower surface brightness radial profile than that of their embedded radio halo, and are characterised by a very steep spectrum ($\alpha \gtrsim 1.6$ between $\sim 50-150$ MHz).

Studying targets with uncommon features can shed light on the physical properties that underlie standard classification schemes (e.g. relaxed and merger) for galaxy clusters, and especially for the aims of this work, on the origin of hybrid sources and the physical and/or evolutionary connection between its radio components. The nearby and massive galaxy cluster Abell 2142 (hereafter A2142) is a favourable target to investigate the role of intricate environmental and dynamical conditions at different scales in the formation and shaping of multi-component radio sources. A2142 is the main member of the A2142-supercluster, has an intermediate dynamical state that challenges the usual classification as relaxed/merging system, is characterised by extremely complex dynamics possibly triggered by many minor merger events, and is known to host an hybrid halo consisting of two radio components. In this work, we report on our study of A2142 by means of new deep LOFAR observations that allowed us to discover an additional, ultra-steep spectrum, radio component of the hybrid halo on large scale. We used complementary multi-frequency radio and X-ray data to probe the uncommon non-thermal and thermal properties of A2142, discuss the origin of the only three-component radio halo discovered so far, and search for evidence of possible emission in the form of a megahalo.

The paper is organised as follows. In Sect. 2, we describe the A2142 galaxy cluster. In Sect. 3, we present the radio and X-ray data and summarise their processing. In Sect. 4, we show the results of our analysis. In Sect. 5, we compare the properties of the hybrid halo in A2142 with those of the 4 megahalos recently discovered, and discuss the origin of its radio components. In Sect. 6, we summarise our work. Throughout this paper we adopted a standard $\Lambda$CDM cosmology with $H_0=70\;\mathrm{km\; s^{-1}\; Mpc^{-1}}$, $\Omega_{\rm M}=0.3$ and, $\Omega_{\rm \Lambda}=0.7$. At the cluster redshift $z=0.0894$, the luminosity distance is $D_{\rm L}=408.6$ Mpc and $1''=1.669$ kpc (or $1'\sim 100$ kpc).

\section{The galaxy cluster Abell 2142}  

\begin{figure*}
	\centering
	\includegraphics[width=0.75\textwidth]{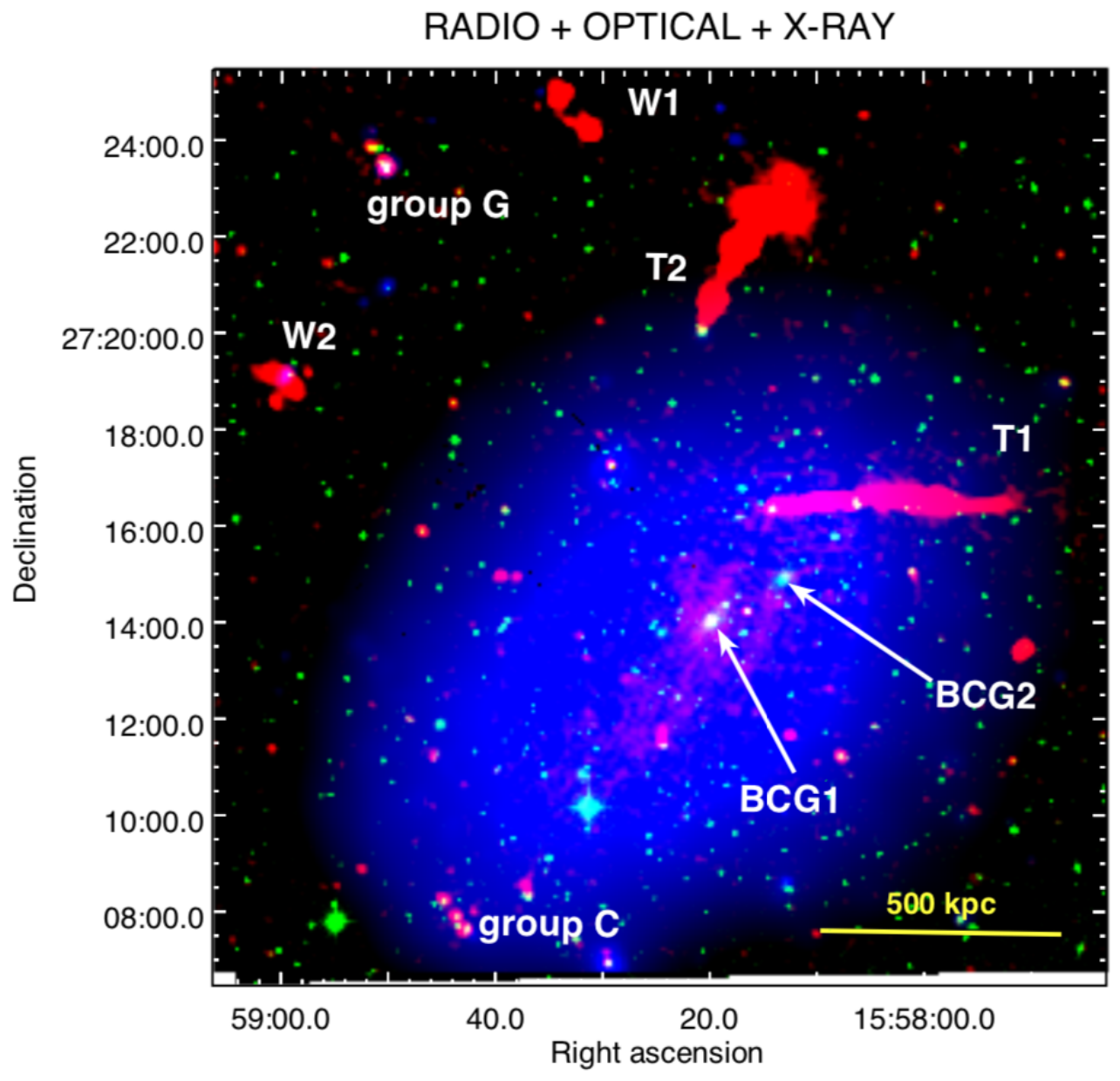}
	\smallskip
	
	\caption{Composite RGB image of A2142: radio (GMRT, 323 MHz) in red, optical (DSS-2, red filter) in green, and X-rays (XMM-Newton) in blue. The discrete sources discussed in the text are labelled as in \cite{venturi17}.}
	\label{ROX}
\end{figure*}

A2142 (RA$_{\rm J2000}=15^{\rm h}58^{\rm m}20^{\rm s}$, Dec$_{\rm J2000} =  27^{\rm o}14'00''$) is a nearby ($z=0.0894$) galaxy cluster of mass $M_{500}=(8.8\pm0.2)\times 10^{14} \;  M_\odot$ within a radius $R_{500}=14.07\pm 0.70 \; {\rm arcmin}$   \citep[$1408.5\pm 70.4$ kpc at the cluster redshift;][]{planckcollaboration16,tchernin16}. The galaxy cluster A2142 is located in the centre of the A2142 supercluster, to which it gives the name \citep{einasto15,gramann15}. The A2142 supercluster is in turn part of an interacting system with the Corona Borealis supercluster \citep{pillastrini&baiesi19}. In Fig. \ref{ROX} we overlay optical, X-ray, and radio images of the target from archival observations (see details below).

As derived from photometric and spectroscopical optical studies, $\sim 900$ galaxies within a radius of $\sim 3.5$ Mpc are confirmed members of A2142, and are hierarchically organised in many structures and sub-structures typically consisting of small groups \citep{owers11,einasto18,liu18}. The main and richest structure hosts the primary brightest cluster galaxy `BCG1', which is located at the centre of the potential well of the cluster \citep{okabe&umetsu08,wang&markevitch18}. The dynamics of A2142 is extremely complex owing to several ongoing minor mergers, as galaxy groups of $\sim 10$ members belonging to various sub-structures are infalling towards the cluster centre \citep[e.g.][]{owers11,eckert17,liu18}. At a projected distance of $\sim 180$ kpc from `BCG1', the secondary brightest cluster galaxy `BCG2' has a high peculiar velocity \citep[e.g.][]{oegerle95} and is likely the main member of another merging group. Even though the cluster is still accreting through these minor mergers, \cite{einasto18} compared the present mass of A2142 and the evolution of simulated cosmic large structures and derived an estimate of the half-mass epoch; according to these authors, half of the current mass of A2142 was likely accreted through merger events occurred $\gtrsim 4$ Gyr ago, thus forming the main structure of A2142.

In the X-ray band, the ICM has an elongated morphology along the NW-SE axis, and is aligned with the filamentary structure of the supercluster. The global temperature is $kT\gtrsim 9-10$ keV, but it moderately decreases towards the central regions, where $kT\lesssim 7$ keV \citep[e.g.][]{henry&briel96}. A2142 has morphological and thermodynamical properties intermediate between those of a relaxed cool core and an unrelaxed merging cluster \citep[e.g.][]{cavagnolo09,rossetti13,tchernin16,cuciti21b}. Even though a well-defined X-ray peak is present as in relaxed clusters, the density, temperature, and entropy of the ICM derived from its spectrum are not consistent with typical values found for cool cores, therefore A2142 was classified by \cite{wang&markevitch18} as a rare case of a `warm' core cluster \citep[see also][for discussion on targets with similar properties]{rossetti&molendi10,molendi23}. 

Cold fronts are contact discontinuities detected in the X-rays as surface brightness edges, which were first discovered in A2142 by Chandra \citep{markevitch00}. In particular, A2142 exhibits a system of three roughly concentric cold fronts which follow a spiral-like path and are located close to the two BCGs \citep{markevitch00,markevitch&vikhlinin07,wang&markevitch18}. The X-ray peak of the warm core shows a prominent offset of 30 kpc with respect to the centre of the cluster \citep{wang&markevitch18}. It was suggested that an intermediate mass-ratio merger did not completely disrupt the core, but displaced a large fraction of gas from the minimum of the gravitational potential; this event caused extreme sloshing motions of the cool gas, which then induced the formation of the cold fronts \citep{rossetti13}. A further cold front was discovered by XMM-Newton in the SE of the cluster, at a distance of $\sim 1$ Mpc from the centre \citep{rossetti13}. The SE cold front may result from the long-term evolution of the central sloshing, thus indicating that it is a phenomenon able to move gas from the core towards relatively larger scales \citep{rossetti13}. 

In the radio band, diffuse synchrotron emission was first detected with the Very Large Array (VLA) in the central regions of the cluster \citep{giovannini&feretti00}, and then with the Green Bank Telescope (GBT) up to $\sim 2$ Mpc \citep{farnsworth13}. Deep follow-up observations at low frequencies with the Giant Metrewave Radio Telescope (GMRT) allowed \cite{venturi17} to classify the diffuse emission as a giant radio halo with two components, which are characterised by different morphological and spectral properties. The most compact, roundish, and brightest component (the `core', `H1') is spatially confined by the inner cold fronts. The faintest, elongated, and largest component (the `ridge', `H2') extends in the direction of the SE cold front. \cite{venturi17} reported spectral indices of $\alpha \sim 1.3$ and $\alpha \sim 1.5$ between 118 and 1780 MHz, for the core and the ridge, respectively. The outermost emission revealed by the GBT at 2 Mpc is not detected with the GMRT, thus its properties are unconstrained. \cite{venturi17} suggested that both H1 and H2 may trace regions of turbulent particle re-acceleration. In this case, turbulence in H1 would be generated by the dissipation of the kinetic energy of gas sloshing in the core, similarly to the scenario invoked for the formation of mini-halos \citep[e.g.][]{mazzotta&giacintucci08,fujita&ohira13,zuhone13}; turbulence in the ridge could have been induced by less energetic mergers (including the same intermediate mass-ratio merger likely associated with the origin of core sloshing), or trace the evolution of the central perturbations at larger scales. Alternatively, A2142 could be an hybrid hadronic/re-acceleration radio halo, in which secondary CRe dominate the emission in H1, whereas the turbulent re-accelerated electrons dominate the emission in H2.

\section{Observations and data reduction}

 \begin{table*}
   \centering
   	\caption[]{Details of the LOFAR LBA (PI: L. Bruno, project code: LC17\_012), LOFAR HBA (pointing name: P239+27 \& PI: F. Vazza, project code: LC14\_018), GMRT (PI: T. Venturi, project code: 23\_017), uGMRT (PI: T. Venturi, project code: 33\_052), and VLA (PI: D. Farnsworth, project code: 11B-156) radio data analysed in this work.}
   	\label{datiRADIO}
   	\begin{tabular}{ccccc}
   	\hline
   	\noalign{\smallskip}
   	Instrument & Band name &  Frequency coverage &  Observation date & On-source time \\
   	&  &  (MHz) & & (h) \\
   	\noalign{\smallskip}
  	\hline
   	\noalign{\smallskip}
   	 LOFAR  & LBA  & 30-78 & 08,17,23-Dec.-2021 & 16.0     \\
   	 LOFAR & HBA & 120-168 & 15-Sept.-2018; 25,31-Oct.-2020; 13-Nov.-2020 & 32.0 \\
%   	$\rm LOFAR_{\rm [LoTSS]}$ & HBA & 144 & 120-168 & 15-Sept.-2018 & 8.0     \\
%   	$\rm LOFAR_{\rm [LC14\_018]}$  & HBA & 144 & 120-168 & 25,31-Oct.-2020, 13-Nov.-2020 & 24.0     \\

%    GMRT & 2 & 234 & 225-240 & 23-Mar.-2013 & 6.0     \\
    GMRT  &  & 307-339 & 27-Mar.-2013 & 5.0     \\
    uGMRT  & band-3 & 300-500 & 15-Mar.-2018 & 3.0     \\
%    GMRT & 4 & 608 & 590-625 & 22-Jun.-2013 & 5.0     \\
%    GMRT & L  & 1440 & 1420-1455 & 20-Mar.-2018 & 1.5 \\
   	$\rm VLA_{\rm [C-array]}$  & L & 1000-2000 & 27-Apr.-2012 & 0.5     \\
   	$\rm VLA_{\rm [D-array]}$  & L & 1000-2000 & 9-Oct.-2011 & 1.5      \\
   	\noalign{\smallskip}
   	\hline
   	\end{tabular}
   \end{table*}

 \begin{table}
 \fontsize{8.}{8.}\selectfont
      \centering
   	\caption[]{Summary of the parameters for LOFAR images discussed in \ref{sect:Radio images}. Col. 1: central frequency ($\nu$). Cols. 2 to 4: minimum baseline ($B_{\rm min}$), robust parameter of the Briggs weighting, and Gaussian tapering. Cols. 5 to 7: restoring beam ($\theta$), beam position angle ($P.A.$), and reached noise ($\sigma$).}
   	\label{mapperadioweightscheme}
   	\begin{tabular}{ccccccc}
   	\hline
   	\noalign{\smallskip}
   	 $\nu$ & $B_{\rm min}$ & Robust & Taper & $\theta$ & $P.A.$ & $\sigma$ \\
   	  (MHz) & ($\lambda$) &  & ($''$) & ($'' \; \times \; ''$) & (deg) & (${\rm mJy \; beam^{-1}}$) \\
   	\noalign{\smallskip}
   	\hline
   	\noalign{\smallskip}
   	 50 & 30  & $-1.0$ & - & $14\times10$ & 86 & 1.9 \\
   	 50 & 30  & $-0.5$ & 15  & $22\times13$ & 84 & 1.6 \\
   	 50 & 30  & $-0.5$ & 30  & $44\times32$ & 84 & 2.7 \\
   	 50 & 30  & $-0.5$ & 60  & $69\times63$ & 50 & 4.4 \\
   	 143 & 80  & $-0.5$ & -  & $9\times6$ & 88 & 0.075 \\
   	 143 & 80  & $-0.5$ & 15  & $21\times20$ & 84 & 0.16 \\
   	 143 & 80  & $-0.5$ & 30  & $38\times35$ & 16 & 0.25 \\
     143 & 80  & $-0.5$ & 60  & $73\times66$ & 288 & 0.40 \\
     143 & 50 & $-0.5$ & 120 & $128\times117$ & 310 & 1.0 \\
     
   	\noalign{\smallskip}
   	\hline
   	\end{tabular}
   \end{table}  
% \fontsize{}{}\selectfont  

      \begin{table}
      \centering
   	\caption[]{Details of the Chandra X-ray data analysed in this work (PI of ObsID 5005: L. van Speybroeck; PI of ObsIDs 15186, 16564, 16565: M. Markevitch.)}
   	\label{datiX}
   	\begin{tabular}{cccc}
   	\hline
   	\noalign{\smallskip}
   	ObsID &  CCDs  &  Observation date &    Clean time \\
   	&  &  & (ks) \\
   	\noalign{\smallskip}
   	\hline
   	\noalign{\smallskip}
   	5005 & S2,I0,I1,I2,I3  & 13-Apr.-2005 &   41.5   \\
   	15186 & S1,S2,S3,I2,I3 & 19-Jan.-2014 &   82.7   \\
   	16564 & S1,S2,S3,I2,I3 & 22-Jan.-2014 &   43.2   \\
   	16565 & S1,S2,S3,I2,I3 & 24-Jan.-2014 &   19.5   \\

   	\noalign{\smallskip}
   	\hline
   	\end{tabular}
   \end{table}  
   
In this Section we present the data analysed in this work and the corresponding data reduction. The details of the radio data are summarised in Table \ref{datiRADIO}, while the details of Chandra X-ray data are summarised in Table \ref{datiX}.

\subsection{LOFAR HBA radio data}

A2142 was observed for 8 hours in September 2018 in the context of the LOFAR Two Meter Sky Survey \citep[LoTSS, pointing P239+27;][]{shimwell17,shimwell19LOTSS,shimwell22LOTSSDR2}. Furthermore, 24 additional hours were spent on A2142 between October and November 2020. The High Band Antenna (HBA) Dutch array operating in the 120-168 MHz frequency range was employed for all the observations, with 23 core stations, 14 remote stations, and 14 international stations (the latter were not included in this work). Data were recorded with an integration time of 1 s and 64 channels (of width 3 kHz each) per sub-band, and then averaged to 16 channels per sub-band after removal of Radio Frequency Interference (RFI). The source 3C 295 was used as flux density scale calibrator.

All the data were processed together by means of the LOFAR Surveys KSP reduction pipelines, which perform direction-independent and direction-dependent calibration using {\tt PREFACTOR}\footnote{\url{https://github.com/lofar-astron/prefactor}} \citep{vanweerenDDFACET16,williams16,degasperin19} and {\tt ddf-pipeline}\footnote{\url{https://github.com/mhardcastle/ddf-pipeline}} v. 2.4, which makes use of {\tt DDFacet} \citep{tasse18} and {\tt KillMS} \citep{tasse2014b,tasse14a,smirnov15} to compute direction-dependent calibration of the whole LOFAR field of view \citep[see also][]{tasse21,shimwell22LOTSSDR2}.

Following the `extraction \& re-calibration' technique described in \cite{vanweeren21}, sources outside a square region of sizes $25.8'\times25.8'$ centred on the target were subtracted from the \textit{uv}-data; we then performed 8 additional cycles of amplitude and phase direction-independent self-calibration on the extracted datasets to improve the quality of the images towards A2142. 

Uncertainties in the beam model of LOFAR HBA and calibration errors can introduce offsets in the flux density scale when amplitude solutions are transferred from the primary calibrator to the target \citep[e.g.][]{hardcastle21}. As in LoTSS \citep{shimwell22LOTSSDR2}, the flux density scale was thus set by cross-matching the LOFAR image with the NRAO VLA Sky Survey \citep[NVSS][]{condon98NVSS}, and by assuming a ratio $S_{\rm 6C}/S_{\rm NVSS}=5.9124$ between the flux density of the 6C radio catalogue \citep{hales88,hales90} and that of NVSS, at 150 MHz and 1.4 GHz, respectively. Due to this procedure, each of our LOFAR HBA images has to be multiplied by a flux scale correction factor of 0.8793.

\subsection{LOFAR LBA radio data}

A2142 was observed by LOFAR with the Low Band Antenna (LBA) in the frequency range 30-78 MHz on December 2021, for a total on-source time of 16 hours. Observations were carried out in {\tt LBA Sparse Even} mode by 24 core stations and 14 remote stations, with two separate beams simultaneously pointing on the target and flux density calibrator 3C 295, respectively. This observing setup both provides a larger field of view and reduces the impact of interfering sources from the first sidelobe with respect to the {\tt LBA\_OUTER} mode. Data were first acquired with an integration time of 1 s and 64 channels (of width 3 kHz each) per sub-band, and then averaged to 4 s and 8 channels per sub-band after removal of RFI and demixing, using a strategy consistent with  \cite{degasperin20b}. 

Calibration was performed by means of the Library for Low-Frequencies (LiLF\footnote{\url{https://github.com/revoltek/LiLF}}) pipelines. Following the process described in \cite{degasperin19}, gain solutions are obtained for the flux density calibrator to correct for polarisation alignment, bandpass, and clock drift, and then transferred to the target. Ionospheric effects on the target are corrected through steps of direction-independent and direction-dependent calibration \citep[see details in][]{tasse18,degasperin20}. This procedure provides an image of the whole field of view ($10^{\rm o}\times10^{\rm o}$) at a resolution of $15''$. 

Similarly to the LOFAR HBA data, the `extraction \& re-calibration' technique was successfully applied on LBA datasets as well \citep[e.g.][]{edler22,pasini22}. We carried out the extraction technique on our data, by testing several extraction regions to maximise the signal-to-noise ratio (${\rm S/N}$) of our images after the full self-calibration process; the best results were obtained with a circular extraction region of diameter $23'$. We then performed 4 rounds of amplitude and phase direction-independent self-calibration.

Differently from LOFAR HBA, the beam model of LOFAR LBA is expected to be more accurate \citep{degasperin23}. Therefore, the flux density scale of LOFAR LBA does not require further corrections.

\subsection{GMRT and uGMRT radio data}

We retrieved archival GMRT observations of A2142 at 307-339 MHz for 5 hours on-source, first presented by \cite{venturi17}. The total bandwidth is 32 MHz (split into 256 channels). The source 3C286 was used as absolute flux density scale calibrator. We reprocessed the data by means of the Source Peeling and Atmospheric Modeling ({\tt SPAM}) automated pipeline \citep{intema09}, which corrects for ionospheric effects and removes direction-dependent gain errors. Bright sources in the field are used to derive directional-dependent gains and fit a phase-screen over the field of view. Finally, images are corrected for the system temperature variations between the calibrators and the target. These procedures allowed us to reach a noise level of $\sim 40 \; \mu {\rm Jy \; beam^{-1}}$ at $9''$, which is noticeably lower (by a factor $\sim 3$) than that reached by \cite{venturi17}, mainly due to a better calibration of the short baselines.

Additional, as yet unpublished, uGMRT observations of A2142 are available in the archive. We retrieved a 3 hour-long observation in the frequency range 300-500 MHz (band-3). The total bandwidth of 200 MHz is split into 4000 channels of 50 kHz each. The sources 3C286 and 1602+334 were used as absolute flux density scale and phase calibrators, respectively. We adopted 
the CAsa Pipeline-cum-Toolkit for Upgraded GMRT data REduction \citep[{\tt CAPTURE}\footnote{\url{https://github.com/ruta-k/CAPTURE-CASA6}};][]{kale&ishwara-chandra21} v.2.0
to perform a standard, fully automated, calibration of the wide-band data and derive delay, bandpass, phase, and amplitude corrections. We then performed 3 cycles of phase-only and 2 cycles of phase plus amplitude self-calibration, which allowed us to reach a noise level of $\sim 30 \; {\rm \mu Jy \; beam^{-1}}$.

\subsection{VLA radio data}
\label{datiJVLA}

We retrieved three VLA pointings on A2142 at 1-2 GHz (L-band) in C and D configurations, for 0.5 and 1.5 hours, respectively. The three pointings are slightly offset from the target, allowing us to mosaic the region with a more uniform sensitivity. The source 3C286 was used as flux density calibrator, whereas J1609+2641 was used as phase calibrator. The data were recorded with 16 spectral windows, each divided into 64 channels.

The data reduction was carried out with the National Radio Astronomy Observatory (NRAO) Common Astronomy Software Applications \citep[{\tt CASA};][]{mcmullincasapaper07} v. 5.1, by performing standard initial phase, bandpass, and gain calibrations. Due to the usual high amount of RFI in this band, $\sim 50\%$ of the bandwidth was flagged. Further rounds of self-calibration were not required. In order to obtain deeper images, we combined the visibilities of the C and D configuration datasets. We finally split the remaining bandwidth into two datasets of $\sim 250$ MHz width each, centred on 1380 and 1810 MHz.

\subsection{Radio imaging and source subtraction}
\label{Sect: Radio imaging and source subtraction}

The imaging process was carried out for GMRT, uGMRT, and LOFAR datasets with {\tt WSClean} v. 2.10 \citep{offringa14,offringa17}, which can account for wide-field, multi-frequency, and multi-scale synthesis. To properly combine the VLA pointings, we imaged these datasets with {\tt tclean} in {\tt CASA} making use of mosaic gridding, as well as multi-frequency and multi-scale synthesis.

We produced images by varying the relative weights of baselines through the robust parameter of the Briggs weighting \citep{briggs95} and Gaussian tapering of the outer \textit{uv}-coverage to study the radio emission on different spatial scales. The weighting schemes adopted for the images discussed in the next sections are summarised in Table \ref{mapperadioweightscheme}.

To accurately measure the flux densities of the diffuse emission, the contribution of the embedded discrete sources needs to be removed. Following \cite{venturi17}, we first obtained models of the discrete sources by imaging the data at high resolution, excluding baselines $<2{\rm k}\lambda$, corresponding to maximum recoverable scales of $\sim 100''$ (i.e. $\sim 170$ kpc at the cluster redshift). This procedure was carried out with {\tt WSClean} for all the datasets, including the VLA datasets, whose pointings were handled separately. Then, the clean components in the model images were subtracted from the \textit{uv}-data. Even though this process allowed us to remove the majority of the sources, residual artefacts associated with the two extended head-tail radio galaxies are still present; moreover, due to calibration imperfections, when decreasing the resolution, very faint residuals from compact sources can also be enhanced and severely contaminate the diffuse emission \citep[e.g.][]{bruno23}. Owing to the complexity of the discrete sources and different quality of the subtraction at each frequency, here we assume no systematic subtraction errors, but consider regions of the target where discrete sources are absent as much as possible when measuring the flux density (see details in Sect \ref{sect: Spectral indices}).

By ignoring the subtraction error contribution, uncertainties $\Delta S$ on the reported flux densities are given by:
\begin{equation}
\Delta S= \sqrt{ \left( \sigma \cdot \sqrt{N_{\rm beam}} \right)^2 + \left(  \xi_{\rm cal} \cdot S \right) ^2}
\label{erroronflux}
\end{equation}
where $\sigma$ is the RMS noise of the image, $N_{\rm beam}$ is the number of independent beams within the considered region, and $\xi_{\rm cal}$ is the calibration error. We adopted $\xi_{\rm cal}=10\%$ for LOFAR HBA \citep{shimwell22LOTSSDR2} and LBA \citep{degasperin21}, $\xi_{\rm cal}= 6\%$ for GMRT and uGMRT band-3 \citep{chandra04}, and $\xi_{\rm cal}=5\%$ for VLA L-band \citep{perley&butler13}. In the following, position angles ($P.A.$) of radio beams and regions are measured north to east.

\subsection{Chandra X-ray data}

We analysed the archival Chandra X-ray data of A2142 that are summarised in Table \ref{datiX}. Observations were carried out in 2005 and 2014, in VFAINT mode, with both ACIS-I and ACIS-S CCDs to recover the full extension of the target. We reprocessed the 4 observations using {\tt CIAO} v. 4.13, with {\tt CALDB} v. 4.9.4. We extracted light curves in source-free regions to filter out soft proton flares with the {\tt lc\_clean} algorithm. This procedure left a total clean time of $186.9$ ks. 

After correcting each pointing for the corresponding point spread function and exposure map, we combined them to produced a single flux image in the 0.5-2 keV band with the task {\tt mergeobs}. Candidate point sources were identified through the {\tt wavdetect} task; after visual inspection, we subtracted the confirmed ones.

Background event files were obtained through the {\tt blanksky} tool, which re-projects blansky pointings in the direction of our observations, and calculates normalisation factors to match the count rates in the 9-12 keV band (with the parameter {\tt weight \textunderscore method=particle-rate}). These event files were used as input for {\tt blanksky\textunderscore image} tool to produce the corresponding background images. 

X-ray spectral analyses were carried out with {\tt XSPEC} \citep{arnaud96xspec} v. 12.10.1. It is worth mentioning that present releases of {\tt XSPEC} are not able to properly deal with spectra extracted from re-scaled background events produced by {\tt blanksky}. To this aim, we therefore manually re-scaled the exposure time of the re-projected blansky pointings to match the count rates of our observations in the 9-12 keV band.

\subsection{XMM-Newton X-ray data}

A2142 was first observed by XMM-Newton in July 2011 (ObsID 0674560201, PI: M. Rossetti), and then followed-up in July 2012 (ObsIDs 0694440101, 0694440501, 0694440601, 0694440201, PI: D. Eckert) as part of the XMM Cluster Outskirts Project \citep[X-COP;][]{eckert17XCOP} to map both the centre and periphery of the cluster. These data consist of a mosaic of five pointings for a total exposure time of 195 ks. We retrieved the final products of these data (count, exposure, background, and background-subtracted flux maps in the 0.7-1.2 keV band) from the public website of the project\footnote{\url{https://dominiqueeckert.wixsite.com/xcop}} \citep[see][for details on observations and data processing.]{rossetti13,tchernin16,ghirardini18}

\section{Results}

\subsection{Radio images}
\label{sect:Radio images}

\begin{figure*}
	\centering
	\includegraphics[width=0.45\textwidth]{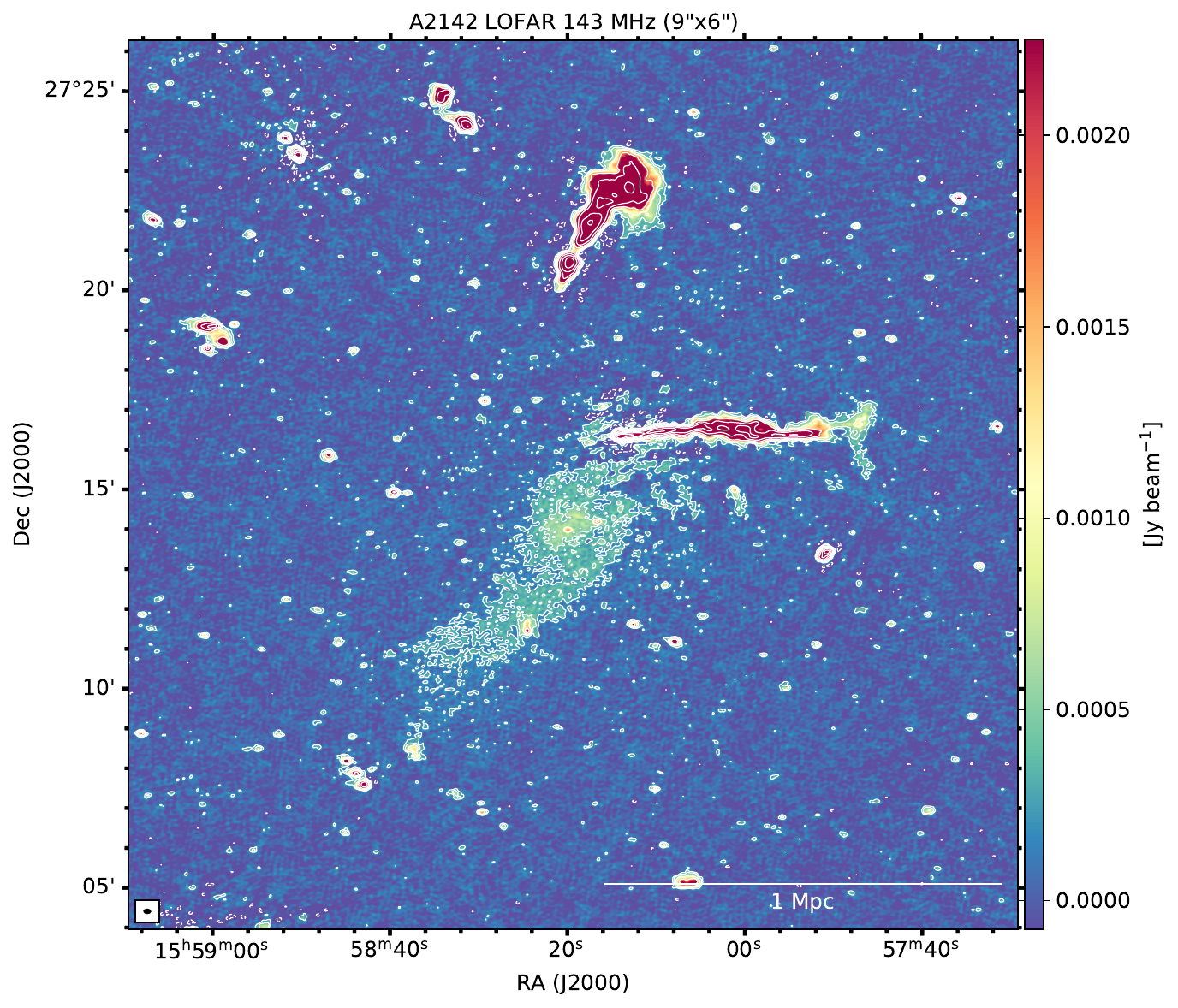}
	\includegraphics[width=0.45\textwidth]{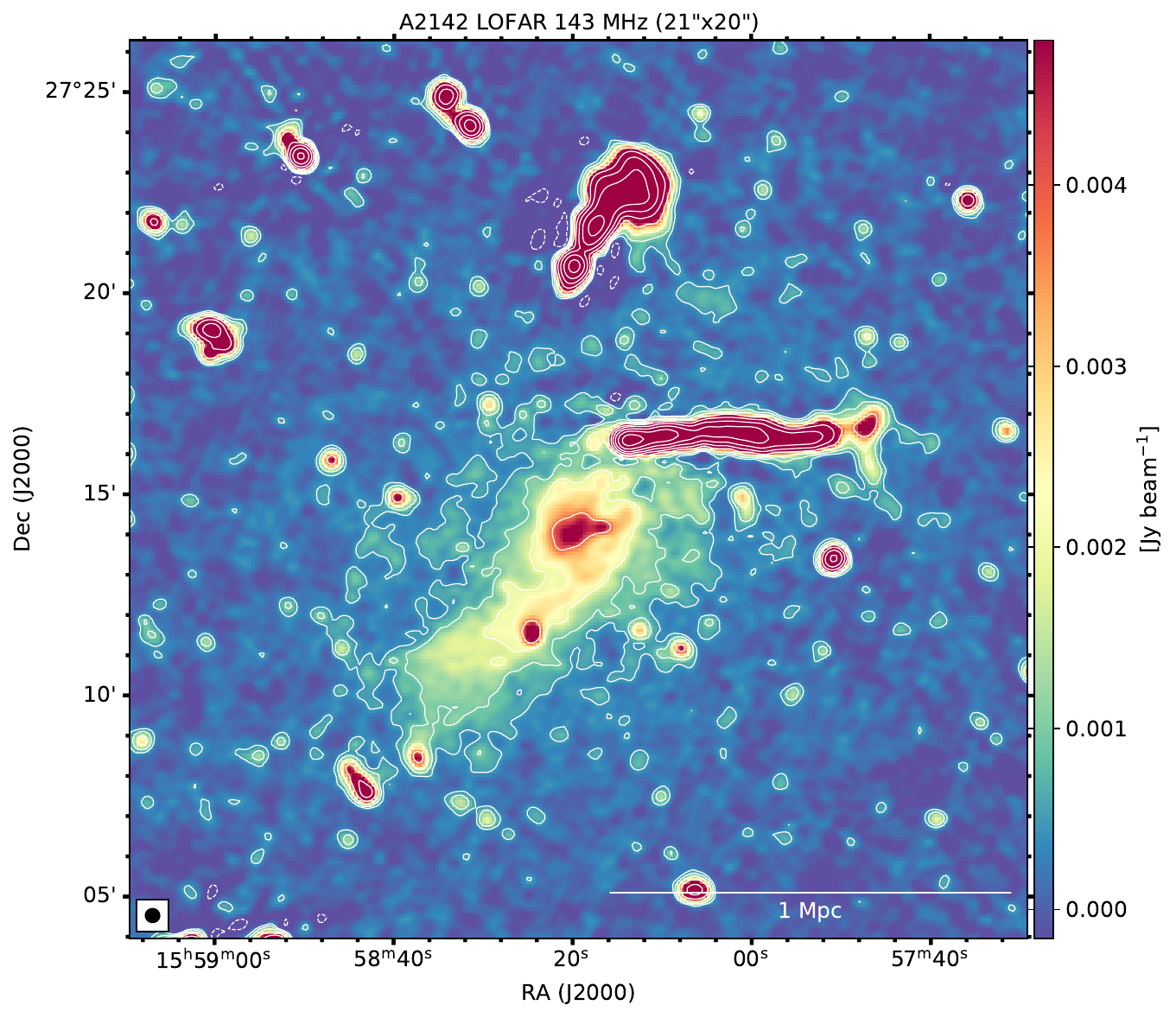}
	\includegraphics[width=0.45\textwidth]{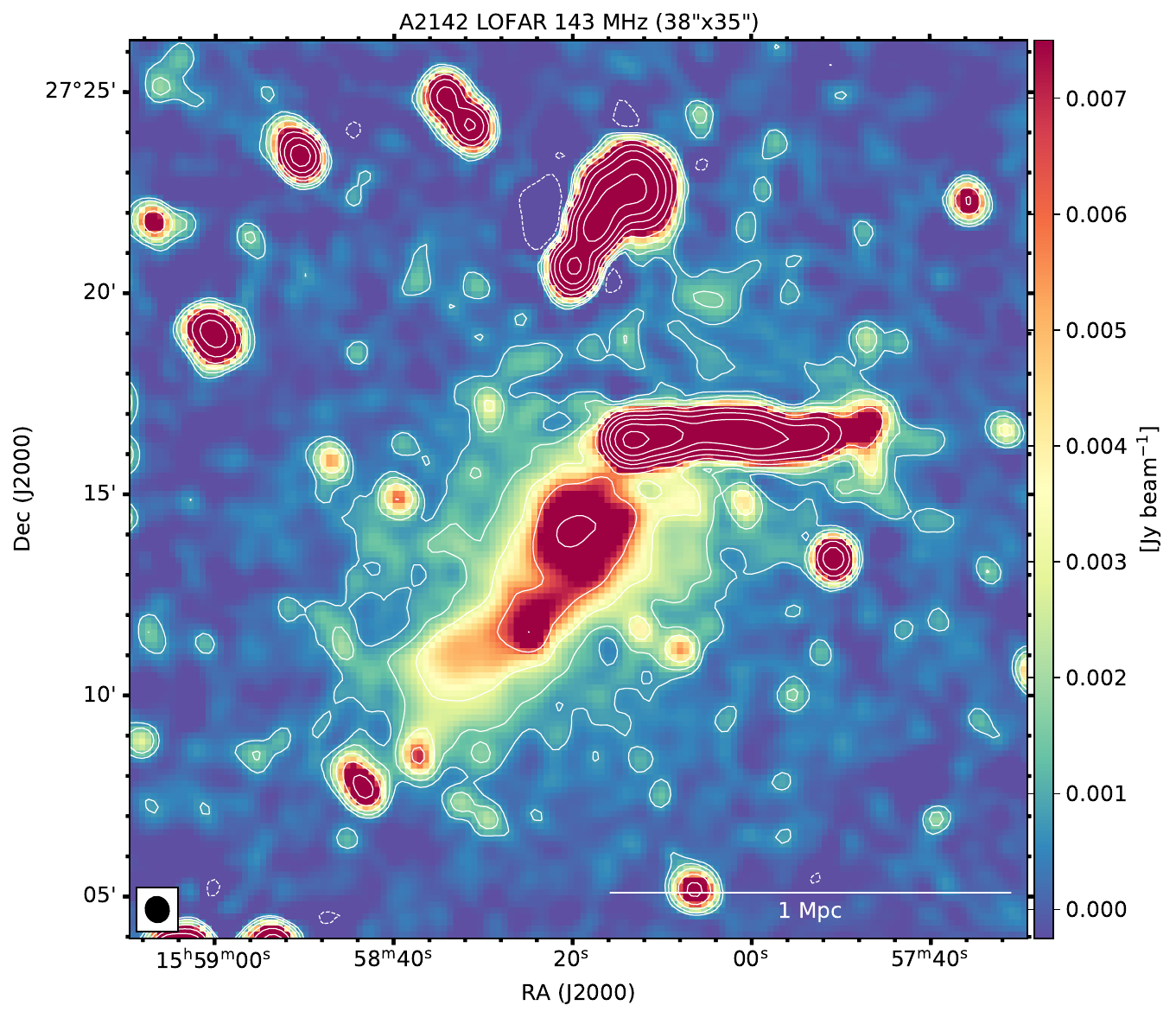}
	\includegraphics[width=0.45\textwidth]{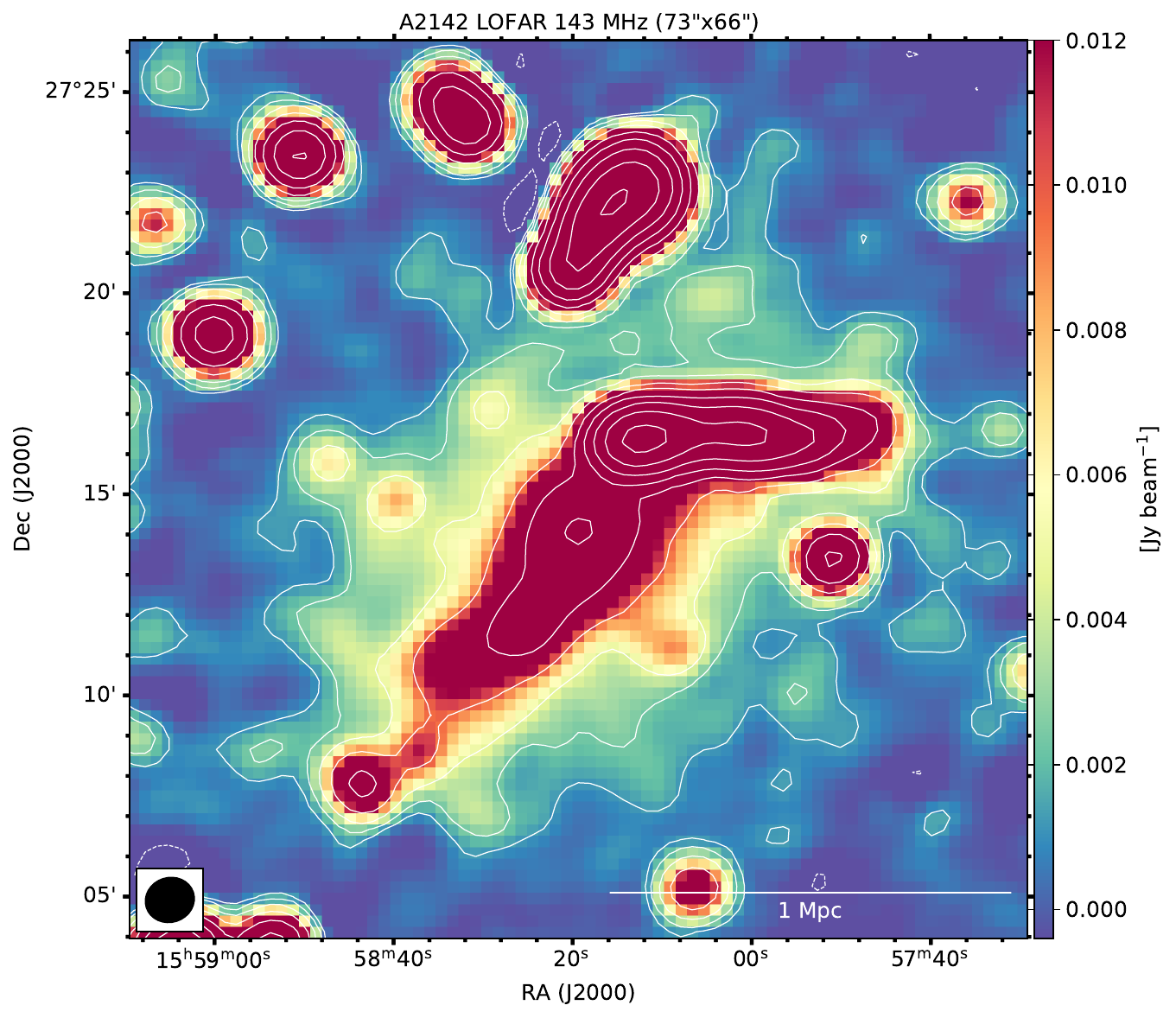}
	
	\smallskip
	
	\caption{A2142 LOFAR HBA radio images at 143 MHz at different resolutions. {\it Top left}: $9''\times6''$ resolution ($\sigma = 0.075 \; {\rm mJy \; beam^{-1}}$). {\it Top right}: $21''\times20''$ resolution ($\sigma = 0.16 \; {\rm mJy \; beam^{-1}}$). {\it Bottom left}: $38''\times35''$ resolution ($\sigma = 0.25 \; {\rm mJy \; beam^{-1}}$). {\it Bottom right}: $73''\times66''$ resolution ($\sigma = 0.40 \; {\rm mJy \; beam^{-1}}$). In all the panels, the contour levels are $[\pm3, \;6, \;12, ...]\times \sigma$.}
	\label{radiomapHBA}
\end{figure*}

\begin{figure*}
	\centering
	\includegraphics[width=0.45\textwidth]{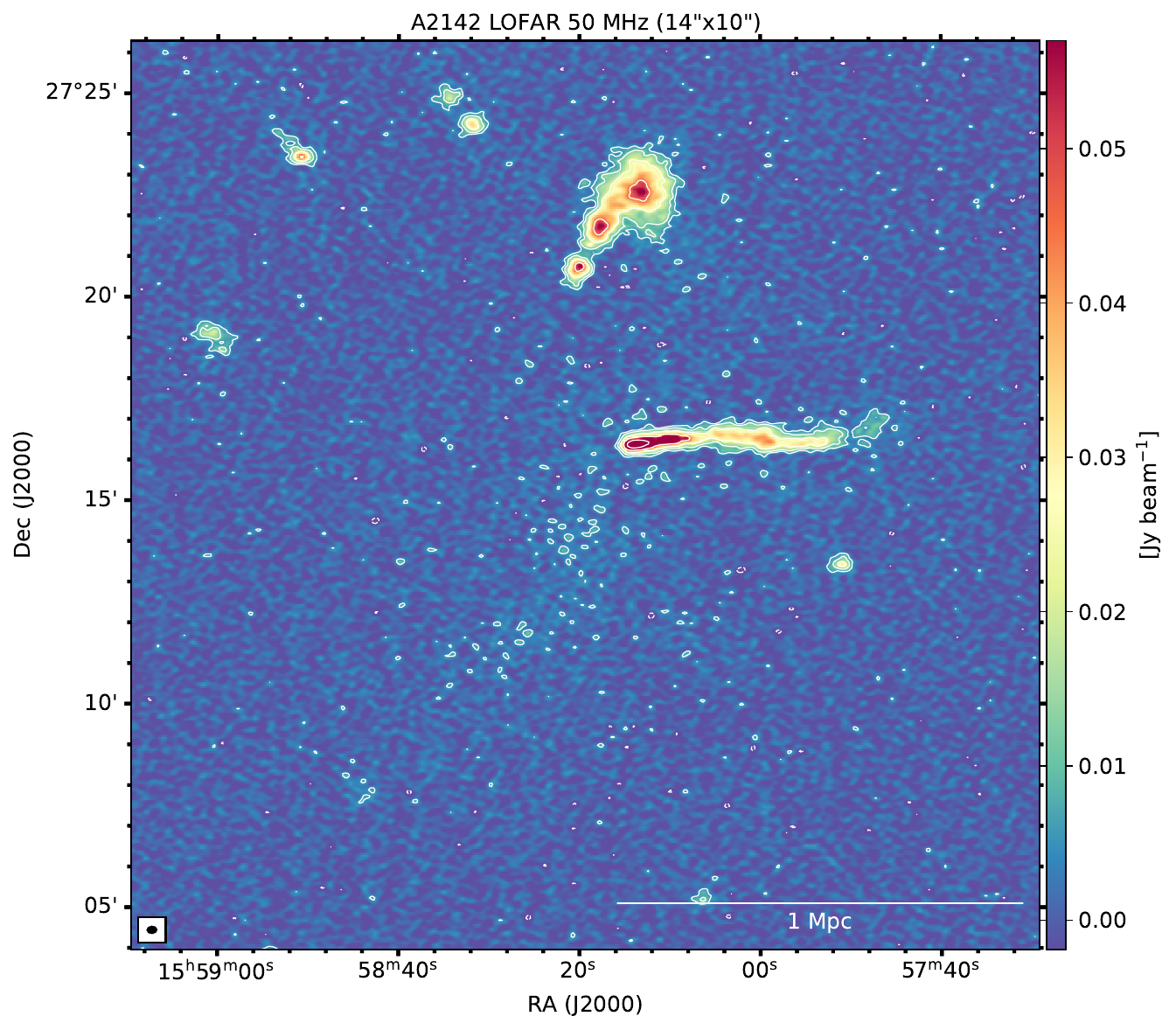}
	\includegraphics[width=0.45\textwidth]{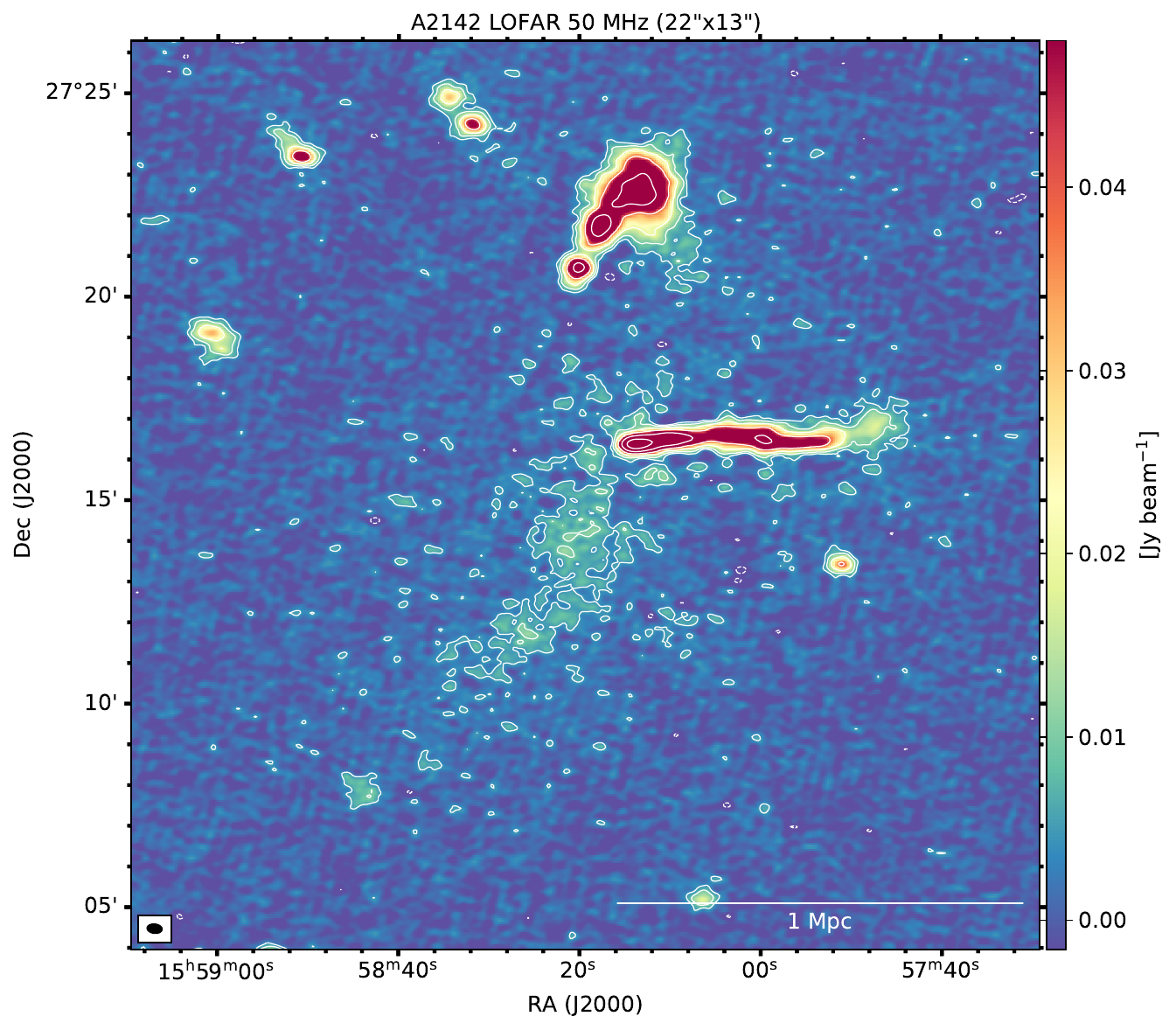}
	\includegraphics[width=0.45\textwidth]{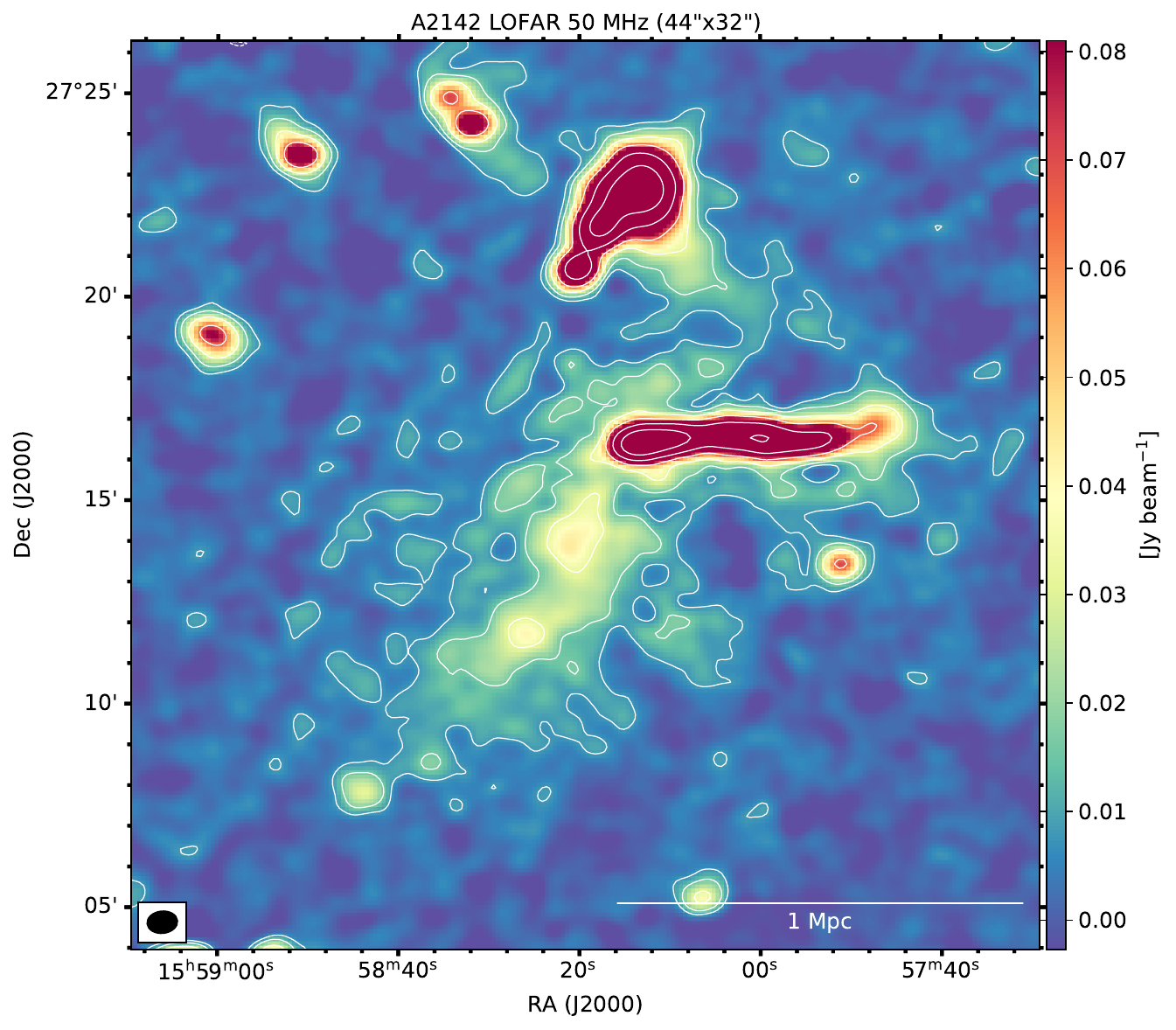}
	\includegraphics[width=0.45\textwidth]{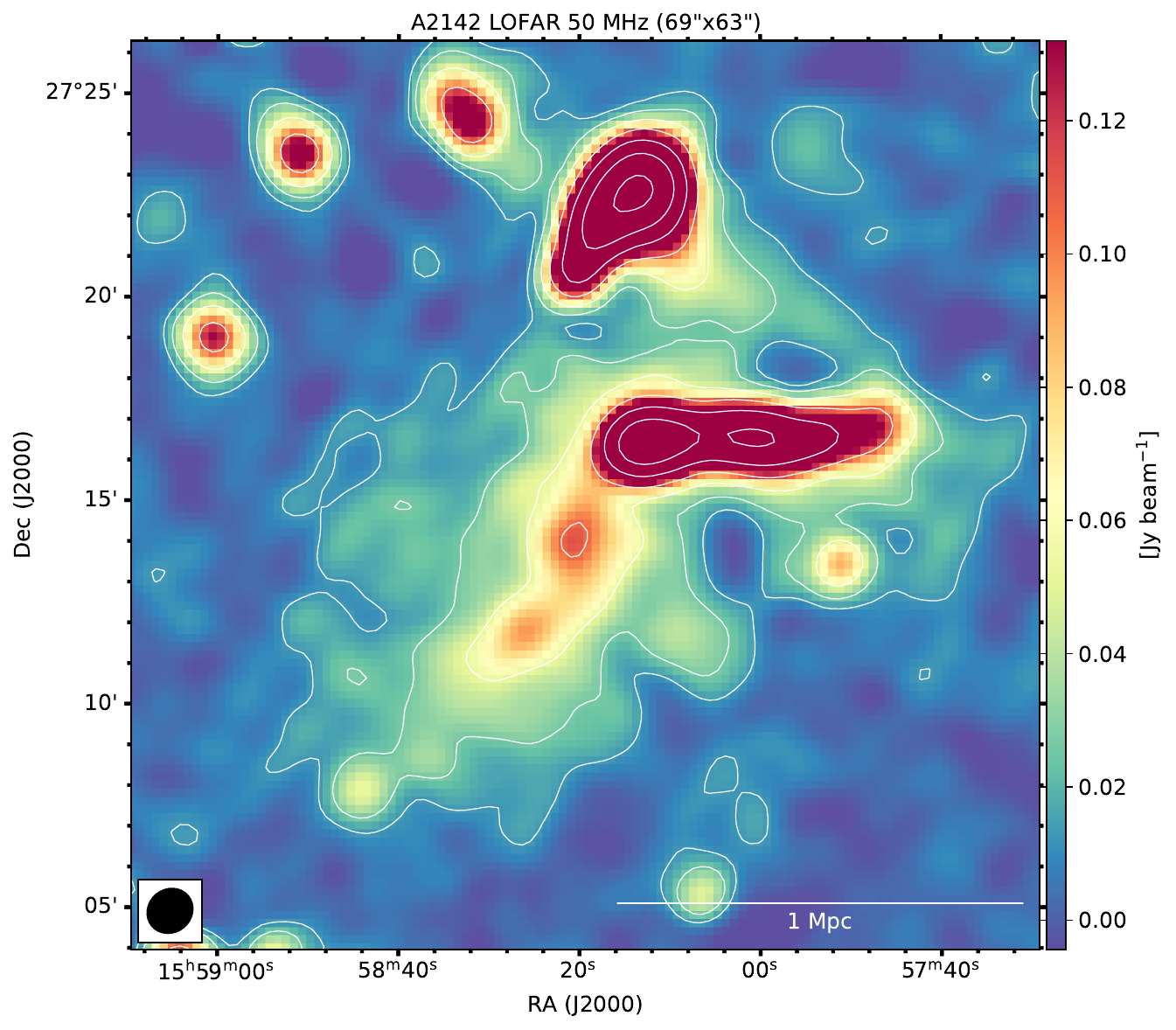}
	
	\smallskip
	
	\caption{A2142 LOFAR LBA radio images at 50 MHz at different resolutions. {\it Top left}: $14''\times10''$ resolution ($\sigma = 1.9 \; {\rm mJy \; beam^{-1}}$). {\it Top right}: $22''\times13''$ resolution ($\sigma = 1.6 \; {\rm mJy \; beam^{-1}}$). {\it Bottom left}: $44''\times32''$ resolution ($\sigma = 2.7 \; {\rm mJy \; beam^{-1}}$). {\it Bottom right}: $69''\times63''$ resolution ($\sigma = 4.4 \; {\rm mJy \; beam^{-1}}$). In all the panels, the contour levels are $[\pm3, \;6, \;12, ...]\times \sigma$.}
	\label{radiomapLBA}
\end{figure*}

\begin{figure*}
	\centering
\includegraphics[width=0.75\textwidth]{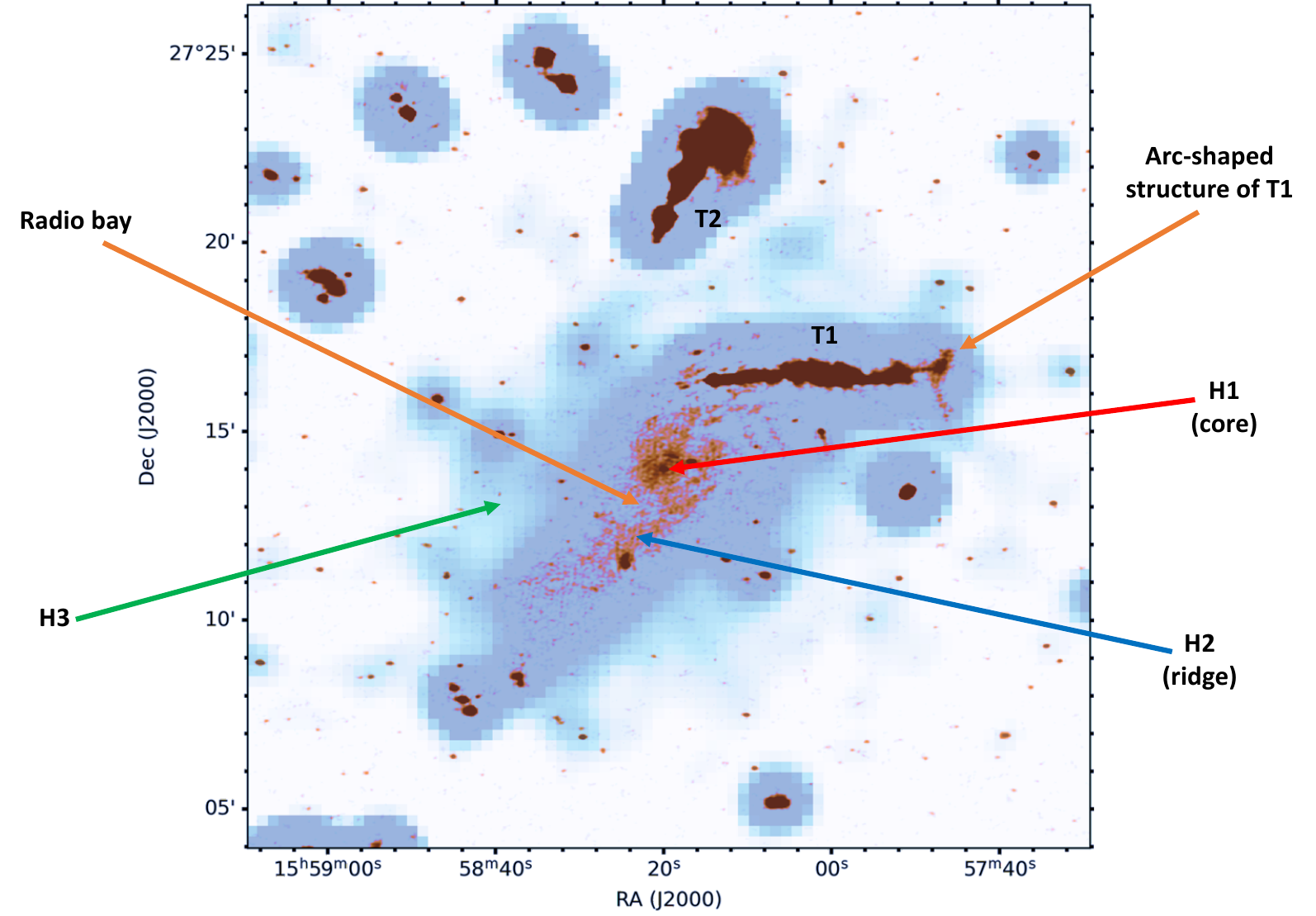}
	\smallskip
	
	\caption{Composite LOFAR HBA image of A2142 at high ($9''\times 6''$, reddish colours) and low ($73''\times66''$, blueish colours) resolution. The radio components of the halo (H1, H2, H3), radio galaxies (T1, T2), and additional features discussed in the text are labelled. }
	\label{T0+T120}
\end{figure*}   

\begin{figure*}
	\centering
	\includegraphics[width=0.96\textwidth]{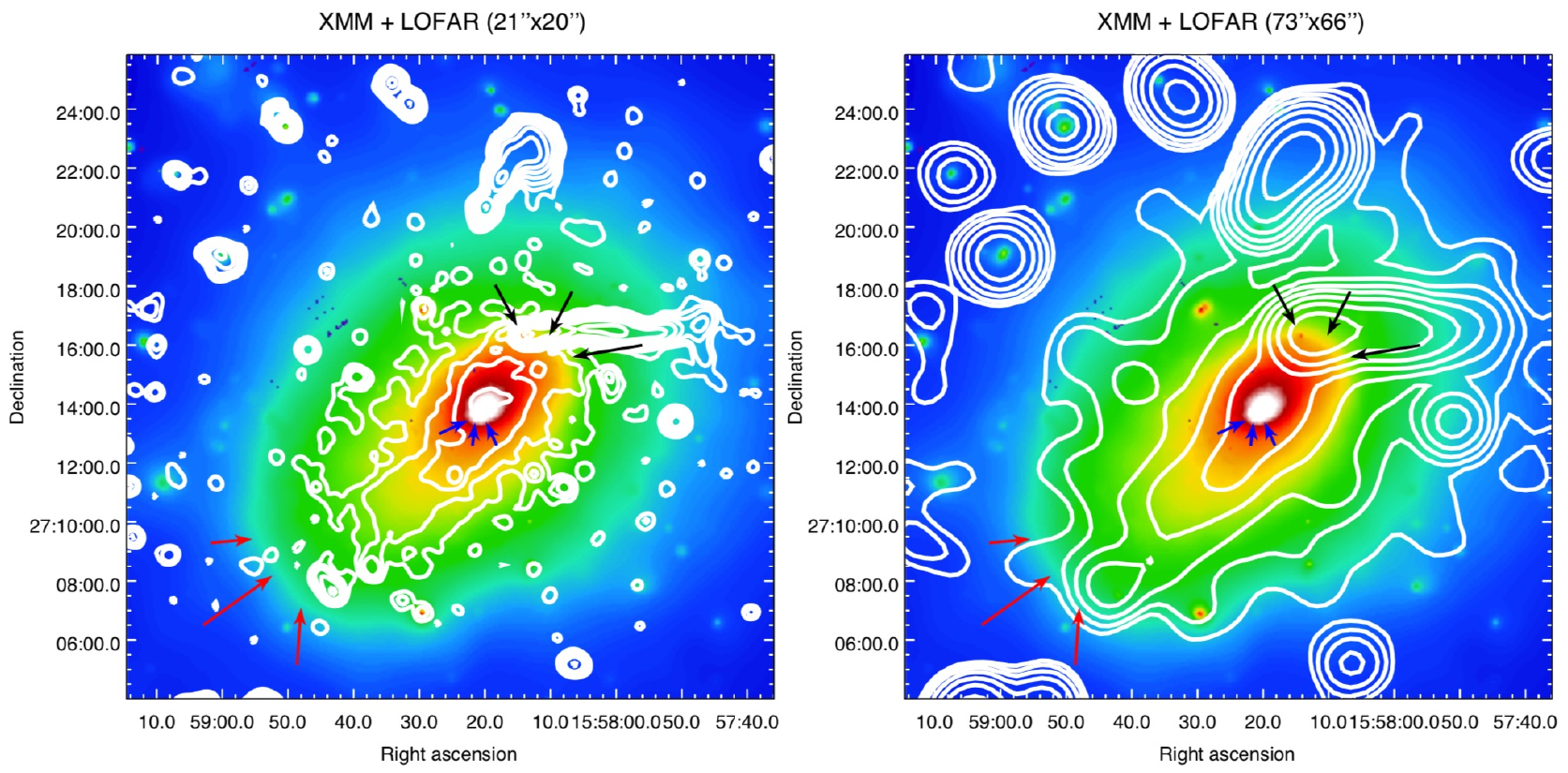}
	\smallskip
	\caption{A2142 LOFAR HBA radio contours from images in Fig. \ref{radiomapHBA} (upper right and lower right panels) overlaid on XMM-Newton X-ray image. The location of the cold fronts are indicated by arrows.}
	\label{radio+x_images}
\end{figure*}

In this Section we present the new LOFAR HBA and LBA images of A2142 shown in Figs. \ref{radiomapHBA} and \ref{radiomapLBA}, respectively. A2142 hosts a number of compact and extended radio galaxies that were discussed in \cite{venturi17}. Here we summarise the most interesting sources, which are labelled in the composite radio (GMRT), X-ray (XMM-Newton), and optical (DSS-2) image reported in Fig. \ref{ROX} as in \cite{venturi17}. The diffuse radio components of the halo are labelled in Fig. \ref{T0+T120}, where LOFAR HBA images at high and low resolutions are overlaid.  

The most spectacular extended radio galaxies are `T1' and `T2', two head-tail galaxies. The head of T1 (i.e. the core of the radio galaxy) is coincident with the NW cold front, and the tail has a projected length of $\sim 5.5'$, corresponding to $\sim 550$ kpc, if measured from the HBA at the highest resolution ($9''\times 6''$). The morphology of the tail is not straight and its width is not constant. Additional radio emission is detected by LOFAR HBA, which shows that the tail ends with a thin filament directly connected to a perpendicular arc-shape structure (see also Fig. \ref{T0+T120}), extending for $\sim 2'$ ($\sim 200$ kpc) in the NS direction. The head-tail galaxy T2 is located at a projected distance of $\sim 650$ kpc from the cluster centre, and extends north-westwards for $\sim 4'$ ($\sim 400$ kpc) in the highest resolution image. The complex morphology and features of T1 and T2 suggest an interplay with ICM motions, which will be the subject of a dedicated analysis in the future. The locations of the two BCGs are reported in Fig. \ref{ROX}. As mentioned earlier, the primary BCG is coincident with the cluster centre, whereas the secondary one is likely the main member of a merging group. Only BCG1 is radio active, hosting a compact radio galaxy. We refer to \cite{venturi17} for information on `W1' and `W2' (two wide-angle tails which are not cluster members), and galaxy groups `G' \cite[in the north-east, see also the X-ray counterpart in][]{eckert17} and `C' (in the south).

The diffuse components H1 (core) and H2 (ridge) are visible at 143 MHz at high resolution (Fig. \ref{radiomapHBA}, top left panel), whereas they are undetected at the $3\sigma$ level in the $14''\times 10''$ LBA image (Fig. \ref{radiomapLBA}, top left panel). With an intermediate resolution ($22''\times 13''$), the core and the ridge are detected at 50 MHz as well (Fig. \ref{radiomapLBA}, top right panel). By decreasing the resolution and increasing the sensitivity to the diffuse emission, further emission is revealed, as suggested by \cite{farnsworth13} with GBT observations. Here we unambiguously confirm the existence of a third radio halo component, `H3', which has an elliptical morphology elongated in the NW-SE direction, and embeds the core, the ridge, and the two head-tail galaxies T1 and T2, as shown in the $73''\times 66''$ and $69''\times 63''$ images at 143 and 50 MHz, respectively (see also Fig. \ref{T0+T120}). 

In Fig. \ref{radio+x_images}, the radio contours of the $21''\times 20''$ and $73''\times 66''$ resolution images at 143 MHz are overlaid on the XMM-Newton image of the cluster, with arrows indicating the position of the cold fronts. The core is confined by the southern central cold front, whereas the northern boundary is not clear owing to the presence of T1. The ridge extends along the NW-SE axis, towards the outermost southern cold front. The region in-between the core and the ridge, south-east of the cold front, shows a depletion of radio emission, which forms a `bay'-like structure that is labelled in Fig. \ref{T0+T120}. The bulk of the volume of the ICM is occupied by H3 and its radio emission follows the spatial distribution of the thermal X-ray emission, as typical of giant radio halos, reaching the outermost cold front at south. We measured the maximum extension of H3 by considering the $2\sigma$ level of a LOFAR HBA image at resolution $128''\times 117''$; we can approximate its morphology as an ellipse of projected axis lengths of $2.4 \; {\rm Mpc} \times 2.0$ Mpc.

The detection of H3 makes A2142 to be the only hybrid radio halo with three distinct components discovered so far. In the next sections we will analyse the properties of the radio halo in details with multi-frequency data to investigate the origins and possible connections between its components.

\subsection{Spectral indices}
\label{sect: Spectral indices}

\begin{table}
 \fontsize{8.5}{8.5}\selectfont
\centering
	\caption[]{Sets of images used for flux density measurements. Col. 1: Frequency of considered datasets. Cols. 2, 3: Adopted \textit{uv}-range and Gaussian taper. Col. 4: Circular beam adopted to convolve the images.}
	\label{table: imaging_for_flux}   
	\begin{tabular}{ccccc}
	\hline
	\noalign{\smallskip}
	$\nu$ & \textit{uv}-range & Taper & $\theta_{\rm conv}$   \\  
	(MHz) &  & (arcsec) & (arcsec)     \\  
    \hline
	\noalign{\smallskip}
    50, 143, 323, 407, 1380, 1810 & $[250\lambda - 15{\rm k}\lambda]$ & - & 25 \\
    50, 143, 323, 407 & $[60\lambda - 18{\rm k}\lambda]$ & 15 & 25 \\
    50, 143, 323, 407 & $[60\lambda - 18{\rm k}\lambda]$ & 60 & 85 \\
    50, 143 & $[50\lambda - 18{\rm k}\lambda]$ & 120 & 134 \\
\noalign{\smallskip}
	\hline
	\end{tabular}  
%	\begin{tablenotes}
%\item    {\small \textbf{Notes}.} 
% \end{tablenotes}	
\end{table}

\begin{table*}
\centering
	\caption[]{Flux densities of H1, H2, and H3 as derived from the regions shown in Fig. \ref{spettriradio}. Column 8 reports the fitted spectral index.}
	\label{fluxtabsource}   
	\begin{tabular}{ccccccccc}
	\hline
	\noalign{\smallskip}
	Source & Region & $S_{50}$ & $S_{143}$ & $S_{323}$ & $S_{407}$ & $S_{1380}$ & $S_{1810}$ & $\alpha$ 
	\\  
	& & (mJy) & (mJy) & (mJy) & (mJy) & (mJy) & (mJy) &   \\  
    \hline
	\noalign{\smallskip}
	H1 & red & 266.4 $\; \pm \;27.6$ & $89.1\; \pm \;8.9$ & $33.6\; \pm \;2.1$ & $27.6\; \pm \;1.8$  & $7.5\; \pm \;0.4$ & $5.1\; \pm \;0.4$ & $1.09\; \pm \;0.02$  \\
	H2 & blue & 414.4$\; \pm \;43.0$ & $117.8\; \pm \;11.8$ & $47.5\; \pm \;2.9$ & $37.0\; \pm \;2.4$ & - & - & $1.15\; \pm \;0.02$  \\
	H3 & green & 360.8$\; \pm \;41.9$ & $73.7\; \pm \;7.6$ & $27.0\; \pm \;2.5$ & $20.3\; \pm \;2.0$ & - & - & $1.36\; \pm \;0.05$  \\
	H3 & purple & 169.3$\; \pm \;29.0$ & $32.5\; \pm \;3.9$ & - & -  & - & - & $1.57 \pm 0.20 $  \\
\noalign{\smallskip}
	\hline
	\end{tabular}  
	\begin{tablenotes}
\item    {\small \textbf{Notes}. The regions defined for H1 and H2 roughly cover their total extensions and flux densities are thus total values. Flux densities of H3 are instead local values, obtained as the sum of the measurements within the considered boxes. Uncertainties on $\alpha$ for H1, H2, and inner H3 are fitting errors; the uncertainty on $\alpha$ for outer H3 is derived from the standard formula for the error propagation as in Eq. \ref{eq:spectralindexerrorformula}. } 
 \end{tablenotes}	
\end{table*}

\begin{figure*}
	\centering

\includegraphics[width=0.4\textwidth]{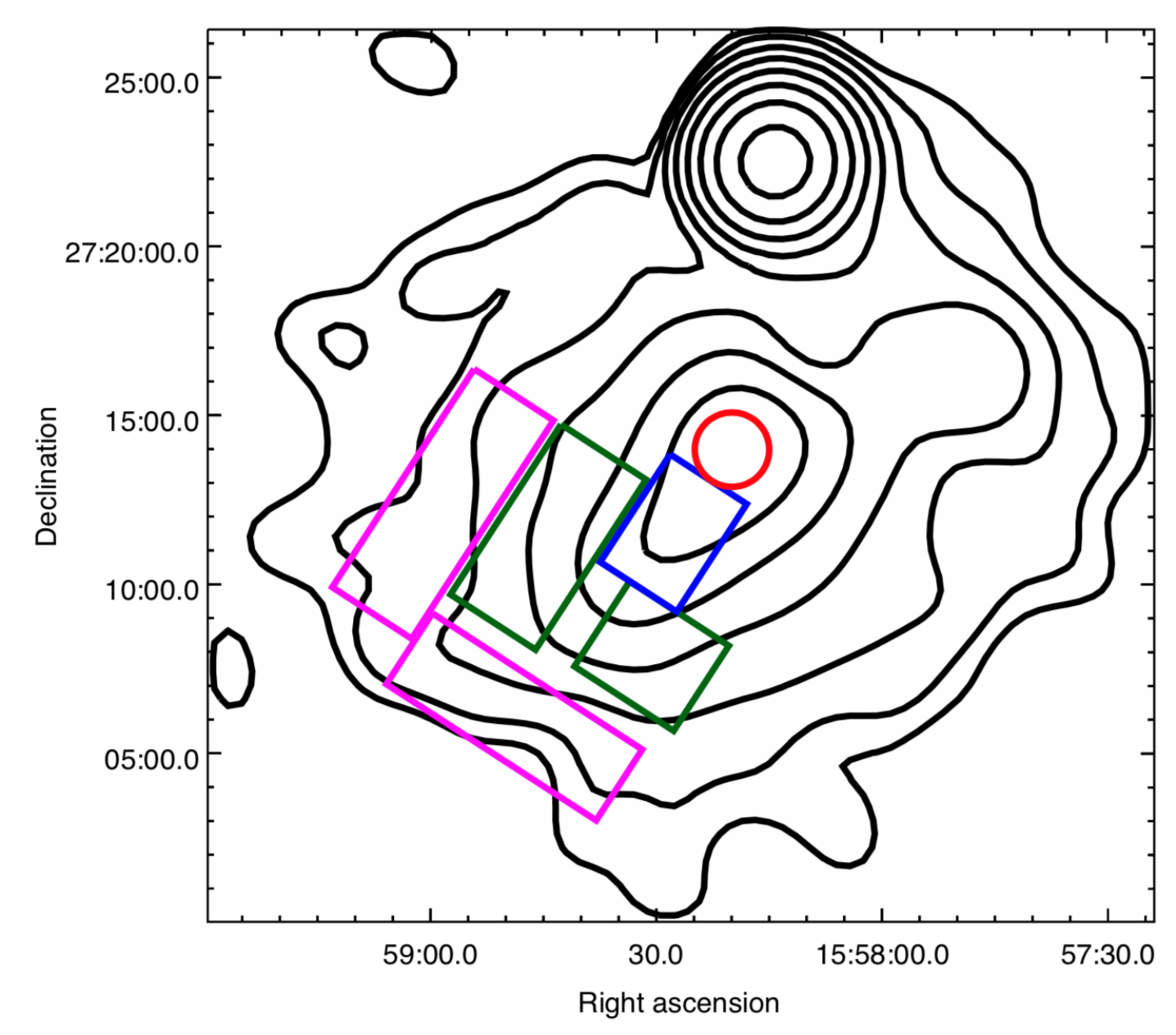}	
\includegraphics[width=0.5\textwidth]{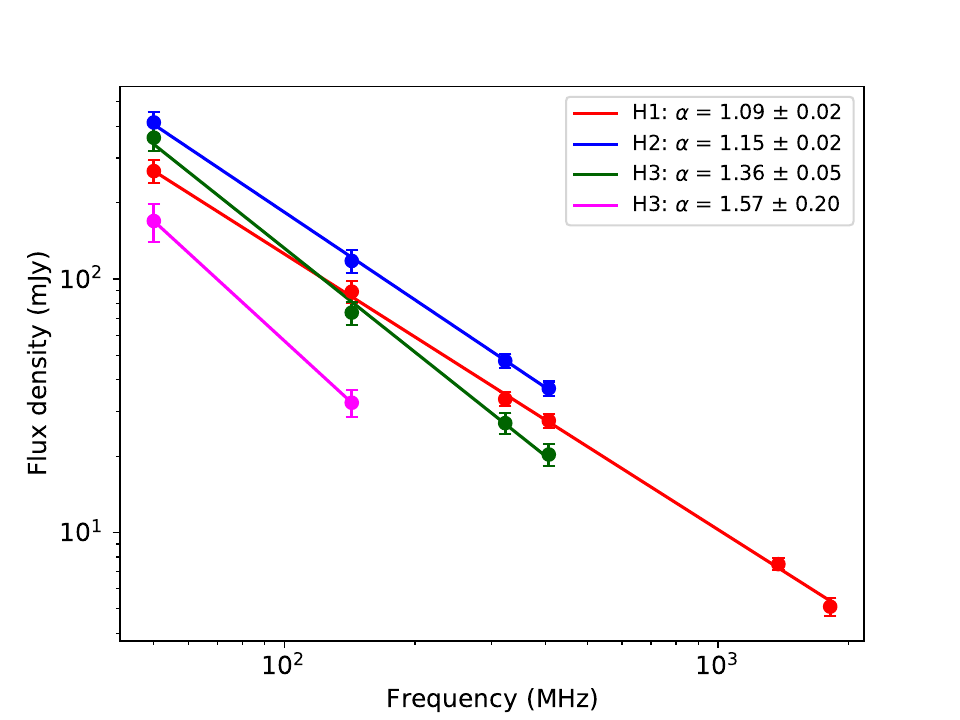}	
	\smallskip
	
	\caption{Radio spectrum of A2142. \textit{Left panel}: Regions used to measure the flux densities of H1, H2, and H3 from various images overlaid on the 143 MHz contours ([$2,\; 4,\; 8,\; ...$]$\times \sigma$) of the $134''\times 134''$ image (see details in the text). \textit{Right panel}:  Radio spectra of H1, H2, H3 fitted with power-laws.    } 
	\label{spettriradio}
\end{figure*}

\begin{figure*}
	\centering
	\includegraphics[width=0.45\textwidth]{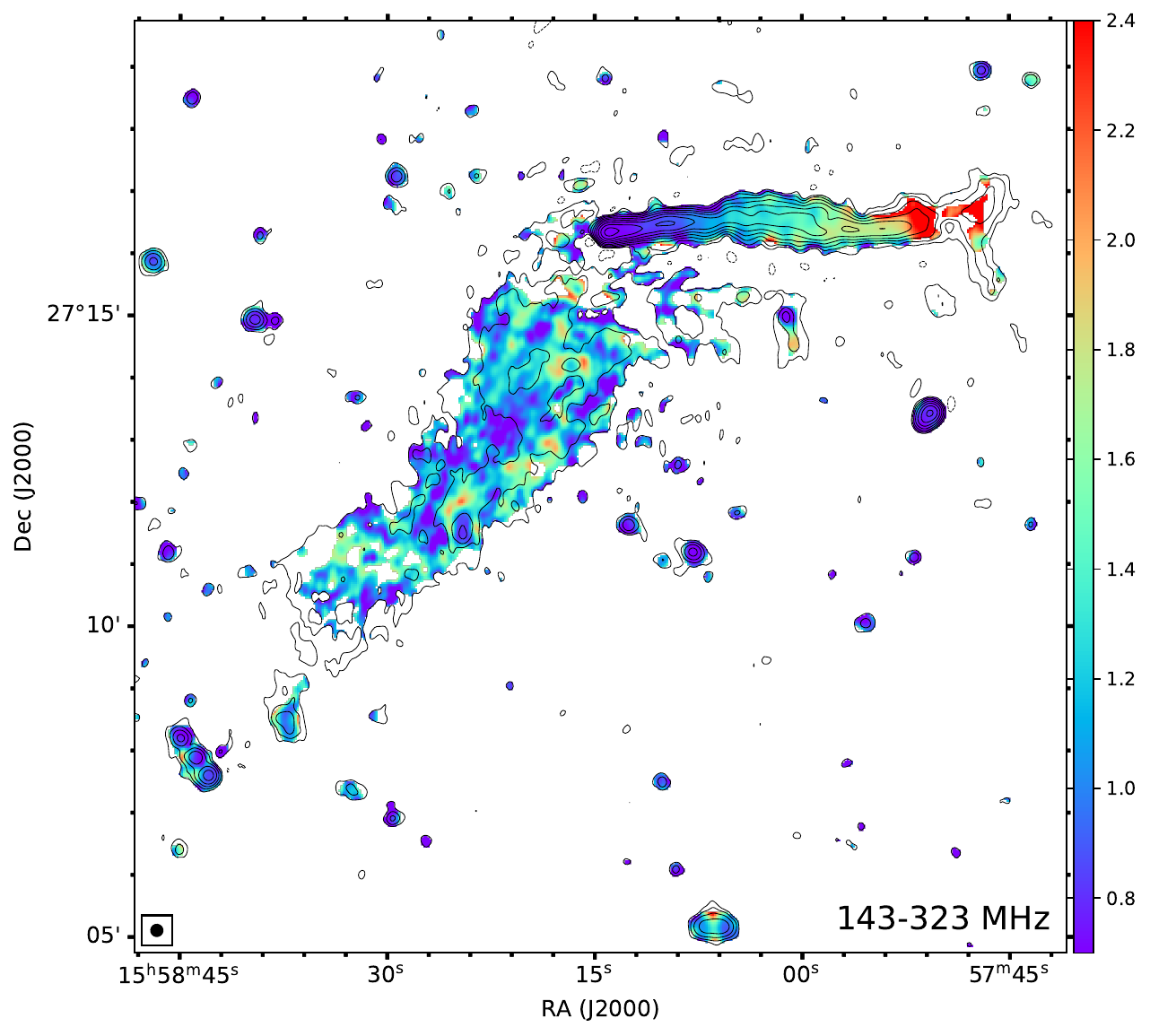}
     \includegraphics[width=0.45\textwidth]{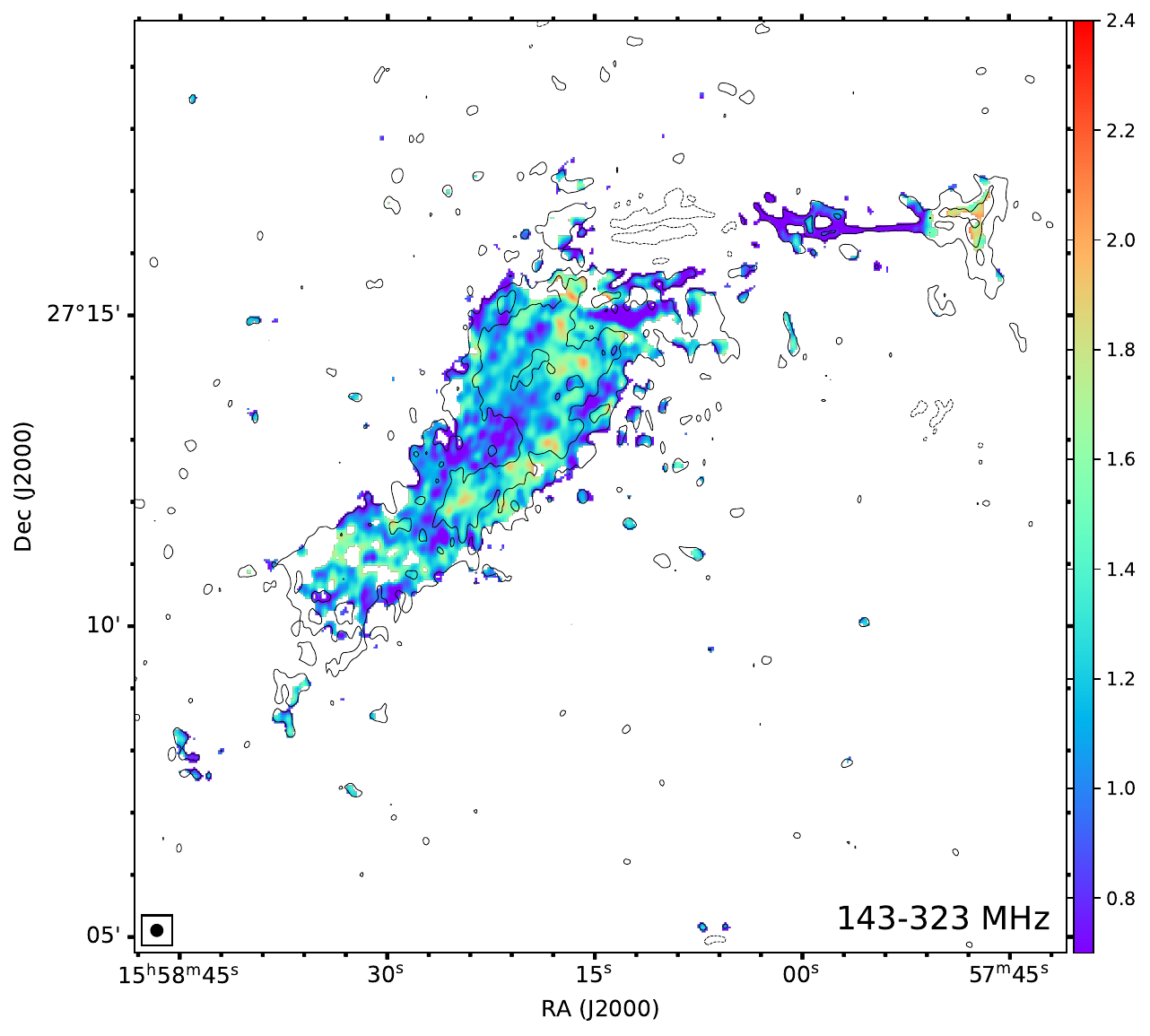}
	\smallskip
	
	\caption{A2142 spectral index maps between 143 and 323 MHz at $11''\times11''$ resolution, before (left) and after (right) the subtraction of the discrete sources. LOFAR contours are drawn in black. } 
	\label{SPIX_lofar-gmrt}
\end{figure*}

\begin{figure*}
	\centering
	\includegraphics[width=0.45\textwidth]{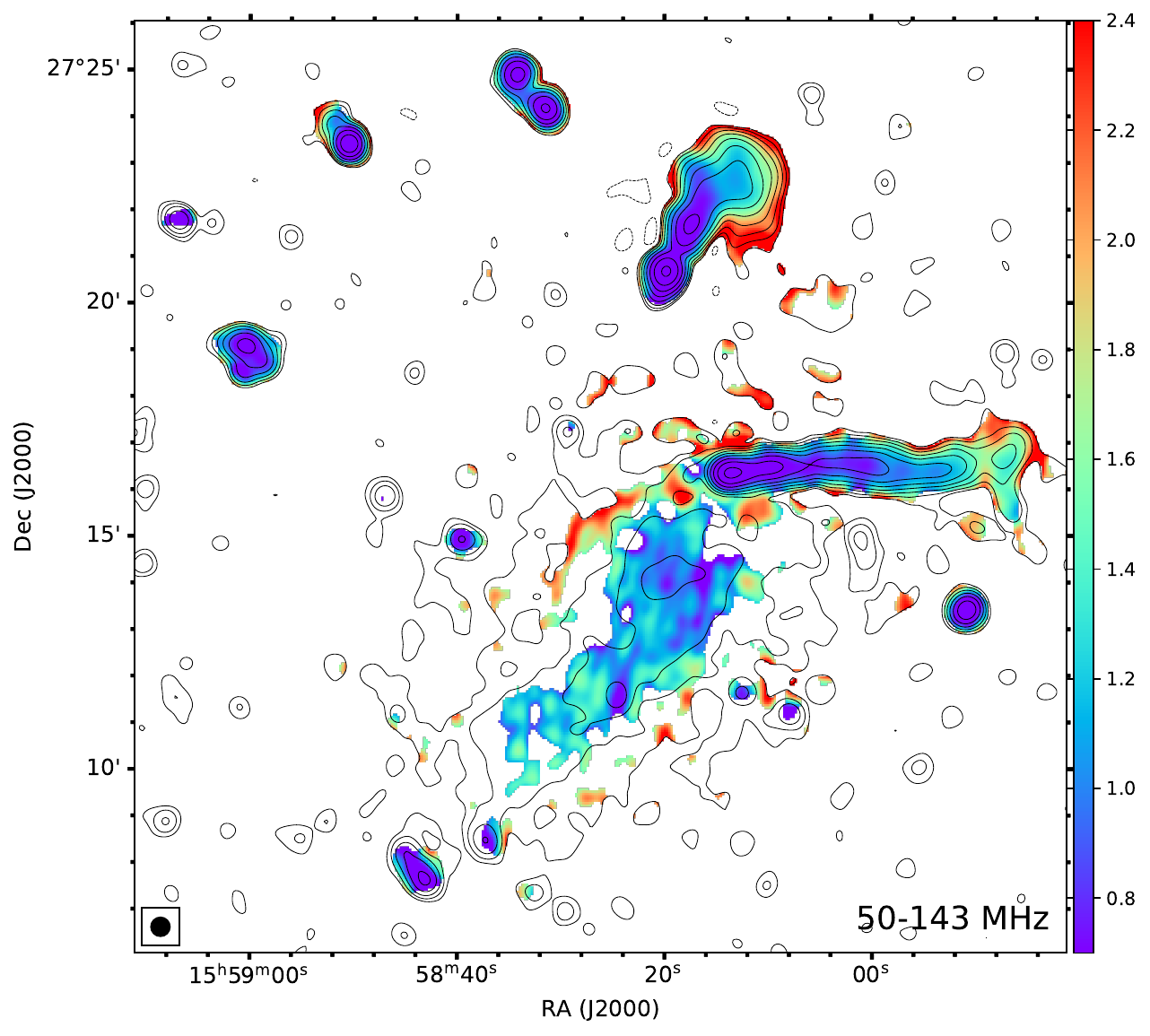}
    \includegraphics[width=0.45\textwidth]{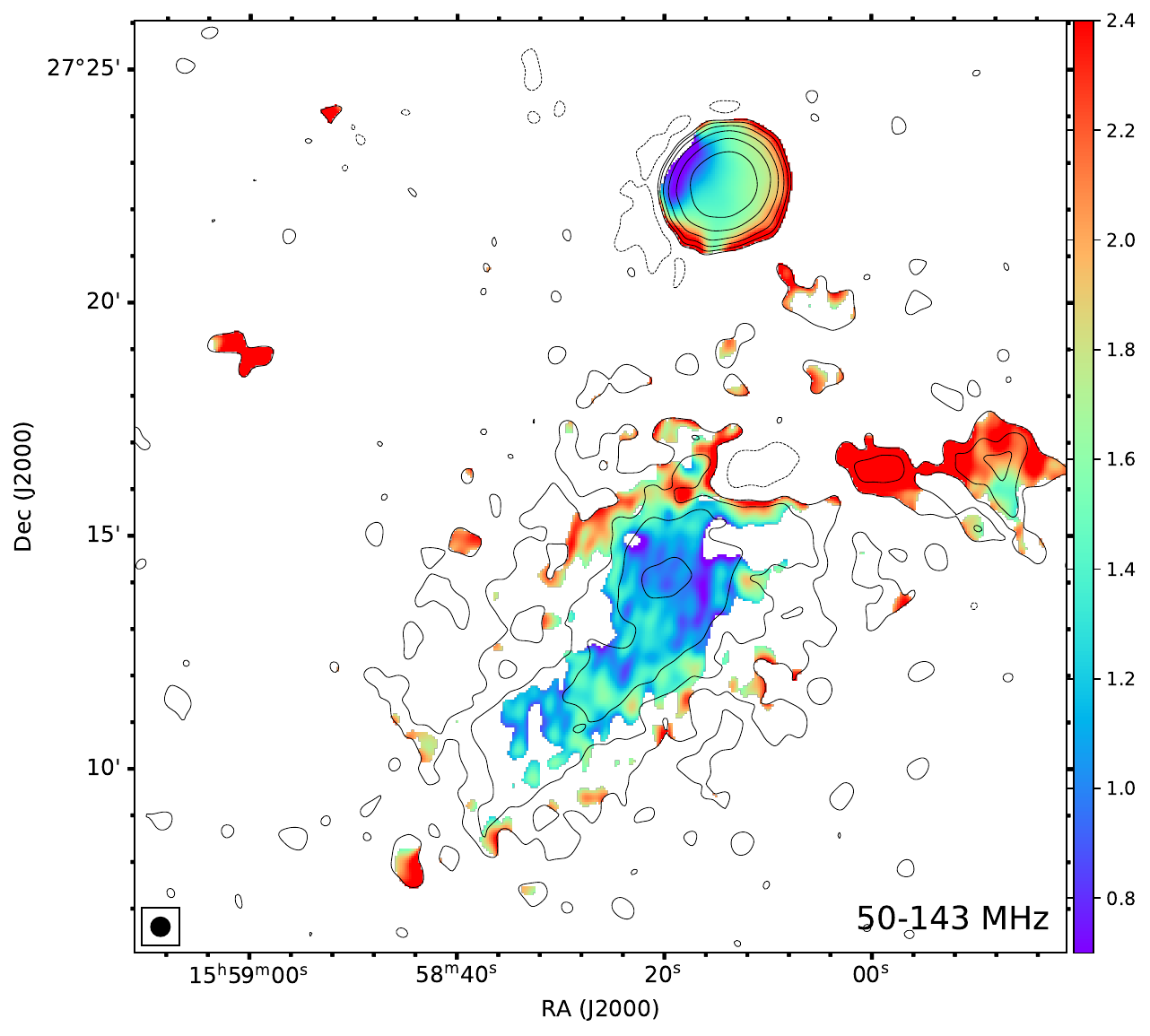}

\includegraphics[width=0.45\textwidth]{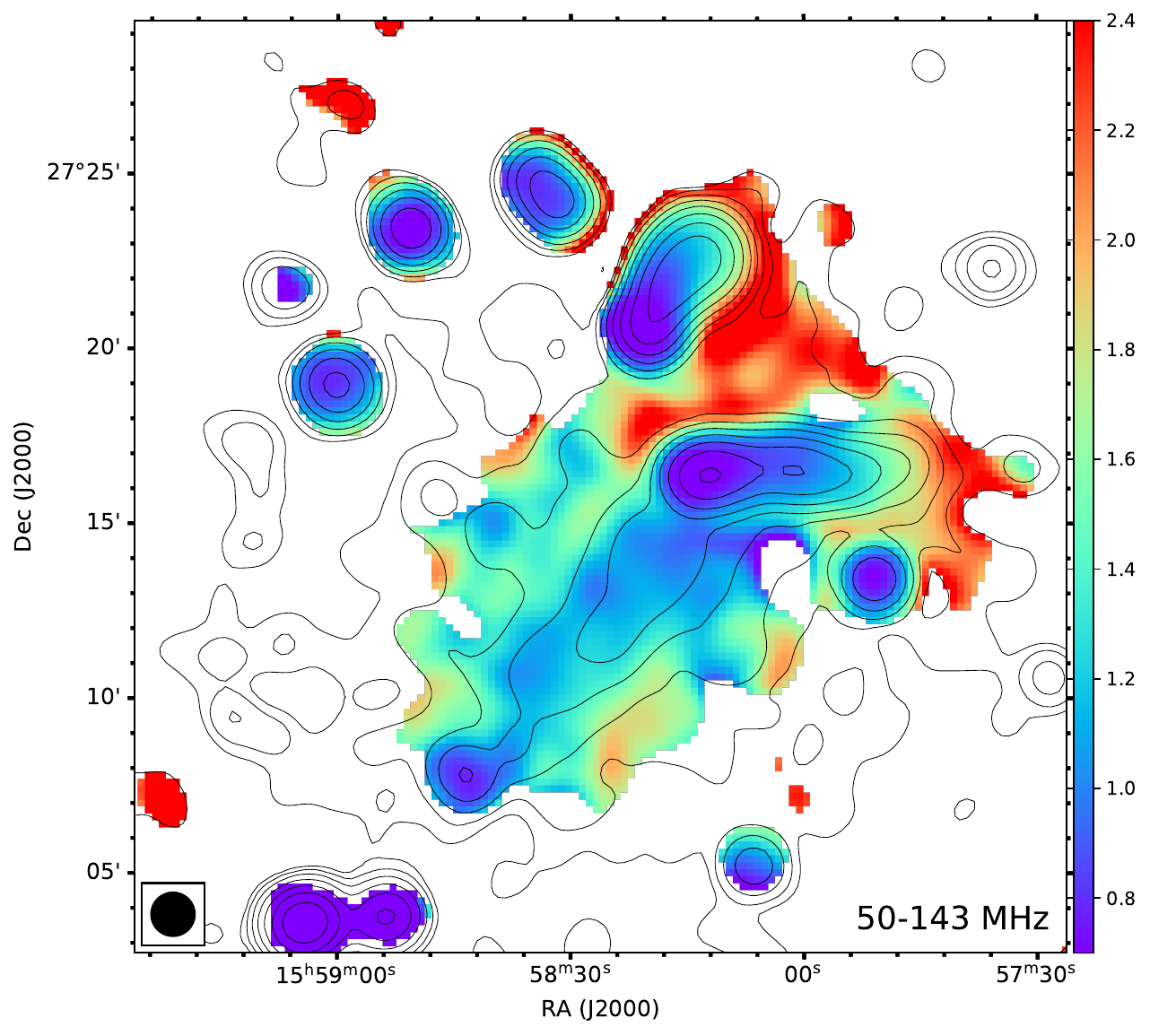}
\includegraphics[width=0.45\textwidth]{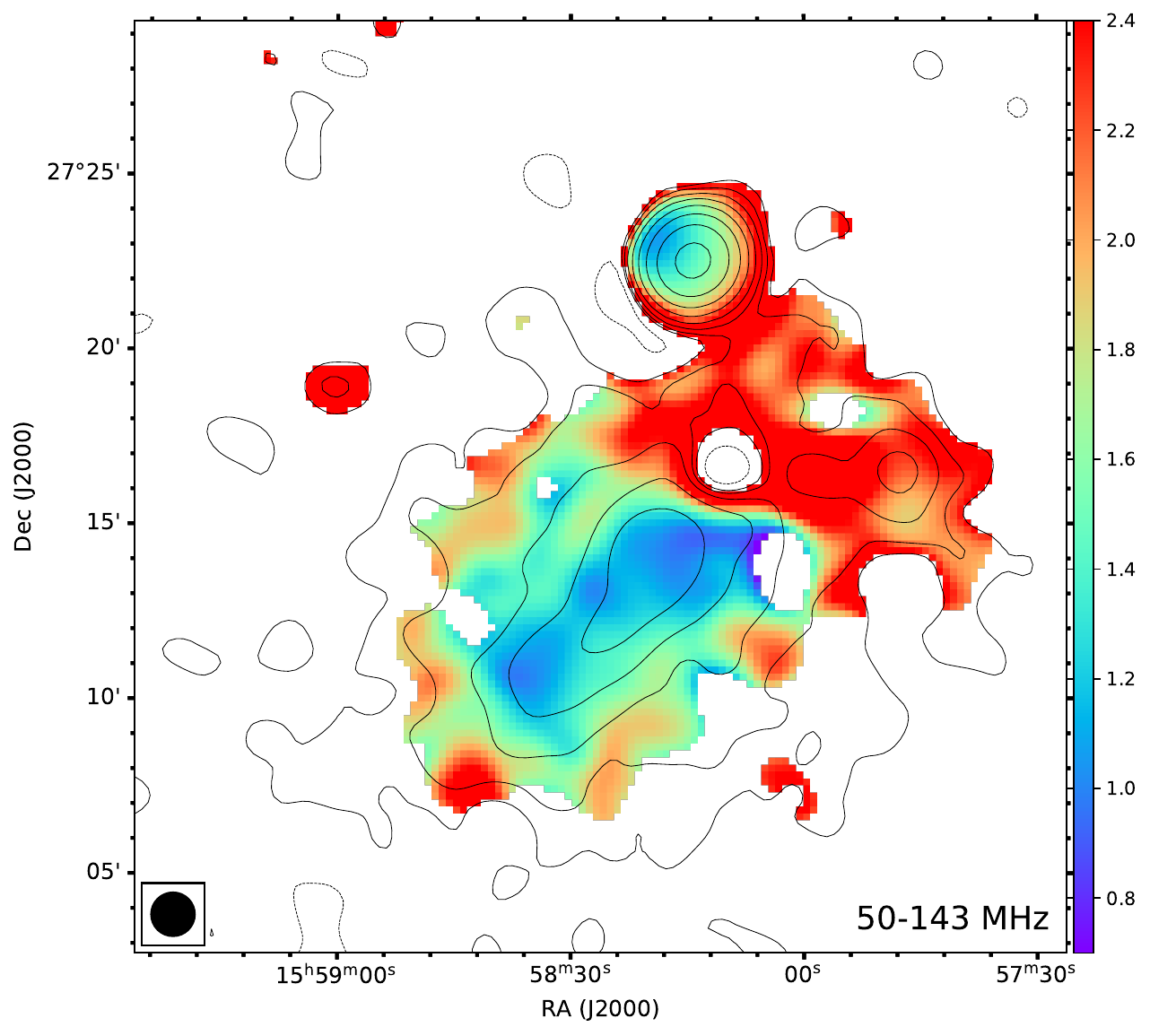}
\includegraphics[width=0.45\textwidth]{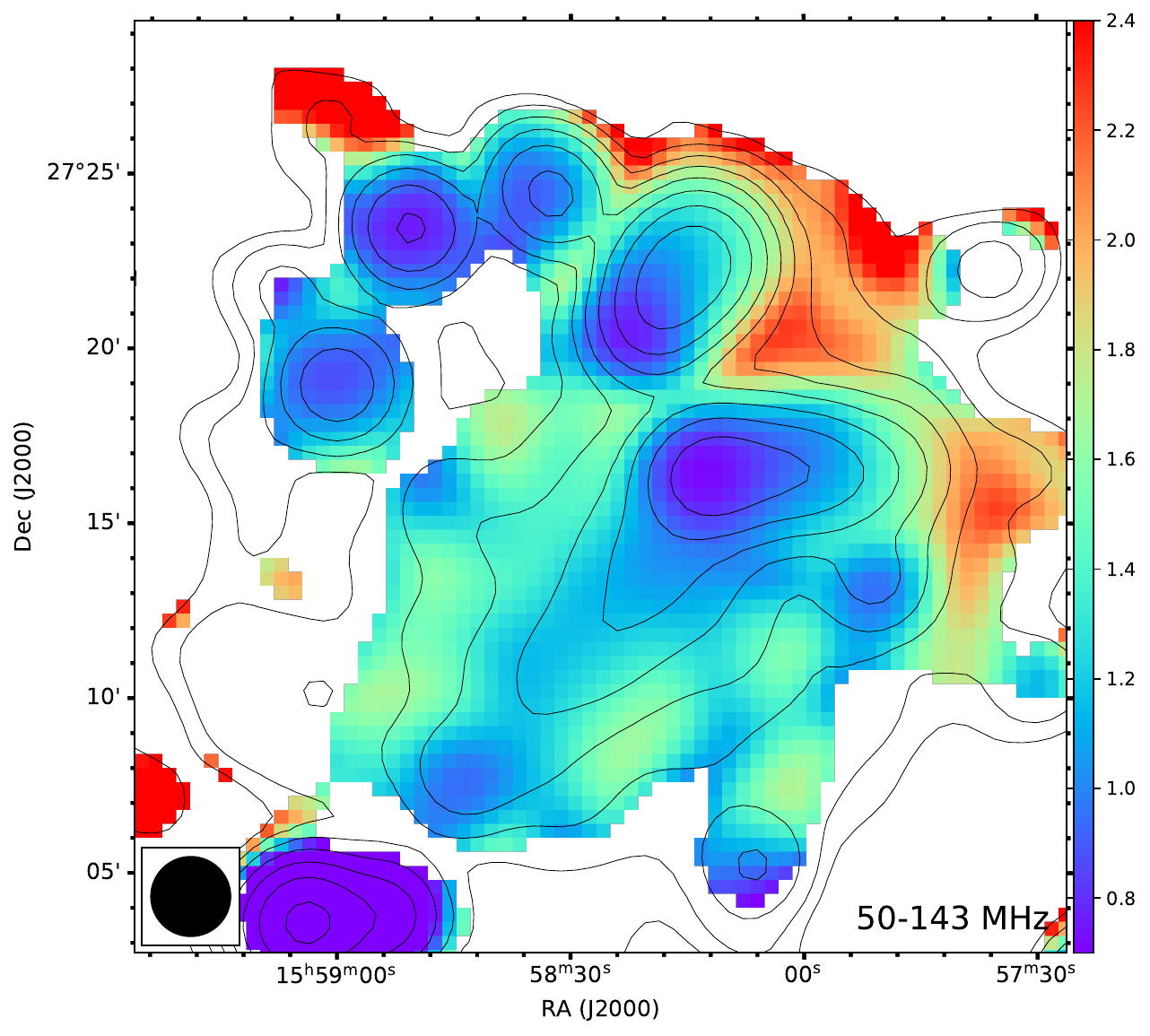}
\includegraphics[width=0.45\textwidth]{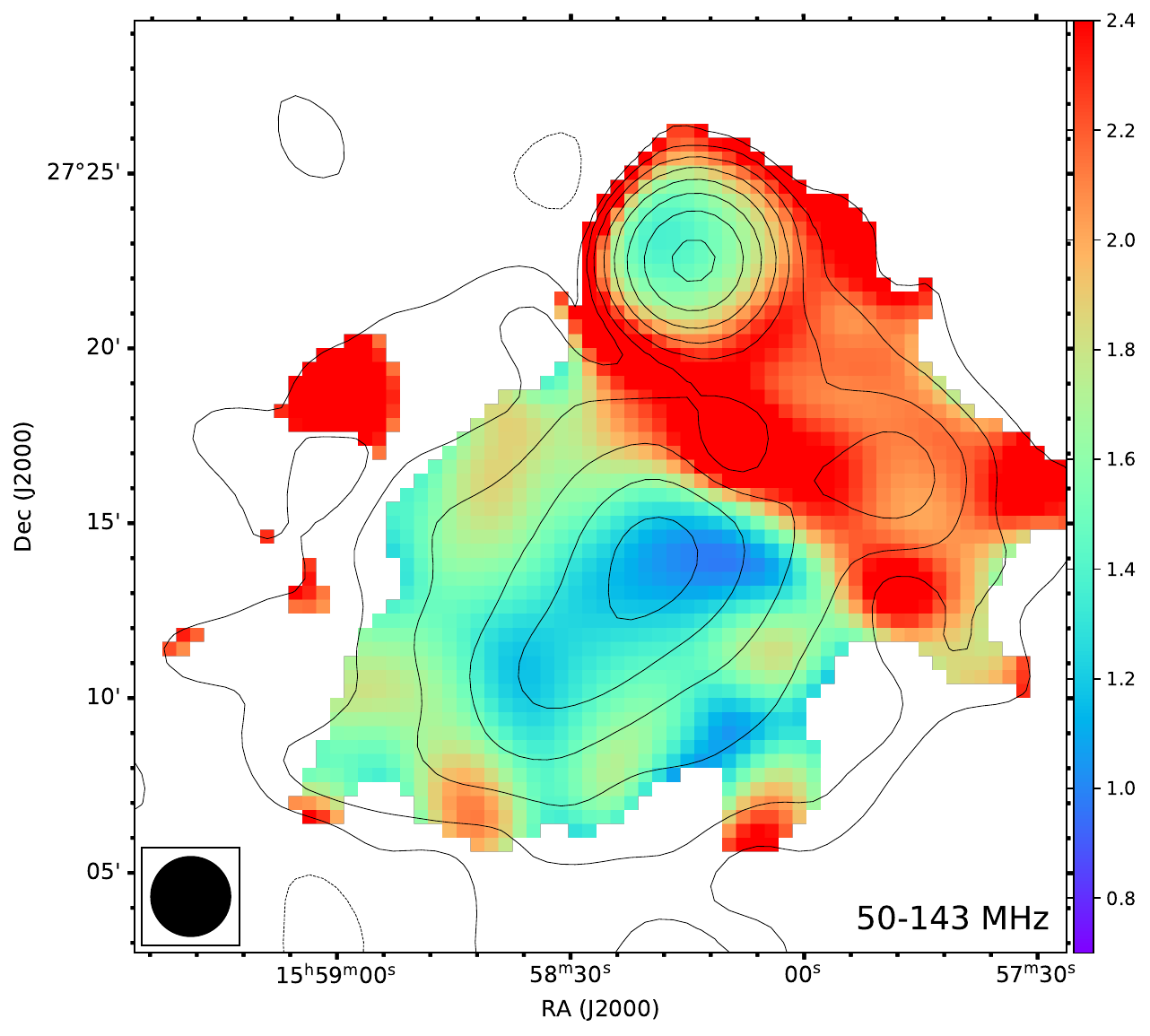}

	\smallskip
	
	\caption{A2142 spectral index maps between 50 and 143 MHz at various resolutions before (left panels) and after (right panels) the subtraction of the discrete sources. \textit{Top}: maps at $24''\times24''$ resolution. \textit{Middle}: maps at $75''\times75''$ resolution. \textit{Bottom}: maps at $136''\times136''$ resolution. In all the panels, the corresponding contours at 143 MHz are drawn in black. }
	\label{SPIX_lofar}
\end{figure*}

The spectral index is a key to probing the mechanisms responsible for the origin of radio halos. Owing to the extremely complex structure of A2142, a straightforward measure of $\alpha$ is not trivial, thus requiring different methods to check for self-consistency of our results. First, to accurately determine the spectral index of the various components, we produced sets of images at different resolutions with common \textit{uv}-ranges that allow us to consistently measure the flux densities and minimise systematics introduced by inhomogeneous \textit{uv}-coverage of the different instruments (see details on the imaging parameters in Table \ref{table: imaging_for_flux}).

The first set of images at $25''$ spans a frequency range from 50 to 1810 MHz. These provide sufficient resolution and sensitivity to measure the flux density of H1. As the northern limit of H1 cannot be easily determined, we considered the core as a sphere centred on BCG1 and confined at south by the inner cold fronts, and thus we measured the flux density in a circle (red circle in Fig. \ref{spettriradio}) of radius $R_{\rm H1}=1.1'\sim 110$ kpc. 

The \textit{uv}-coverage of VLA is less dense at short spacings than those of LOFAR and GMRT. To avoid flux density losses due to missing short baselines that would bias the spectral index towards steeper values, we excluded the VLA data from the measures in H2. As a compromise between resolution and sensitivity, we considered again images convolved to $25''$. Similarly to \cite{venturi17}, we considered a box (blue box in Fig. \ref{spettriradio}) of sides $3.8''\times2.7''$ (corresponding to $380 \; {\rm kpc} \times 270$ kpc) and $P.A.=33^{\rm o}$ that roughly encompasses the $3\sigma$ level of the ridge.

A third set of images from 50 to 407 MHz was obtained with a resolution of $85''$. At such low resolution, residuals from the discrete source subtraction are enhanced and have a non-negligible contribution when compared with the faint diffuse emission. As explained in \ref{Sect: Radio imaging and source subtraction}, obtaining quantitative estimates on the quality of the subtraction of each dataset is not trivial. Therefore, to avoid the contamination from residual of discrete sources as much as possible, we measured the local flux densities of H3 in two boxes (green boxes in Fig. \ref{spettriradio}) of sides $6'\times 3'$ ($600 \; {\rm kpc} \times300$ kpc) and $3.5'\times 3'$ ($350 \; {\rm kpc} \times 300$ kpc), where the halo is detected at $3\sigma$ at each frequency. 

Due to insufficient sensitivity at 323 and 407 MHz, we are able to recover the outer emission of H3 only with LOFAR. We thus obtained images at 50 and 143 MHz, and convolved them to $134''$ to further enhance the ${\rm S/N}$. We considered two boxes (purple boxes in Fig. \ref{spettriradio}) of sides $7.7'\times 2.8'$ ($770 \; {\rm kpc} \times280$ kpc) and $7.4'\times 2.5'$ ($740 \; {\rm kpc} \times 250$ kpc) free from source contamination, where the halo is detected at the $2\sigma$ level of both images.

All the regions defined as above are overlaid on the contours of the $134''\times134''$ HBA image of Fig. \ref{spettriradio}. We used the measured flux densities reported in Table \ref{fluxtabsource} to fit the radio spectra with power-laws that well described the data points, as shown in Fig. \ref{spettriradio}. The fitted spectral indices are $\alpha_{\rm H1}^{[50-1810]}=1.09 \pm 0.02$ for H1, $\alpha_{\rm H2}^{[50-323]}=1.15 \pm 0.02$ for H2, $\alpha_{\rm H3}^{[50-323]}=1.36 \pm 0.05$ in the inner parts (green boxes) of H3, and $\alpha_{\rm H3}^{[50-143]}=1.57\pm 0.20$ in the outer parts (purple boxes) of H3 (see also the derivation of $\alpha$ of H3 from the total flux density in Sect. \ref{sect: Spectral index of H3}).

In Fig. \ref{SPIX_lofar-gmrt} we present the spectral index maps (the corresponding error maps are shown in Fig. \ref{errSPIX_lofar-gmrt}) at $11''\times11''$ between 143 and 323 MHz (with and without discrete sources). The spectral index maps between 50 and 143 MHz, at resolutions $24''$, $75''$, and $136''$ are shown in Fig. \ref{SPIX_lofar} (the corresponding errors maps are shown in Fig. \ref{errSPIX_lofar}). The spectral indices inferred from these maps are globally consistent with the fitted spectra. The high resolution 143-323 MHz spectral index maps (Fig. \ref{SPIX_lofar-gmrt}) show that the spectral index distribution, especially along the ridge, is not uniform; interestingly, a significant flattening ($\alpha\sim 0.9$) is observed in the region coincident with the radio bay. Outside H2 (Fig. \ref{SPIX_lofar}), the spectrum further steepens up to $\sim 1.6$. In the northern part, unreliable spectral indices $\alpha \gtrsim 2$ are the result of subtraction artefacts from T1 and T2. On the other hand, it is plausible that the tails of these galaxies release very old populations of electrons that are then re-accelerated by turbulence, and contribute to the steepest regions of the halo in the north, as suggested by the non-source-subtracted  spectral index maps.  

We point out that our integrated spectra are flatter than $\alpha\sim 1.3$ and $\alpha\sim 1.5$, as reported by \cite{venturi17} for H1 and H2, respectively. Discrepancies are likely caused by a combination of several factors. Indeed, we considered a smaller region for H1 to avoid possible contamination from H2 and/or residuals of subtraction from T1, improved the calibration of the GMRT data at 323 MHz, have a more accurate LOFAR HBA flux density value (for instance, uncertainties in \cite{venturi17} are $\sim 35\%$ for the LOFAR measurement), and considered only the datasets with the densest \textit{uv}-coverage. Through our procedures, we obtained less scattered data-points to be fitted than those reported by \cite{venturi17}.

\subsection{Radio surface brightness profile}
\label{sect: Spectral index of H3}

\begin{figure}
	\centering
	\includegraphics[width=0.4\textwidth]{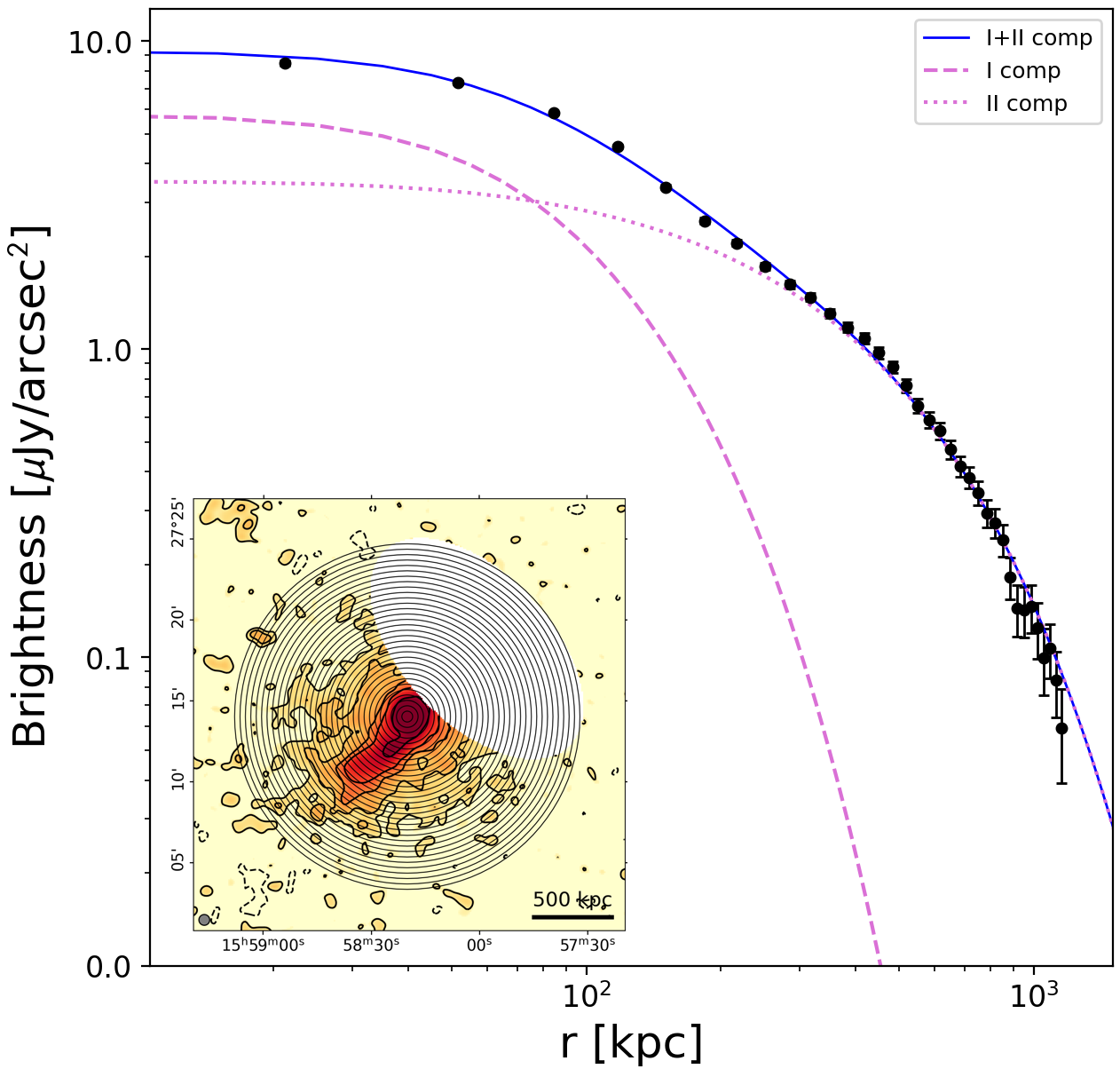}
	\includegraphics[width=0.4\textwidth]{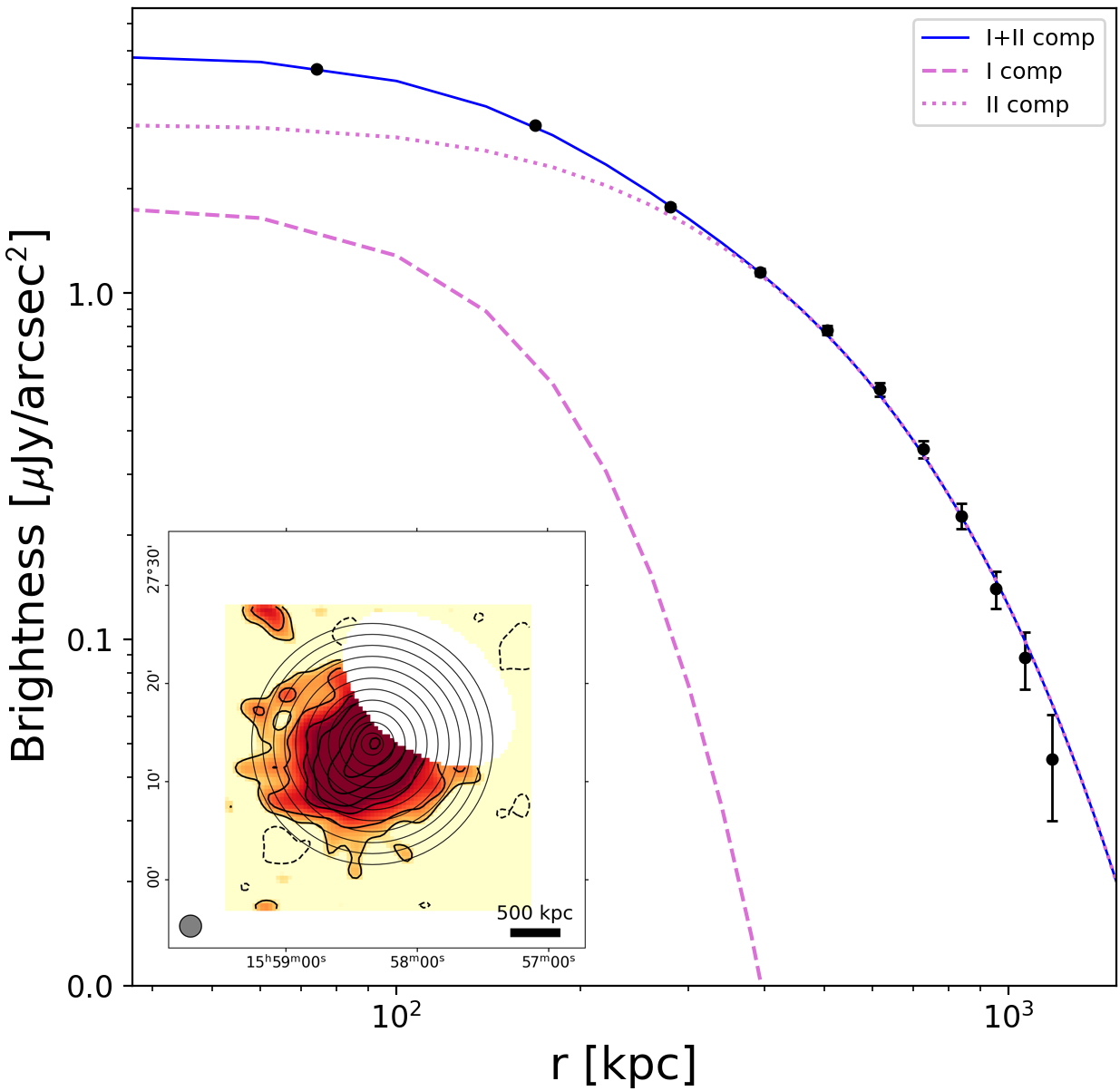}
	\smallskip
	\caption{Fit of the surface brightness of LOFAR HBA images at $40''\times40''$ (top, see inset) and $134''\times134''$ (bottom, see inset) with a double spherically-symmetric exponential model (Eq. \ref{eq:doubleexponential_profile}). In both cases, residuals from T1 and T2 were masked and the width of the sampling annuli is half of the beam FWHM. } 
	\label{radioprofileFIT}
\end{figure}

\begin{table*}
\centering
	\caption[]{Results of the outer component in the double exponential fit of Eq. \ref{eq:doubleexponential_profile} (see Fig. \ref{radioprofileFIT}). Cols. 1, 2: frequency and resolution of the input image. Cols. 3, 4: fitted central surface brightness and \textit{e}-folding radius. Cols. 5, 6: corresponding flux density ($S_{\rm fit}=0.8\times 2 \pi I_{\rm 0} r_{\rm e}^2$) integrated up to $3r_{\rm e,out}$ and error contribution (in percentage) from the fitting uncertainties (see details in the table notes).  }
	\label{Table: double fit results}   
	\begin{tabular}{cccccc}
	\hline
	\noalign{\smallskip}
	$\nu$ & $\theta$ & $I_{\rm 0,out}$ & $r_{\rm e,out}$ & $S_{\rm out,fit}$ & $\xi_{\rm fit}$
	\\  
	 (MHz) &  & (${\rm  \mu Jy \; arcsec^{-2}}$)  & ($ {\rm kpc}$) & (${\rm mJy}$) &  ($\%$) \\  
    \hline
	\noalign{\smallskip}
    50 & $46''\times 46''$ & $19.4\; \pm \;1.1$ & $299.0\; \pm \;11.8$ & $3129.7\; \pm \;628.6$ & 10\\
     50 &  $132''\times 132''$ & $19.5\; \pm \;1.1$ & $304.9\; \pm \;15.5$ & $3271.2\; \pm \;528.3$ & 12 \\
    143 & $40''\times 40''$ &  $3.9\; \pm \;0.2$ &   $303.8\; \pm \;7.3$ & $649.5\; \pm \;110.6$ & 7\\
	143 & $134''\times 134''$  & $4.7\; \pm \;0.3$ & $272.2\; \pm \;8.7$ & $628.4\; \pm \;85.0$ & 9
     \\
\noalign{\smallskip}
	\hline
	\end{tabular}  
	\begin{tablenotes}
\item    {\small \textbf{Notes}. Errors on the fitted flux density are computed as $\Delta S_{\rm fit}= \sqrt{ \left( \sigma \cdot \sqrt{N_{\rm beam}} \right)^2 + \left(  \xi_{\rm cal} \cdot S_{\rm fit} \right)^2 + \left(\xi_{\rm fit} \cdot S_{\rm fit} \right)^2 } \label{erroronfluxFIT}$ to include the uncertainties on the fitted parameters;  $\xi_{\rm fit}$ is obtained from the standard formula for the error propagation as $\xi_{\rm fit} =  \sqrt{\left( \frac{\Delta I_{\rm 0}}{I_{\rm 0}} \right)^2 + 4\left( \frac{\Delta r_{\rm e}}{r_{\rm e}} \right)^2}$}. 
 \end{tablenotes}	
\end{table*}

In Sect. \ref{sect: Spectral indices} we obtained local measures of the flux density and spectral index for H3. We now aim to determine global values by means of the average surface brightness and its radial profile.

As a first check, we masked the regions of H1, H2, and residuals from subtraction, and then obtained an estimate of the average surface brightness of H3 over an ellipse of $2.4 \; {\rm Mpc}\times 2.0 \; {\rm Mpc}$ . By assuming that H3 uniformly fills the masked regions with the same average brightness, we estimated total flux densities of $S_{\rm 50}=3116.6\pm 316.0$ mJy and $S_{\rm 143}=628.3\pm63.0$ mJy.

More accurate values can be derived by fitting the surface brightness profile of the halo. Under simple assumptions of spherical symmetry, the surface brightness profile of both mini-halos and radio halos can be typically reproduced by an exponential law \citep{murgia09} as:
\begin{equation}
I(r)=I_{\rm 0}e^{-\frac{r}{r_{\rm e}}} \;
\label{eq:exponential_profile}
\end{equation}  
where $I_{\rm 0}$ is the central brightness and $r_{\rm e}$ is the \textit{e}-folding radius. However, Eq. \ref{eq:exponential_profile} is not able to properly fit the surface brightness of A2142 owing to its complex multi-component structure. We can assume that both H1 and H3 have spherical symmetry\footnote{Even though an ellipsoidal geometry would be more reliable, we do not expect that our assumption introduces
significant biases in the fitted parameters, since the minor to major axis length ratio (1.2) is relatively small.}, and that H3 fills a fraction of the volume of H1 and H2. We therefore considered a double exponential function as: 
\begin{equation}
I(r)=I_{\rm 0, in}e^{-\frac{r}{r_{\rm e,in}}} + I_{\rm 0, out}e^{-\frac{r}{r_{\rm e,out}}} \;,
\label{eq:doubleexponential_profile}
\end{equation}  
where the contributions of the inner (H1) and outer (H3) components are summed. As we did not attempt to model the emission of the ridge due to its complex morphology, its contribution will thus be included in the fitted flux density of the outer component.

As shown in the examples in Fig. \ref{radioprofileFIT} for LOFAR HBA images at $40''\times 40''$ and $134''\times134''$ resolution, the double exponential model can well reproduce the profile of A2142. In Table \ref{Table: double fit results} we report the central brightness and \textit{e}-folding radius for the outer component H2+H3 which we are interested in. The flux density is computed by integrating the surface brightness in annuli up to a radial distance of $\hat{r}=3r_{\rm e}$, which typically sets the finite extension of radio halos; this radial cut-off provides the $80\%$ of the flux density that would be obtained by integrating the surface brightness up to $\hat{r}=\infty$, thus yielding $S_{\rm fit}=0.8\times 2 \pi I_{\rm 0} r_{\rm e}^2 $. We found discrepancies in the fitted $I_{\rm 0}$ and $r_{\rm e}$, depending on the input image; this is likely due to the (unmodelled) emission of the ridge, which contaminates the fits with different weights. On the other hand, the corresponding flux densities are well in agreement within the errors, and with the rough estimate provided above by considering the elliptical geometry and assuming a uniform distribution of the average brightness, thus confirming the reliability of our approach. 

By considering $S_{\rm 50}=3271.2 \pm 528.3$ mJy and $S_{\rm 143}=628.4 \pm 85.0$ mJy, we obtain a global $\alpha_{\rm H3}^{\rm [50-143]}=1.57\pm 0.20$, which is perfectly consistent with local values inferred in Sect. \ref{sect: Spectral indices}. This confirms that H3 is an ultra-steep spectrum radio halo.

\subsection{Radio power}
\begin{table*}
\centering
	\caption[]{Radio powers of H1, H2, and H3 at 50, 143, 323, 407, 1380, and 1810 MHz from measurements reported in Sects. \ref{sect: Spectral indices} (for H1 and H2) and \ref{sect: Spectral index of H3} (for H2+H3). Values marked with `*' and `**' are obtained by extrapolation with $\alpha=1.15$ and $\alpha=1.57$, respectively.}
	\label{Table:radiopower}   
	\begin{tabular}{ccccccc}
	\hline
	\noalign{\smallskip}
	Source & $P_{50}$ & $P_{143}$ & $P_{323}$ & $P_{407}$ & $P_{1380}$ & $P_{1810}$ 
	\\  
	 & ($10^{24} {\rm \; W \; Hz^{-1}}$) & ($10^{24} {\rm \; W \; Hz^{-1}}$) & ($10^{23} {\rm \; W \; Hz^{-1}}$) & ($10^{23} {\rm \; W \; Hz^{-1}}$) & ($10^{23} {\rm \; W \; Hz^{-1}}$) & ($10^{22} {\rm \; W \; Hz^{-1}}$)   \\  
    \hline
	\noalign{\smallskip}
	H1  &  5.1$\; \pm \;0.5$ & $1.7\; \pm \;0.2$ & $6.4\; \pm \;0.4$ & $5.3\; \pm \;0.3$ & $1.4\; \pm \;0.1$ & $9.7\; \pm \;0.8$ \\
	H2  &  7.9$\; \pm \;0.8$ & $2.3\; \pm \;0.2$ & $9.1\; \pm \;0.6$ & $7.1\; \pm \;0.5$ & 1.7$\; \pm \;0.2$* & 12.1$\; \pm \;1.4$* \\
    H2+H3  &  64.8$\; \pm \;10.5$ & 12.5$\; \pm \;1.7$ & 34.7$\; \pm \;7.4$** & 24.1$\; \pm \;6.0$** & 3.5$\; \pm \;1.7$** & 23.2$\; \pm \;12.2$** \\
\noalign{\smallskip}
	\hline
	\end{tabular}  
%	\begin{tablenotes}
%\item    {\small \textbf{Notes}.} 
% \end{tablenotes}	
\end{table*}

By considering the flux densities and spectral indices of H1, H2, and H3, we computed the corresponding radio powers as:
\begin{equation}
P_{\rm \nu}= 4 \pi D_{\rm L}^{2}S_{\rm \nu}(1+z)^{{\alpha-1}} \; .
\label{radiopower}
\end{equation}
If a flux density measurement was not provided for H2 or H3 at a given frequency, we extrapolated it from the fitted spectrum by assuming $S_{\rm 143}$ as reference. The radio powers are reported in Table \ref{Table:radiopower}. 

As mentioned, the flux density derived from the surface brightness profile of H3 in Sect. \ref{sect: Spectral index of H3} also includes the contribution from H2. We scaled the corresponding radio power $P_{\rm 143}$ to 150 MHz ($P_{\rm 150}=(1.2\pm 0.2)\times 10^{25} \; {\rm W \; Hz^{-1}}$). As typical of ultra-steep spectrum radio halos \citep[e.g.][]{cassano13,donnert13,cuciti21b}, A2142 is located well below (by a factor of $\sim 3$) the scatter of the correlation between radio power and host mass reported in \cite{cuciti23}.

\subsection{Thermal properties and non-thermal emission}

\begin{figure*}
	\centering

	\includegraphics[width=1.0\textwidth]{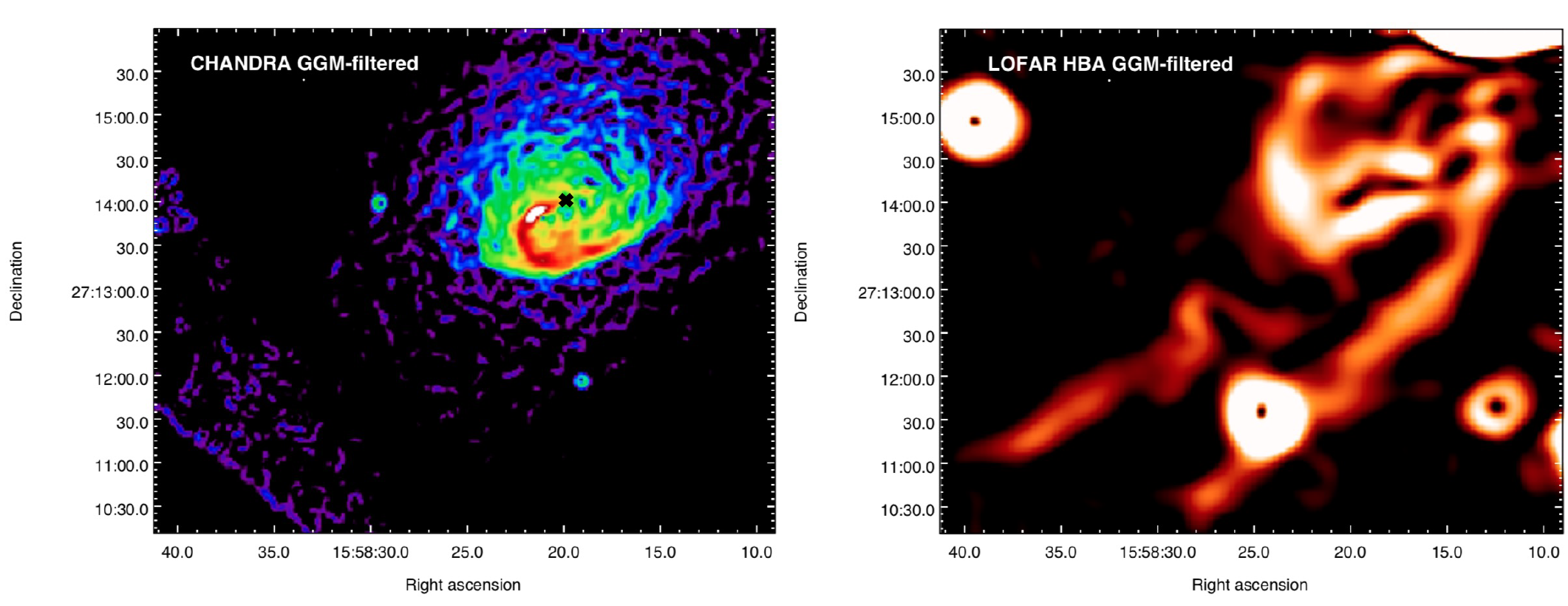}

	\smallskip
	\caption{GGM-filtered images of A2142. Discrete sources were not subtracted and appear as ring structures due to GGM-filtering. \textit{Left panel}: Chandra X-ray flux image filtered with $\sigma_{\rm GGM}=3''$. The black cross marks the location of the BCG. The emission visible in the lower left corner is artificial and associated with the ACIS chip boundaries. \textit{Right panel}: LOFAR HBA radio image filtered with $\sigma_{\rm GGM}=12''$. The radio emission of H1 follows the spiral pattern of the X-ray cold fronts.}
	\label{ggm}
\end{figure*}

\begin{figure*}
	\centering

	\includegraphics[width=0.95\textwidth]{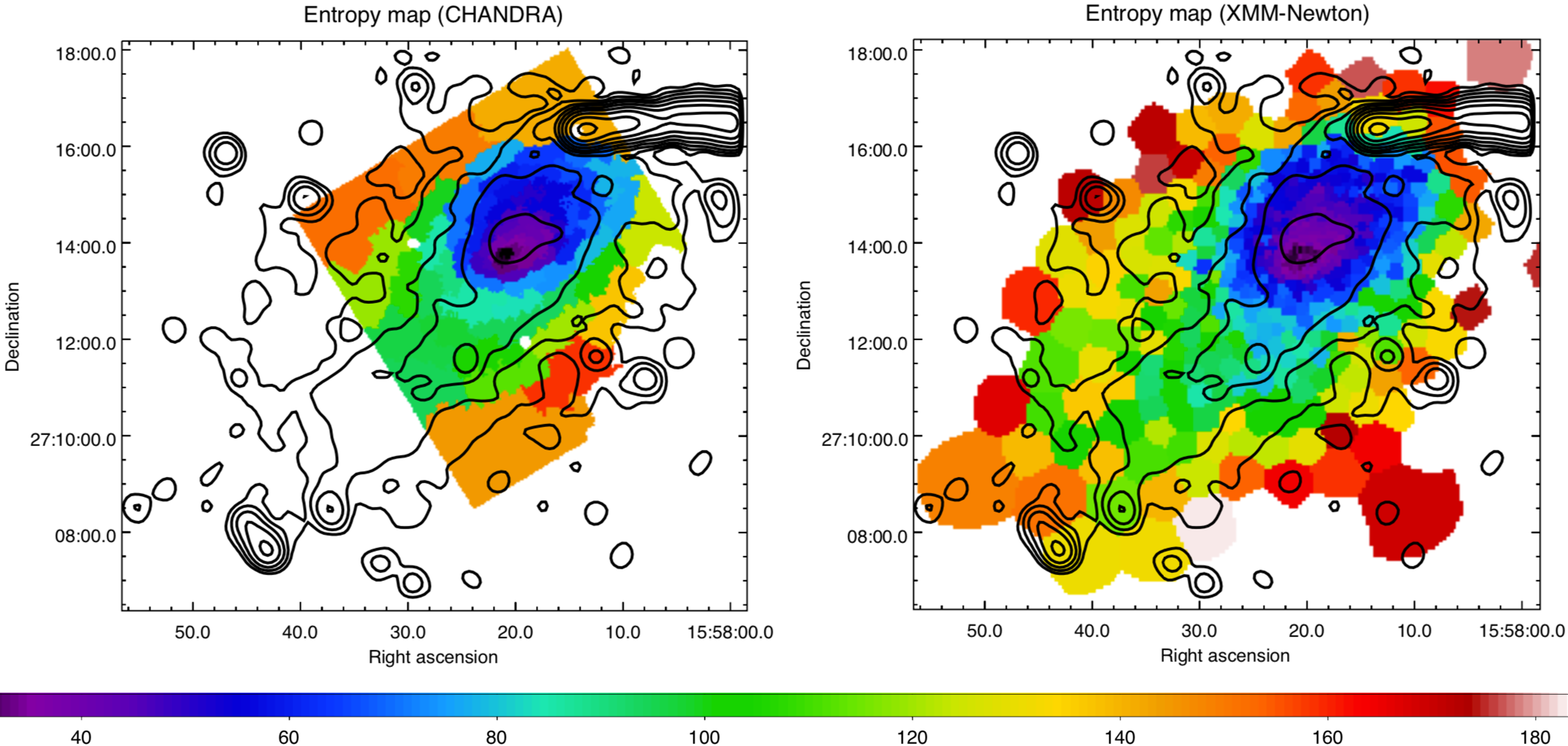}

	\smallskip
	\caption{Projected entropy maps of A2142 derived from Chandra (left) and XMM-Newton (right) in units of ${\rm keV \; cm^{5/3} \; arcmin^{-2/3}}$. Black contours are those of Fig. \ref{radiomapHBA}, upper right panel.}
	\label{entropy}
\end{figure*}

In this Section we compare the thermal and non-thermal properties derived from our X-ray and radio data, respectively.

Radio halos are generally observed to have smooth surface brightness profiles, as discussed in Sect. \ref{sect: Spectral index of H3}. Nonetheless, \cite{botteon23} have recently shown that substructures of the diffuse radio emission, such as surface brightness discontinuities, are detectable when observing radio halos with sufficient high resolution and sensitivity, and that these can be co-spatial with substructures and discontinuities in the X-ray emission of the thermal ICM. Following a  similar approach, we applied the Gaussian gradient magnitude \citep[GGM;][]{sanders16a,sanders16b} filtering to enhance surface brightness discontinuities and sub-structures in A2142. We adopted $\sigma_{\rm GGM}=2$ pix and $\sigma_{\rm GGM}=8$ pix for our Chandra and LOFAR HBA images (where 1 pix $=$ 1.5 arcsec in both cases) as the width of the derivative of the Gaussian filter function. The resulting GGM-filtered images are shown in Fig. \ref{ggm}. In the left panel, we observe some gas that departs from the BCG\footnote{The X-ray peak is found at a distance of $21''= 35$ kpc from BCG1, in line with \cite{wang&markevitch18}.} (black cross) and produces the inner cold fronts following a clear spiral path. In the radio counterpart (right panel), the brightest spots spatially correspond to the boundaries of the cold fronts. The western filamentary structure enhanced in the GGM-filtered radio image might suggest that the ridge is directly connected with the core, thus hinting at a common origin of the two radio components; on the other hand, even by filtering the Chandra image with wider $\sigma_{\rm GGM}$, we do not find a corresponding X-ray structure that could confirm this hypothesis.

By using XMM-Newton X-ray images in different energy bands \citep[see details of this procedure in][]{rossetti07}, \cite{rossetti13} produced projected entropy maps of A2142, and found that the ridge follows a trail of low-entropy ICM phase. We investigated this trend with Chandra data as well. Through {\tt CONTBIN}\footnote{\url{ https://github.com/jeremysanders/contbin}} \citep{sanders06}, we adaptively binned our 0.5-2 keV exposure-corrected image in regions with an high signal-to-noise ratio of ${\rm S/N}=100$. We extracted the spectra in each region of the event and blanksky files, subtracted the corresponding background contribution from the ICM emission, and simultaneously fitted the resulting spectra in {\tt XSPEC} with an absorbed thermal plasma component ({\tt phabs} $\times$ {\tt apec}), by keeping the hydrogen column density and metal abundance fixed at values of $N_{\rm H}=3.8\times10^{20} \; {\rm cm^{-2}}$ and $Z=0.28 \; Z_{\odot}$ \citep{markevitch00}. The Cash statistics \citep[Cstat;][]{cash79} was adopted for the fits; the chosen high ${\rm S/N}$ ensured ${\rm Cstat/d.o.f}$ in the range 0.90-1.14 for each region, thus indicating the goodness of our fits. The procedure and systematics related to the derivation of the entropy map by using the fitted temperature and {\tt apec} normalisation are described in Appendix \ref{appendix:Thermodynamical maps}. In Fig. \ref{entropy} we both show our projected entropy map (left panel) and that produced by \cite{rossetti13} (multiplied by the pixel area to match our units). Despite the different procedures, resolution, and ${\rm S/N}$, the two maps are well in agreement in the common area that covers a radius of $\sim 350$ kpc. The LOFAR HBA contours at $21''\times 20''$ resolution (Fig. \ref{radiomapHBA}, upper right panel) are overlaid on the maps. In the radio core, the ICM has entropy values $s \lesssim 60 \;  {\rm keV \; cm^{5/3} \; arcmin^{-2/3}}$. As claimed by \cite{rossetti13}, we confirm that the ridge is co-spatial with a trail of gas that extends in the NW-SE direction and is characterised by lower entropy values ($s \sim 100 \; {\rm keV \; cm^{5/3} \; arcmin^{-2/3}}$) than those found along different directions at the same radial distance.

\subsection{Point-to-point radio vs X-ray analysis}
\label{sect: Point-to-point radio vs X-ray}

\begin{figure}
	\centering
	\includegraphics[width=0.45\textwidth]{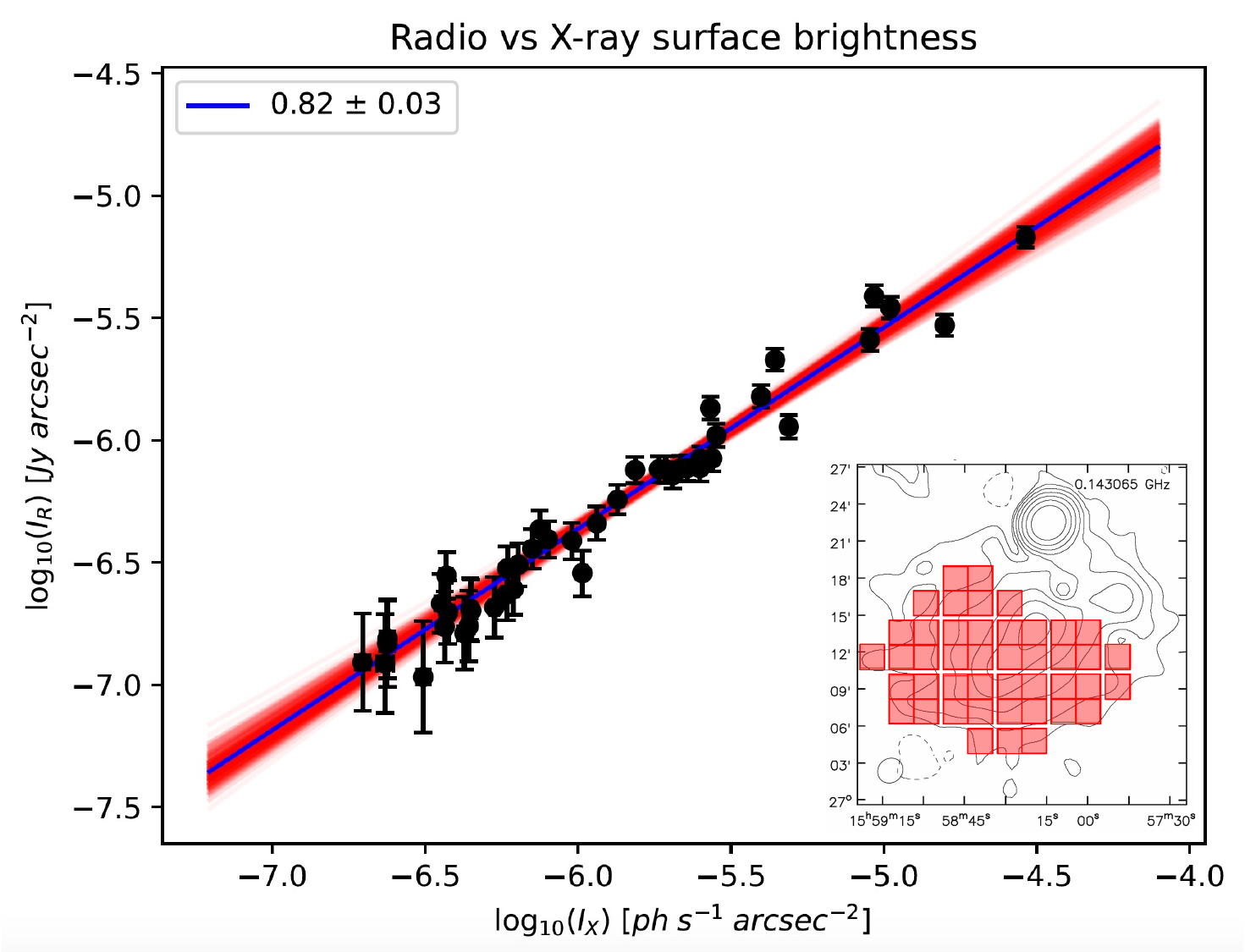} 
	
	\smallskip
	
	\caption{Point-to-point radio (LOFAR HBA) vs X-ray (XMM-Newton) analysis. The linear best fit provides a slope $k=0.82\pm0.03$. The sampling square boxes of sides $128''$ are overlaid on the source-subtracted LOFAR HBA contours at $128''\times 117''$, which are shown in the inset.}
	\label{PTP_T120}
\end{figure}

\begin{figure*}
	\centering
    \includegraphics[width=0.4\textwidth]{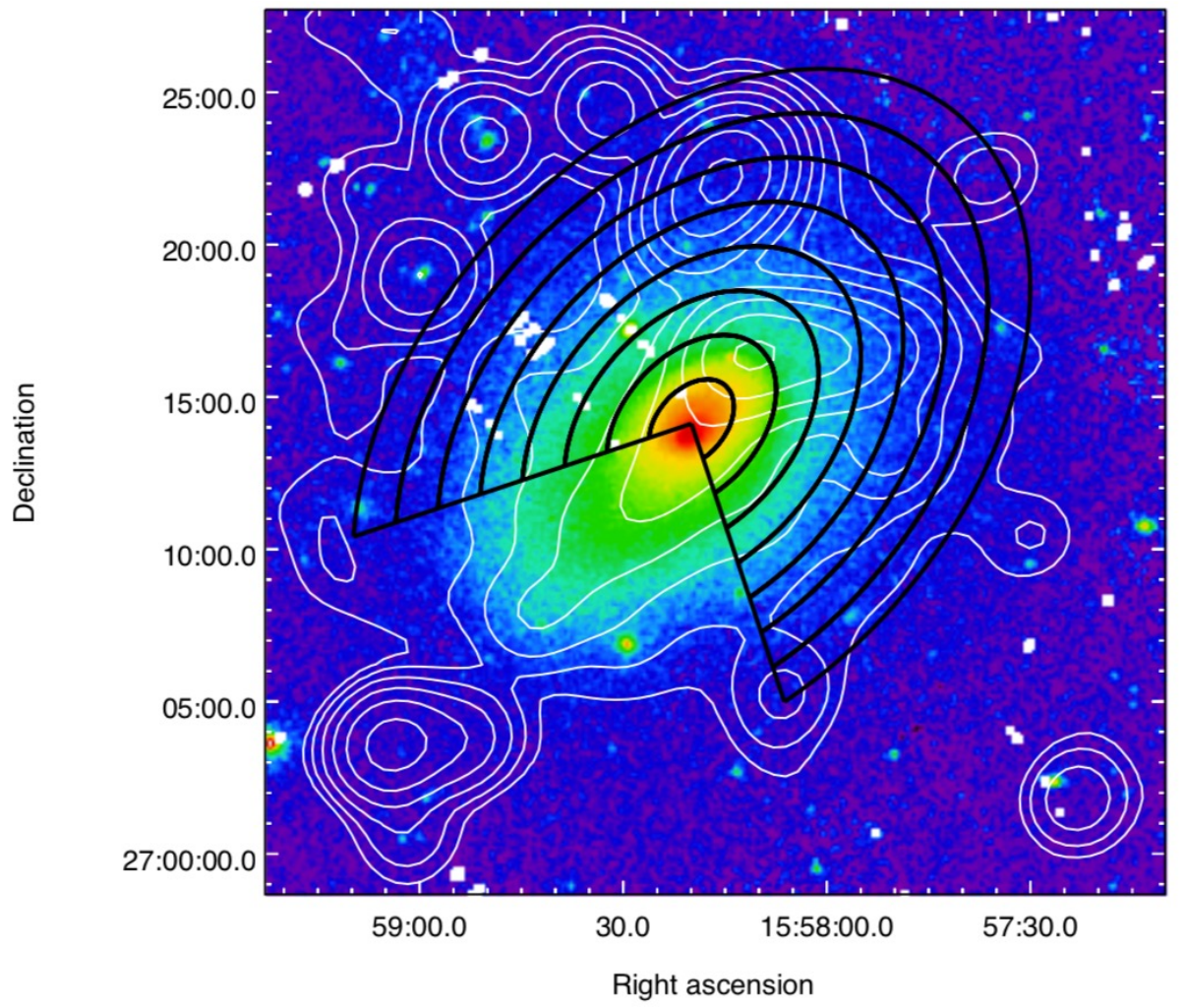}
	\includegraphics[width=0.5\textwidth]{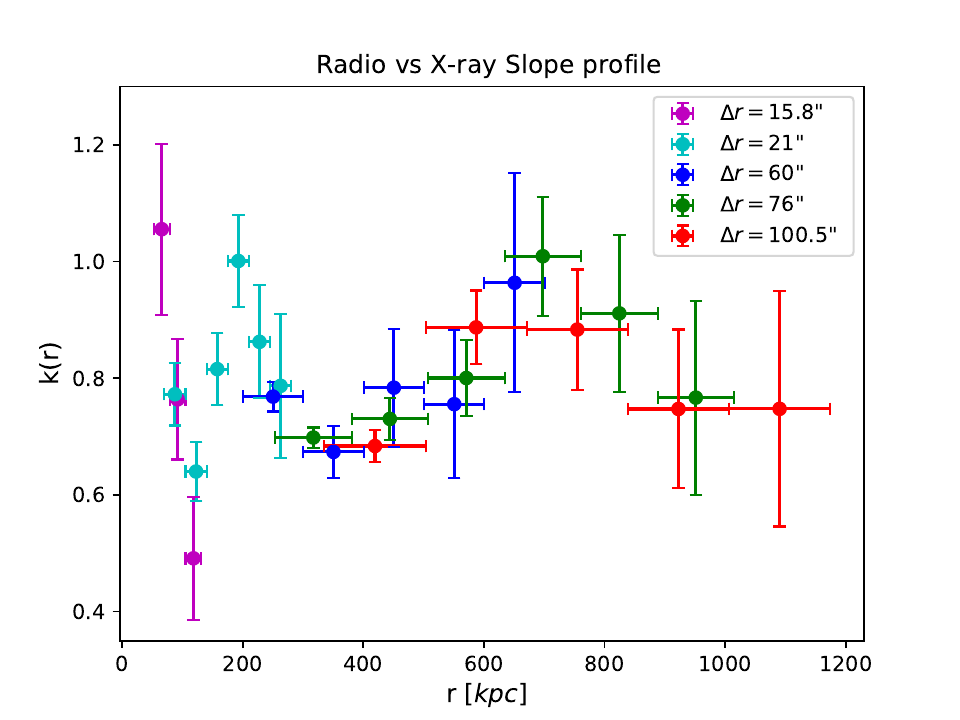} 
	\smallskip
	
	\caption{Radial trends of the radio vs X-ray correlation. \textit{Left}: Example of images (LOFAR HBA contours at $134''$ are overlaid on the XMM-Newton image) and elliptical sectors used to measure the surface brightnesses; for inspection purposes, the shown images include the discrete sources, whereas they were subtracted (or masked in case of residuals) during analysis. \textit{Right}: Radio vs X-ray correlation slope as a function of the radial distance from BCG1. The widths of the concentric elliptical sampling sectors depend on the beam of the radio image and the required S/N, and are reported in the legend. The global trend $k(r)$ is sub-linear, albeit with two narrow peaks.}
	\label{multiresPTP}
\end{figure*}

The radio surface brightness of both halos and mini-halos and the X-ray surface brightness of the thermal ICM are found to be spatially correlated as $I_{\rm R}\propto I_{\rm X}^{k}$ \citep[e.g.][]{govoni01}. The slope $k$ of this correlation indicates the relative spatial distributions of thermal and non-thermal components, and can be thus used to probe the origin of the diffuse radio emission. Super-linear slopes $k>1$ are associated with a narrower distribution of the emitting electrons than that of the thermal particles; conversely, sub-linear $k<1$ slopes are associated with narrower thermal and broader non-thermal distributions, respectively, thus  
supporting a scenario where particle are accelerated and transported by turbulence generated on large scales. Typically, radio halos have been found to exhibit sub-linear slopes \citep[e.g.][]{hoang19,bruno21,duchesne21,rajpurohit21,pasini22}, while super-linear slopes have been found in mini-halos \citep[e.g.][]{govoni09,ignesti20,biava21,riseley22}. 

We investigated the radio vs X-ray correlation in A2142 through the Point-to-point TRend EXtractor code\footnote{\url{https://github.com/AIgnesti/PT-REX}} \citep[PT-REX;][]{ignesti22}, by gridding the regions of interest with beam-size square boxes. The extracted surface brightness values were fitted in a logarithmic plane as $\log_{10} I_{\rm R}=k \log_{10} I_{\rm X} + c$ (where $c$ is the intercept) with {\tt linmix} \citep{kelly07}, which can account for the errors of the measurements and intrinsic scatter of the linear regression through a Bayesian statistical approach. We considered as inputs LOFAR HBA and XMM-Newton images at resolution of $128''\times 117''$, which allow us to investigate the faintest regions of the radio halo and the outskirts of the ICM, and considered a radio surface brightness threshold of $2\sigma$. As shown in Fig. \ref{PTP_T120}, the data points are well correlated and fitted by a sub-linear power-law of slope $k=0.82\pm 0.03$ (Spearman and Pearson ranks are $\rho_{\rm S}=0.92$ and $\rho_{\rm P}=0.93$, respectively), as typical of usual radio halos.

In spite of the solid sub-linear correlation that we found, the slope might not be constant over the whole radio halo due to the presence of multiple radio components \citep[see also the case of Coma in][]{bonafede22}, whose origin may be associated with different physical mechanisms. To investigate the radial behaviour of the slope $k=k(r)$, we considered concentric elliptical sectors that approximately follow the distribution of the ICM to measure $I_{\rm R}$ and $I_{\rm X}$. The slope as a function of the radial distance is thus computed as:
\begin{equation}
k(r)= \frac{\Delta(\ln{I_{\rm R}})}{\Delta(\ln{I_{\rm X})}} \;,
\label{eq:k-radius}
\end{equation}
where $\Delta(\ln{I})$ is the difference of the logarithms of the radio or X-ray brightness evaluated in two consecutive sectors. Owing to its complex geometry, we ignored the region of H2. We considered radio images at various resolutions; the width of the annuli was set depending on the resolution and required signal-to-noise ratio. An example of the sampling sectors is reported in the left panel of Fig. \ref{multiresPTP}.

In the right panel of Fig. \ref{multiresPTP} we report the derived radial profiles $k(r)$ for different resolutions and S/N. We found a possible slightly super-linear correlation within the innermost regions ($\sim 50-100$ kpc) of H1; on the other hand, the rapid drop of $k$ with the distance may more likely suggest that this trend is caused by the offset between the radio and X-ray peaks, and not associated with a specific physical process. For $r\gtrsim 100$ kpc, $k(r)$ is on average sub-linear, thus indicating that the distribution of the non-thermal components is broader than that of thermal components, and the slope varies in the range $\sim 0.7-0.8$. Moreover, we observe two narrow peaks of $k(r)$, for $r \sim 150$ kpc and $r \sim 650$ kpc. These peaks might be related with the transitions H1-H2 and H2-H3, or trace edges in the radio and/or X-ray surface brightness. Even though the width of the sampling annuli is an important factor to ensure sufficient S/N on the different scales, the discussed trends are barely dependent on it, thus not invalidating our conclusions. In Sect. \ref{sect: Probing the hybrid halo origin} we will further discuss the slope and its radial behaviour to constrain the possible formation scenario of each radio component.

\section{Discussion}
\label{Sect:Discussion}

In this Section we discuss the origin of the three components of the halo and the possible existence of a megahalo in A2142.

\subsection{Probing the hybrid halo origin}
\label{sect: Probing the hybrid halo origin}

The results collected in the previous Sections can be exploited to discuss the origin and connection of each component of the puzzling hybrid radio source. 

The core has a roundish morphology, is confined by multiple cold fronts on scales $\sim 200$ kpc, has a moderately steep spectrum $\alpha\sim 1.1$, and the slope of the radio vs X-ray correlation is sub-linear (at least on scales $\sim 100-200$ kpc, see Fig. \ref{multiresPTP}). These properties may support the hypothesis that H1 is a mini-halo originating from turbulence induced by sloshing of cold gas \citep[e.g.][]{zuhone13} or from hadronic collisions between protons confined in the cold fronts \citep[e.g.][]{zuhone15}. Nevertheless, the event that triggered the central sloshing is unclear. \cite{rossetti13} and \cite{venturi17} proposed the hypothesis of an intermediate mass-ratio merger (possibly with the group of BCG2 being involved). Such an event is less disruptive than major mergers and would likely not affect the global properties and morphology of the ICM. On the other hand, strong sloshing motions could be induced by this kind of merger, being able to move cold gas away from the centre, where then mixing with the hotter gas phase is favoured \citep{wang&markevitch18}. This could lead to the formation of the prominently offset warm core from a former cool core; moreover, BCG1 would be deprived of most of the available fuel, thus explaining the absence of a powerful central AGN as in classical relaxed clusters.

The extreme sloshing motions in A2142 likely involve not only the inner regions of the cluster, but also larger scales, as witnessed by the presence of the southernmost cold front at $\sim 1$ Mpc from the centre \citep{rossetti13} and the trail of low-entropy gas. In this case, sloshing could be partly responsible for the formation of the ridge, which extends for $\sim 400$ kpc along the NW-SE axis. The spectral index of the ridge is similar ($\alpha\sim 1.2$) to that of the core, and the projected entropy distribution of the ICM hints at a common origin for H1 and H2. Turbulent motions induced by sloshing may trigger the radio emission of the ridge by re-accelerating particles and amplifying magnetic fields along the low-entropy trail. It is worth noticing that turbulence in the ridge might be also supplied by the infalling groups along the NW-SE axis, which is the direction of the main accretion filament of the A2142-supercluster \citep{einasto15,einasto18,einasto20}; numerical simulations are needed to investigate this scenario.

The observed properties of H3 (large size, $\alpha \sim 1.6$, exponential radial profile, co-spatial thermal and non-thermal emission, sub-linear $k$) allowed us to classify it as a giant ultra-steep spectrum radio halo. Concerning the origin of H3, we discuss two possible scenarios, connected to either (i.) an old merger or (ii.) multiple minor mergers.

The first scenario for the origin of H3 is related to the half-mass epoch \citep[estimated to be $\sim 4$ Gyr ago;][]{einasto18}, when A2142 formed at the centre of the supercluster through energetic mergers. At the present epoch, the turbulence injected by such event may have not been completely dissipated \citep[see also][]{donnert13}. Therefore, the origin of H3 could be associated with this old merger, and its ultra-steep spectrum could be the consequence of re-acceleration in regions where turbulence is significantly dissipated. Furthermore, the hypothesis of an extremely old evolutionary stage of the halo may open to the possibility of an intriguing scenario to explain the peculiar thermal properties of A2142. We might speculate that a former cool core was first disrupted by the old merger, and over the last few Gyr after this event it has started to reform again; nevertheless, this process is hampered by the perturbing minor mergers, thus explaining the presence of a warm core rather than a classical cool core.

Alternatively to the old merger scenario, the ultra-steep spectrum of H3 could be originated by minor mergers \citep[e.g.][]{cassano06}. It is unlikely that a single intermediate mass-ratio merger (as discussed for H1 and H2) is able to inject enough turbulent energy via sloshing to uniformly re-accelerate electrons on the large scales ($\gtrsim 2$ Mpc) of H3. However, a high number of minor mergers occurring over the cluster volume may be able to continuously supply turbulence and power the radio emission of H3, by re-accelerating electrons with a lower efficiency than turbulence induced by major mergers.

The uncommon thermal and non-thermal properties of A2142 offer the possibility to discuss the hypothesis of an evolutionary connection between H1, H2, and H3. By considering the model proposed by \cite{zandanel14}, H1 and H3 might have an hadronic and turbulent re-acceleration origin, respectively. Secondary CRe might significantly contribute to the emission of H1, but, in this case, the sub-linear value of $k$ on scales $\sim 100-200$ kpc would indicate particular conditions, namely a flat CRp distribution in the core and a high ($B>B_{\rm CMB}$, where $B_{\rm CMB}\sim 4 \; {\rm \mu G}$ is the equivalent magnetic field of the Cosmic Microwave Background at the cluster redshift) magnetic field \citep{brunetti&jones14,ignesti20}. Conversely, the observed properties (size, correlation slope, and spectral index) of H2 and H3 indicate that the emission in these regions is less contributed from secondary CRe, and the re-acceleration scenario is thus strongly favoured. In addition, H2 might have been the inner part of H3 in the past, hence implying the same origin of the two components; in this case, ICM turbulence could have then reshaped the inner emission into the present ridge.

Finally, A2142 is an interesting target to search for the origin of seed relativistic particles. Indeed, the possibility that (aged) particles originating in the core are transported and re-accelerated by sloshing turbulence along the ridge \citep{brunetti&jones14} is a reasonable scenario that would be in line with thermal features, such as the low-entropy trail and the southernmost cold front. Furthermore, the seed particles to be re-accelerated in H2 and H3 may be primary CRe injected in the ICM by AGN and galaxy members of the groups infalling in A2142, as hinted by the prominent tails of T1 and T2.

\subsection{A megahalo in A2142 ?}

\begin{figure}
	\centering
	\includegraphics[width=0.45\textwidth]{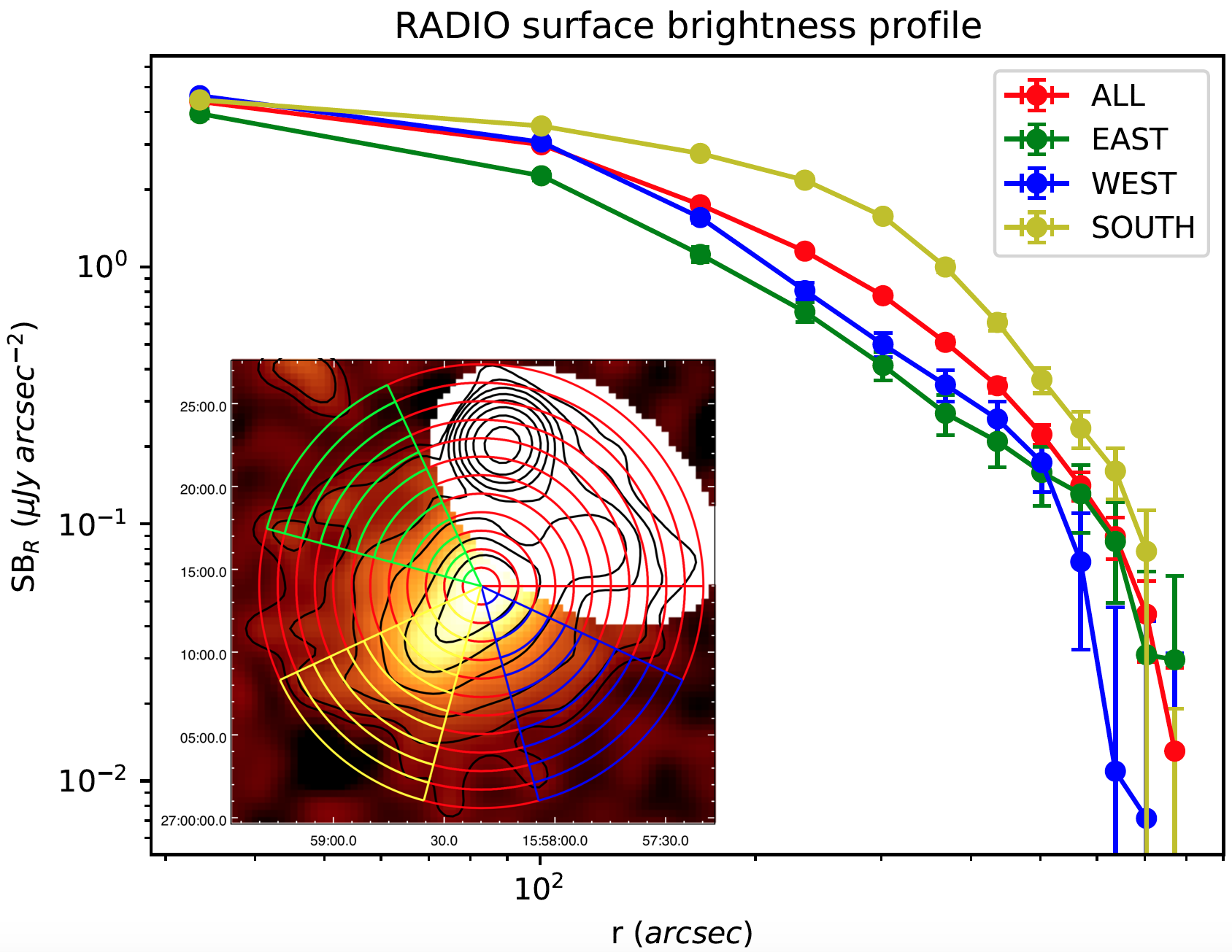}
	\smallskip
	\caption{Azimuthally-averaged surface brightness profile of A2142 at $134''\times 134''$ resolution, up to $0.95R_{\rm 500}$. The inset reports the four sectors considered to extract the profiles: $0^{\rm o}-360^{\rm o}$ (red), $115^{\rm o}-165^{\rm o}$ (green), $205^{\rm o}-255^{\rm o}$ (yellow), $285^{\rm o}-335^{\rm o}$ (blue).  } 
	\label{radioprofileT120}
\end{figure} 

\begin{figure}
	\centering
	\includegraphics[width=0.45\textwidth]{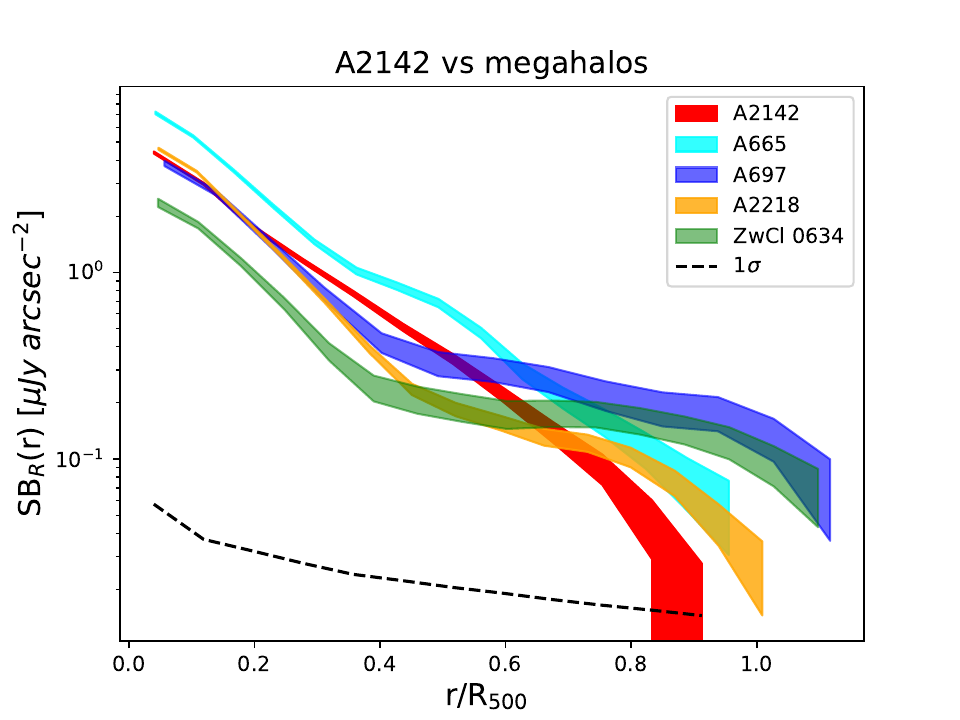}
	\smallskip
	\caption{Surface brightness profiles of A2142 and known megahalos \citep[from][]{cuciti22} as a function of the radial distance (normalised by $R_{500}$). The dashed line indicates the $1\sigma$ detection level of each bin for our observations. Different shapes are observed for A2142 and the other clusters.  } 
	\label{confrontomegahalo}
\end{figure}

Megahalos are diffuse sources exhibiting a radial decline of the radio brightness that is shallower than that of their embedded radio halos \citep{cuciti22}. In addition, the 4 megahalos discovered so far share a large extension ($\sim R_{\rm 500}$), an ultra-steep spectrum, and are found in disturbed clusters. Although the origin of megahalos is still poorly constrained, preliminary numerical simulations suggest that a baseline level of large scale turbulence, induced by the continuous accretion of matter onto the cluster, may be common to all clusters, regardless of their dynamical state. Since all megahalos were detected in massive ($M_{500}\sim 6-11\times 10^{14} \;M_\odot$) clusters at intermediate-low redshift ($z\sim 0.17-0.28$), we searched for the presence of a megahalo in A2142, which is a very massive and nearby cluster as well.

%In this respect, A2142 appears to be a favoured candidate to host the first megahalo in a system that is in an intermediate dynamical state because (i.) it is the main member of a supercluster and prominent accretion phenomena are expected in its outskirts, (ii.) it has a favourable combination of mass and redshift (see Fig. \ref{M-z_plane}), and (iii.) the radio emission has some properties (large extension, multiple components, ultra-steep spectral index) in common with the 4 known megahalos.

To investigate the presence of a megahalo in A2142, throughout this Section we consider the 143 MHz, source-subtracted image (residuals from T1 and T2 were masked) at $134''$ resolution. We extracted the surface brightness profile in circular annuli of width $67''$ (i.e. half FWHM of the restoring beam) up to the $2\sigma$ contour level in different directions to check for possible biases associated with geometrical effects and/or residuals from subtraction. The extraction sectors and computed profiles are reported in Fig. \ref{radioprofileT120}. These do not show the typical shallower trend associated with megahalos.

In Fig. \ref{confrontomegahalo} we compared the surface brightness as a function of the radial distance (normalised by $R_{500}$) of A2142 and the 4 known megahalos. This plot shows that the shallower component arises on scales $r\sim 0.4-0.5R_{500}$ for megahalos. In contrast, the profile of A2142 can be reproduced by two exponential laws (see Sect. \ref{sect: Spectral index of H3}) describing H1 and H3; the observed discontinuities in the profile at $r\sim 0.12R_{500}$ and $r\sim 0.20R_{500}$ mark the transitions from H1 to H2 and from H2 to H3, respectively. A possible megahalo in A2142 may be have been detected by our deep observations beyond the exponential cutoff ($r\sim 0.9R_{500}$) with a significance of $\sim 2 \sigma$. In summary, Figs. \ref{radioprofileT120},  \ref{confrontomegahalo} provide no evidence for a megahalo in A2142. As a sanity check, we also measured the surface brightness profile of A2142 from elliptical sectors (not shown); the shapes of the profiles from circular and elliptical annuli are consistent, thus further supporting our conclusions.

It is worth discussing whether the undetection of a megahalo is the result of calibration issues, or a combination of inadequate \textit{uv}-coverage at short spacings and noise sensitivity. As shown by simulations in \cite{shimwell22LOTSSDR2}, faint and extended emission may be partially lost if unmodelled during calibration steps. On the other hand, faint radio emission in nearby clusters, extending on even larger angular scales than those of A2142 has been well recovered \citep[e.g.][]{rajpurohit21b,bonafede22,botteon22b}. Therefore, it is unlikely that in our target possible emission in the form of a megahalo was completely lost during calibration. For imaging, we adopted a minimum baseline of $50\lambda$, which can recover sources in the sky of size $\sim 1.15^{\rm o}$, corresponding to $\sim 7$ Mpc at the cluster redshift. Based on the results derived in \cite{bruno23} with a standard minimum baseline of $80\lambda$, we should safely observe radio emission up to $\sim 4$ Mpc with losses $\ll 20\%$ at most. Therefore, we can exclude the possibility that a megahalo on scales $\gg R_{\rm 500}$ is not detectable due to missing short spacings.

\section{Summary and conclusions}

The main member of the A2142-supercluster is the nearby and massive galaxy cluster A2142, whose dynamical state is intermediate between a classical major merger and a relaxed cool core. It was known to host an hybrid radio halo with two distinct components \citep{venturi17}, namely the core (H1) and the ridge (H2). The core is roundish and confined by a system of cold fronts within the region of the warm X-ray core, whereas the ridge is elongated and roughly co-spatial with a trail of low-entropy thermal gas that ends in an extremely distant cold front ($\sim 1$ Mpc from the centre). Furthermore, the ridge is aligned with the main accretion filament of A2142. An intermediate mass-ratio merger was invoked to explain the uncommon thermal features of A2142 \citep{rossetti13}. This event could lead to sloshing involving regions beyond the cluster core, without affecting the global morphology of the ICM.

In this work we presented new deep LOFAR HBA and LOFAR LBA observations, which allowed us to detect a new radio component (H3), whose emission follows the X-ray thermal distribution of the ICM up to scales $\gtrsim 2$ Mpc. We investigated the properties, origin, and physical connection of the three components with complementary archival GMRT/uGMRT, VLA, Chandra, and XMM-Newton data. Our main results can be summarised as follows. 
\begin{enumerate}
  \item  The properties of H1 (including the measured spectral index $\alpha=1.09\pm 0.02$) support the possibility that radio emission in this region is powered by sloshing in the core, which can generate turbulence and magnetic fields that confine CRp and/or re-accelerate relativistic electrons, as thought for usual mini-halos. 
  \item The spectral index ($\alpha=1.15\pm 0.02$), morphology, and connection with the low-entropy gas trail suggest that the ridge can be powered by turbulence from large-scale sloshing, which could be also able to move and shape gas beyond the core. Additional supplies of turbulence in H2 might come from merging groups along the main accretion filament of A2142. 
  \item The size (2.4 Mpc $\times$ 2.0 Mpc), spectral index ($\alpha=1.57\pm 0.20$), exponential surface brightness profile, and sub-linear radio/X-ray spatial correlation ($I_{\rm R}\propto I_{\rm X}^k$, where $k\sim 0.7-0.8$) allowed us to classify H3 as a giant ultra-steep spectrum radio halo. We proposed two different scenarios for the origin of H3. In the first case, old mergers ($\sim 4$ Gyr ago) that originated the main structure of A2142 could have disrupted the former cool core and triggered the radio halo, which at present is observed in an advanced evolutionary phase. In this scenario,  minor mergers may contribute by hampering the restoring of a classical cool core. As an alternative possibility, turbulence could be injected over the cluster volume by continuous minor mergers with small groups of galaxies. These mergers would inefficiently re-accelerate particles, thus leading to ultra-steep spectra as predicted by turbulent re-acceleration models. In both these two scenarios, H2 could have been the inner part of H3, which has then been reshaped by turbulence into the present ridge.  
  \item We investigated the possibility that A2142 hosts a megahalo, owing to a favourable combination of mass and redshift that is similar to those of the 4 confirmed cases reported by \cite{cuciti22}. The analysis of the surface brightness profiles in A2142 and in the literature megahalos does not support the presence of a megahalo in our target.
\end{enumerate}

Our work primarily falls within the context of complex non-thermal phenomena occurring on the very large scales of the Universe, but is also relevant to probe the evolution of galaxy clusters. Indeed, we showed that the hybrid halo emission of A2142 can be interpreted in terms of a particular stage of the development of the A2142-supercluster system. Our analysis suggests that various dynamical processes operating not only on different spatial scales, but also on different time scales, can be responsible for the formation and evolution of multi-component halos. In this respect, radio observations can offer a valid perspective to investigate the complex dynamical history of galaxy clusters, from their innermost regions up to their outskirts.

Thanks to the high sensitivity provided by present and next-generation radio facilities, the number of hybrid halos is increasing \citep[][Biava et al. in prep.]{lusetti23}. By means of radio observations of clusters in an intermediate dynamical state, the relative roles of major/minor mergers and primary/secondary CRe in shaping thermal and non-thermal properties can be further constrained, and possibly tested by numerical simulations that follow the evolution of large-scale structures.

\begin{acknowledgements}
We thank the referee for comments and suggestions that have improved the presentation of our results. 
VC acknowledges support from the Alexander von Humboldt Foundation. FdG, GB, FG, MR, RC, SG, SdG acknowledge support from INAF mainstream project “Galaxy Clusters Science with LOFAR”. RJvW acknowledges support from the ERC Starting Grant ClusterWeb 804208. NB and ABonafede acknowledge support from the ERC through the grant ERC-StG DRANOEL n. 714245.
LOFAR \citep{vanhaarlem13} is the Low Frequency Array designed and constructed by ASTRON. It has observing, data processing, and data storage facilities in several countries, which are owned by various parties (each with their own funding sources), and that are collectively operated by the ILT foundation under a joint scientific policy. The ILT resources have benefited from the following recent major funding sources: CNRS-INSU, Observatoire de Paris and Universit\'e d’Orl\'eans, France; BMBF, MIWF- NRW, MPG, Germany; Science Foundation Ireland (SFI), Department of Business, Enterprise and Innovation (DBEI), Ireland; NWO, The Netherlands; The Science and Technology Facilities Council, UK; Ministry of Science and Higher Education, Poland; The Istituto Nazionale di Astrofisica (INAF), Italy. This research made use of the Dutch national e-infrastructure with support of the SURF Cooperative (e-infra 180169) and the LOFAR e-infra group. The J\"ulich LOFAR Long Term Archive and the German LOFAR network are both coordinated and operated by the J\"ulich Supercomputing Centre (JSC), and computing resources on the supercomputer JUWELS at JSC were provided by the Gauss Centre for Supercomputing e.V. (grant CHTB00) through the John von Neumann Institute for Computing (NIC). This research made use of the University of Hertfordshire high-performance computing facility and the LOFAR-UK computing facility located at the University of Hertfordshire and supported by STFC [ST/P000096/1], and of the Italian LOFAR IT computing infrastructure supported and operated by INAF, and by the Physics Department of Turin University (under an agreement with Consorzio Interuniversitario per la Fisica Spaziale) at the C3S Supercomputing Centre, Italy. This reaserch made use of the {\tt HOTCAT} cluster \citep{bertocco20,taffoni20,taffoni22} at Osservatorio Astronomico di Trieste and {\tt Rosetta} \citep{russo22}, a container-centric science platform for interactive data analysis available for accessing {\tt HOTCAT}. This reaserch made use of APLpy, an open-source plotting package for Python \citep{robitaille&bressert12APLPY}, Astropy, a community-developed core Python package for Astronomy \citep{astropycollaboration13,astropycollaboration18}, Matplotlib \citep{hunter07MATPLOTLIB}, Numpy \citep{harris20NUMPY}. %, and TOPCAT \citep{taylor05TOPCAT}. 
The National Radio Astronomy Observatory is a facility of the National Science Foundation operated under cooperative agreement by Associated Universities, Inc. We thank the staff of the GMRT that made these observations possible. GMRT is run by the National Centre for Radio Astrophysics of the Tata Institute of Fundamental Research. The scientific results reported in this article are based on observations made by the Chandra X-ray Observatory data obtained from the Chandra Data Archive. This research has made use of data and/or software provided by the High Energy Astrophysics Science Archive Research Center (HEASARC), which is a service of the Astrophysics Science Division at NASA/GSFC. 

\end{acknowledgements}

\bibliographystyle{aa}
\bibliography{bibliografia}

\begin{appendix}

\section{Spectral index error maps}
%\FloatBarrier

\begin{figure*}
	\centering
	\includegraphics[width=0.45\textwidth]{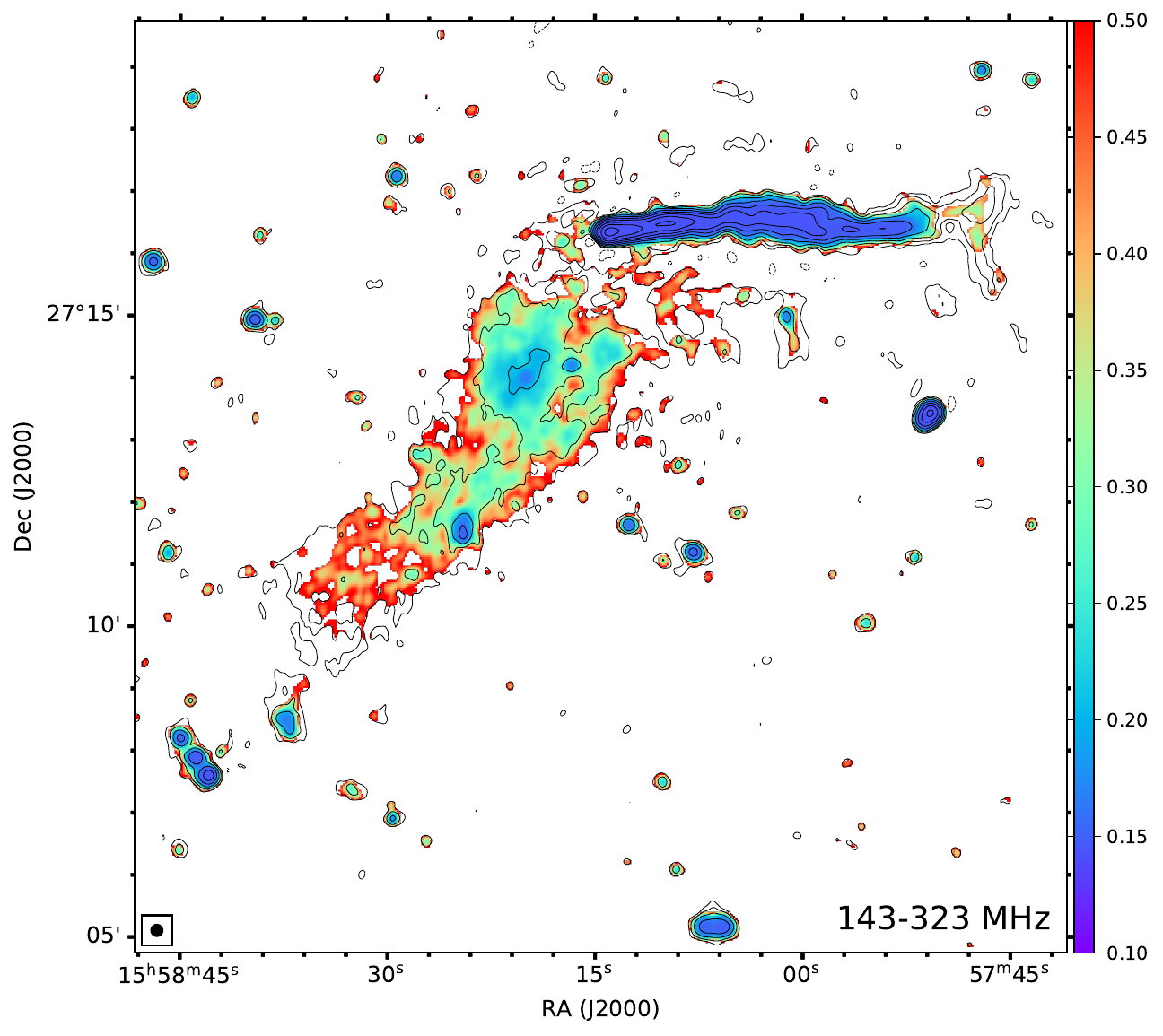}
     \includegraphics[width=0.45\textwidth]{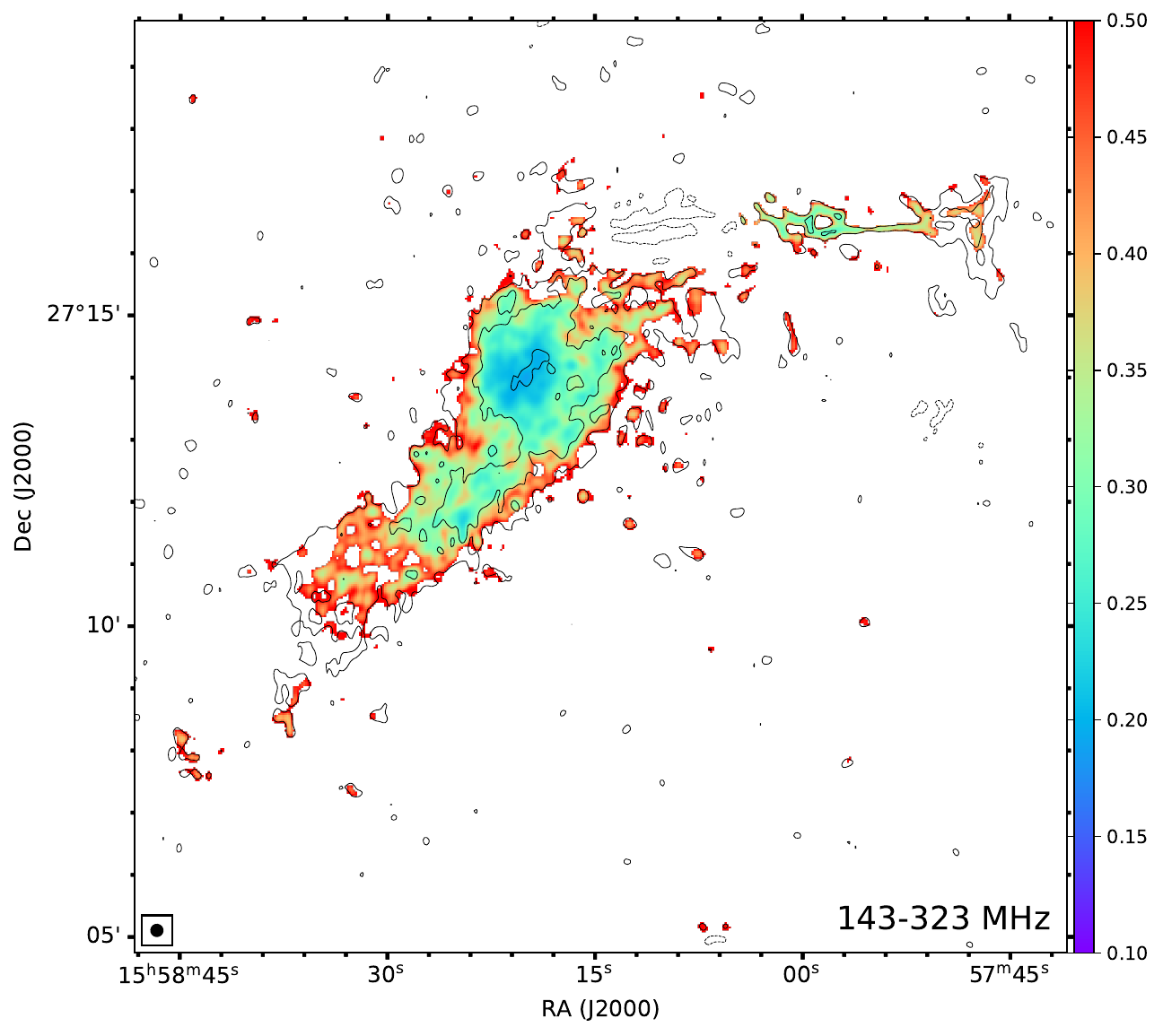}
	\smallskip
	
	\caption{Spectral index error maps between 143 and 323 MHz corresponding to the maps reported in Fig. \ref{SPIX_lofar-gmrt}. } 
	\label{errSPIX_lofar-gmrt}
\end{figure*}

\begin{figure*}
	\centering
	\includegraphics[width=0.45\textwidth]{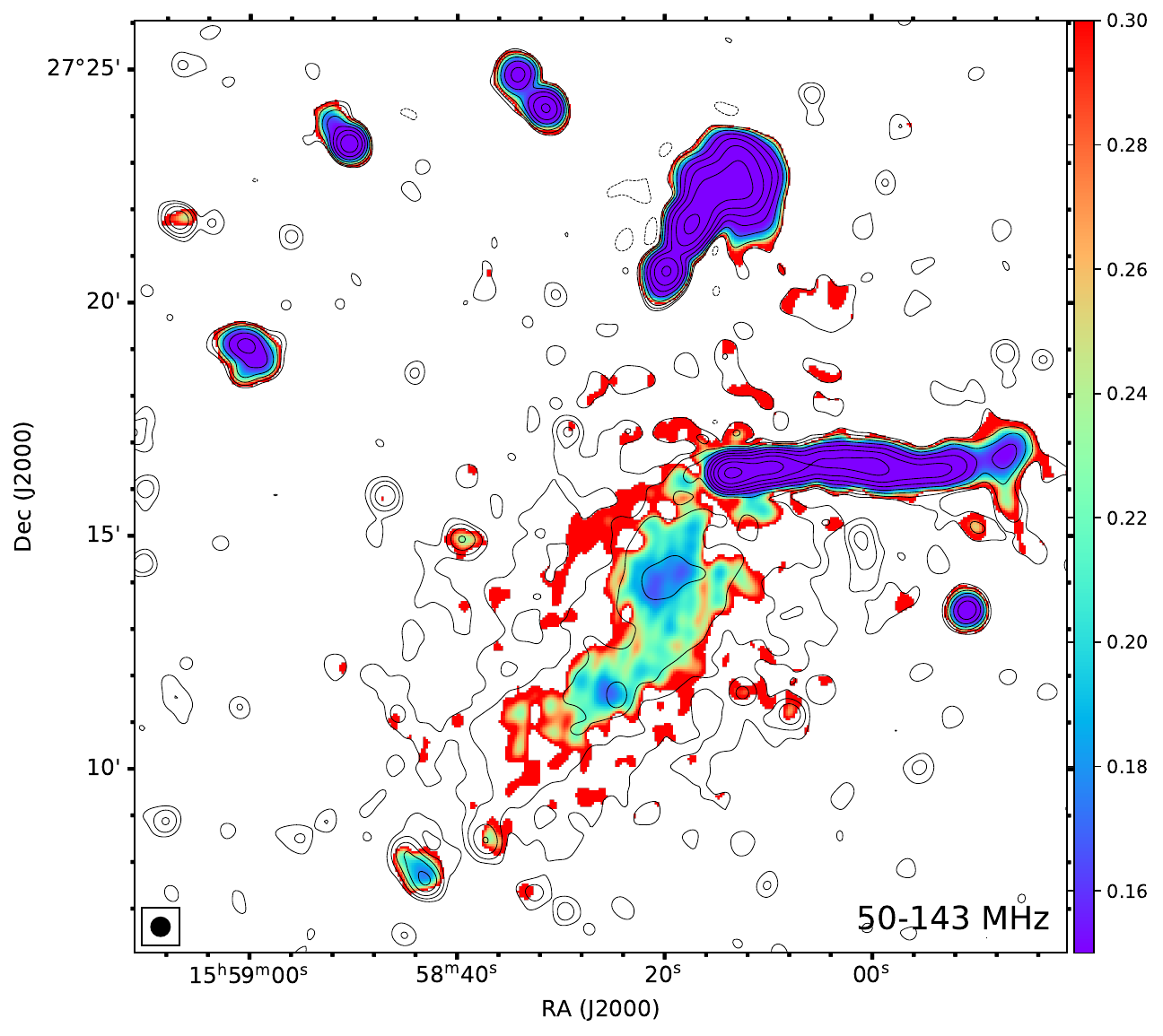}
    \includegraphics[width=0.45\textwidth]{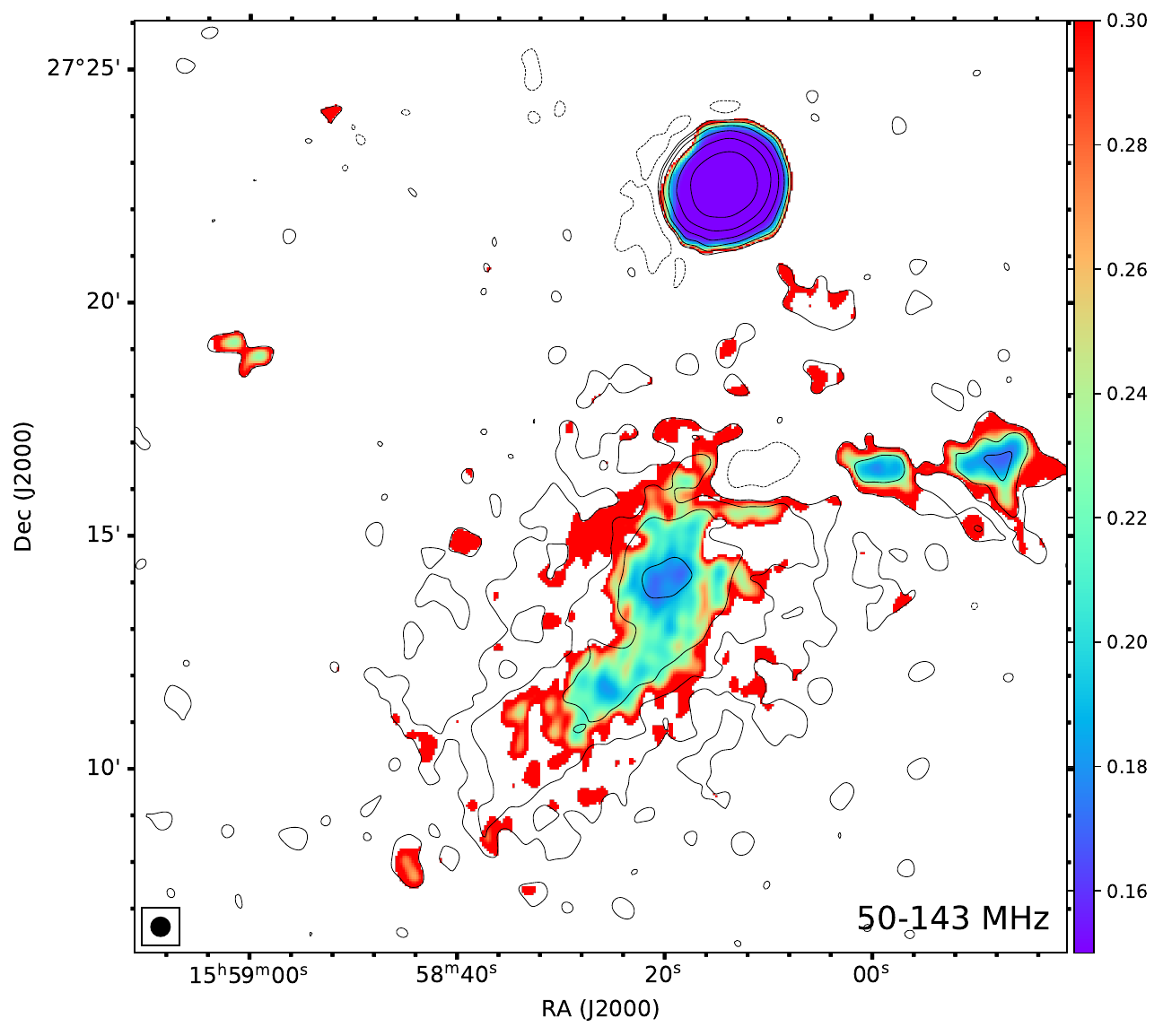}

    \includegraphics[width=0.45\textwidth]{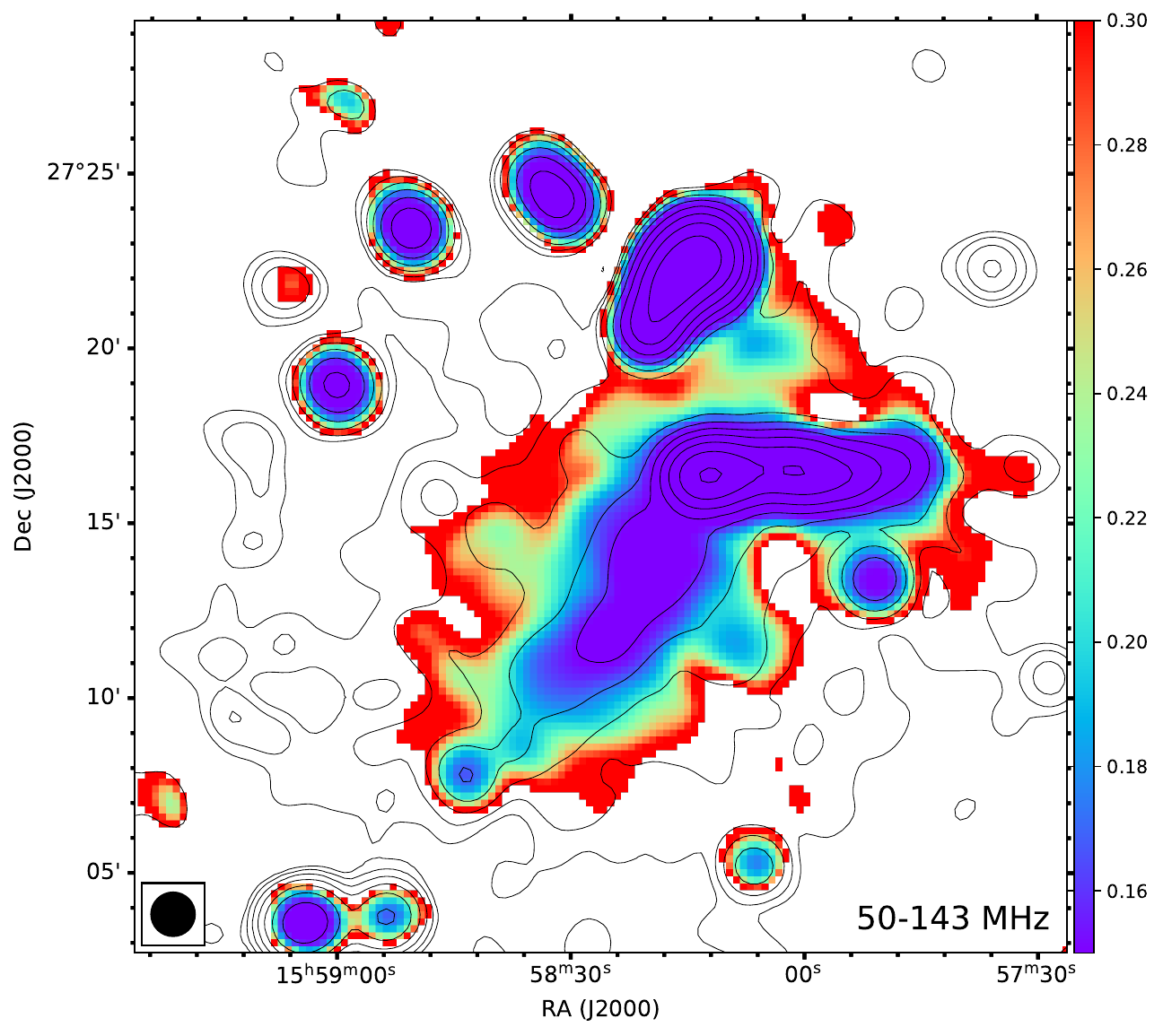}
    \includegraphics[width=0.45\textwidth]{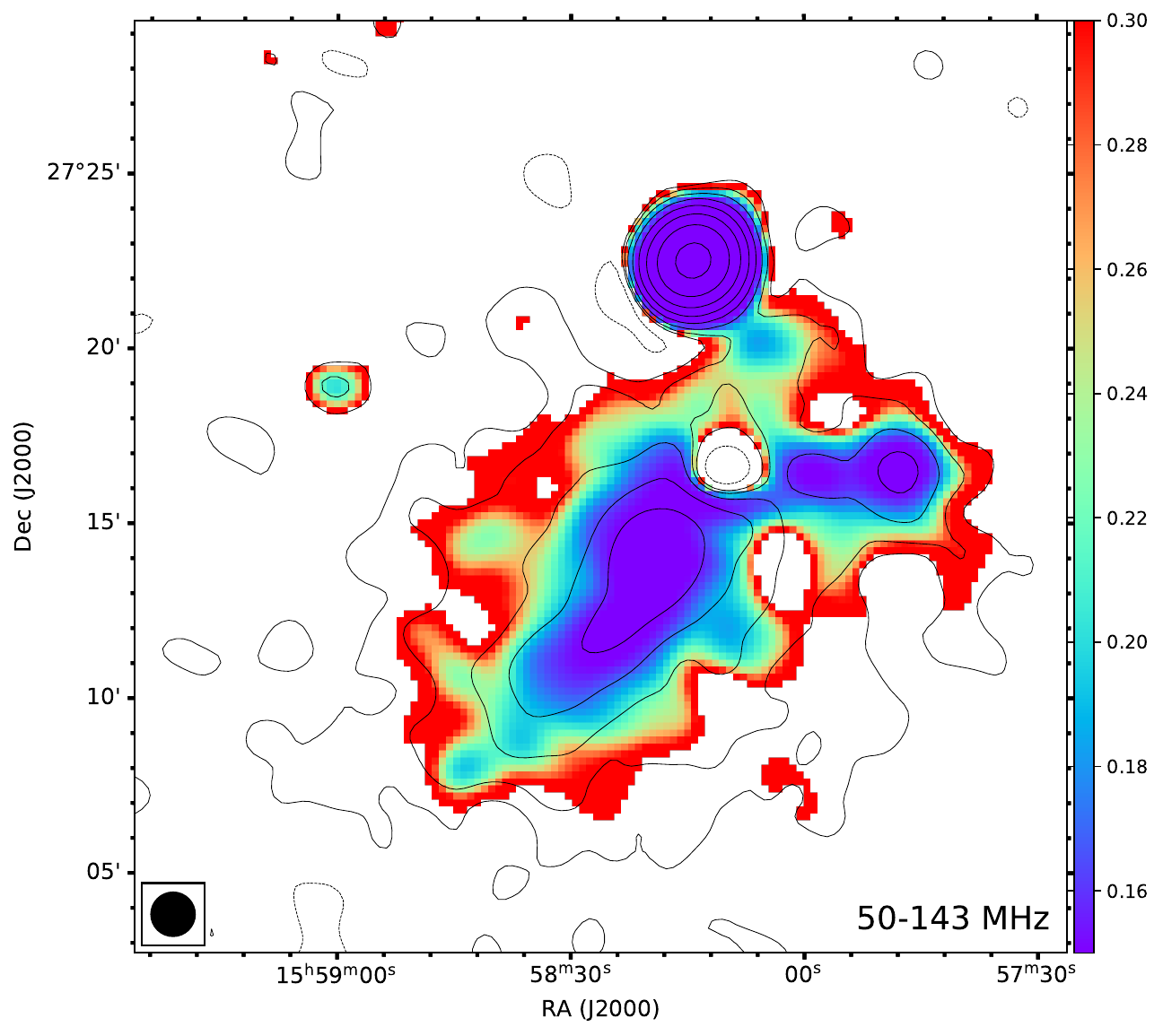}
    
    \includegraphics[width=0.45\textwidth]{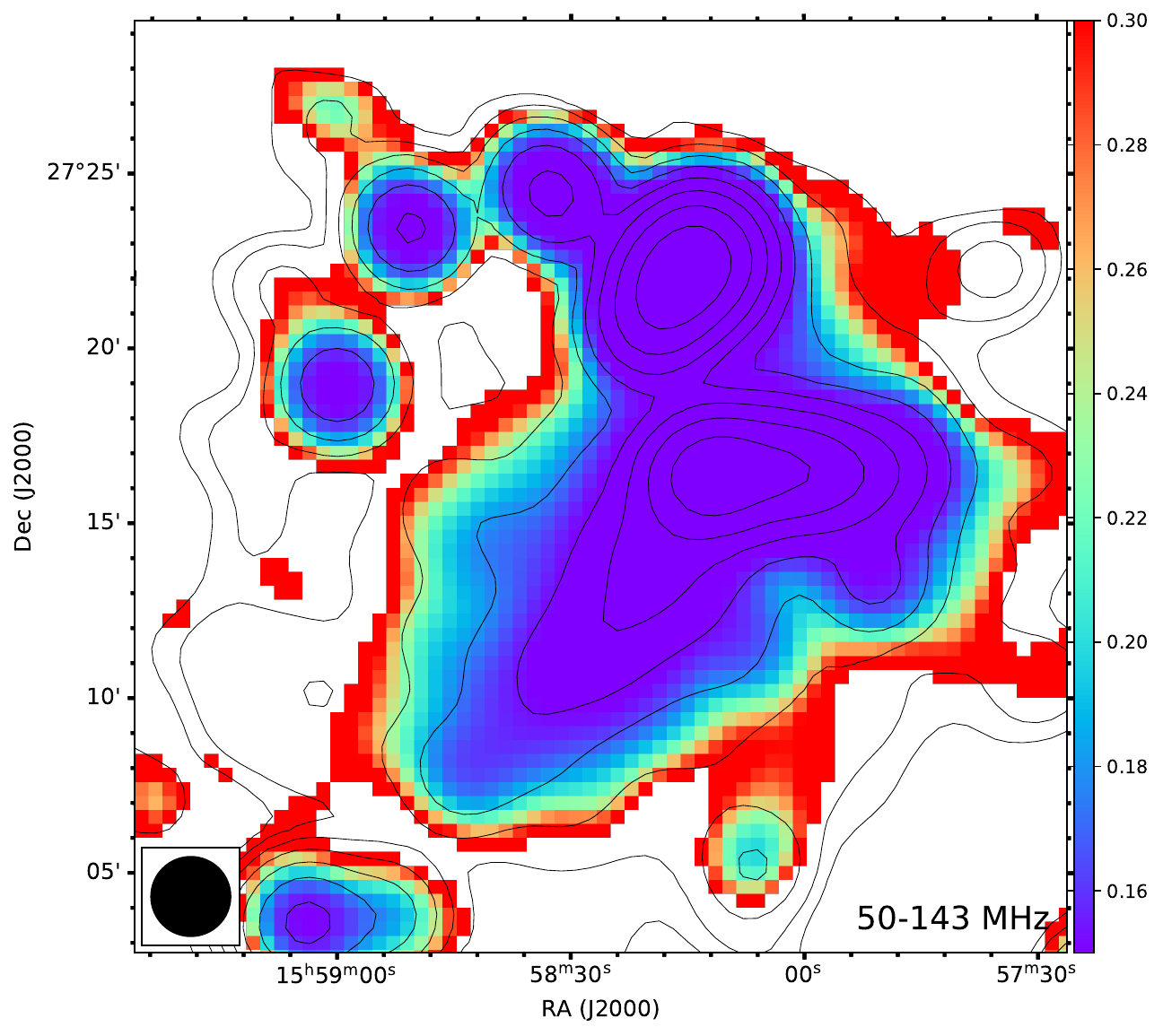}
    \includegraphics[width=0.45\textwidth]{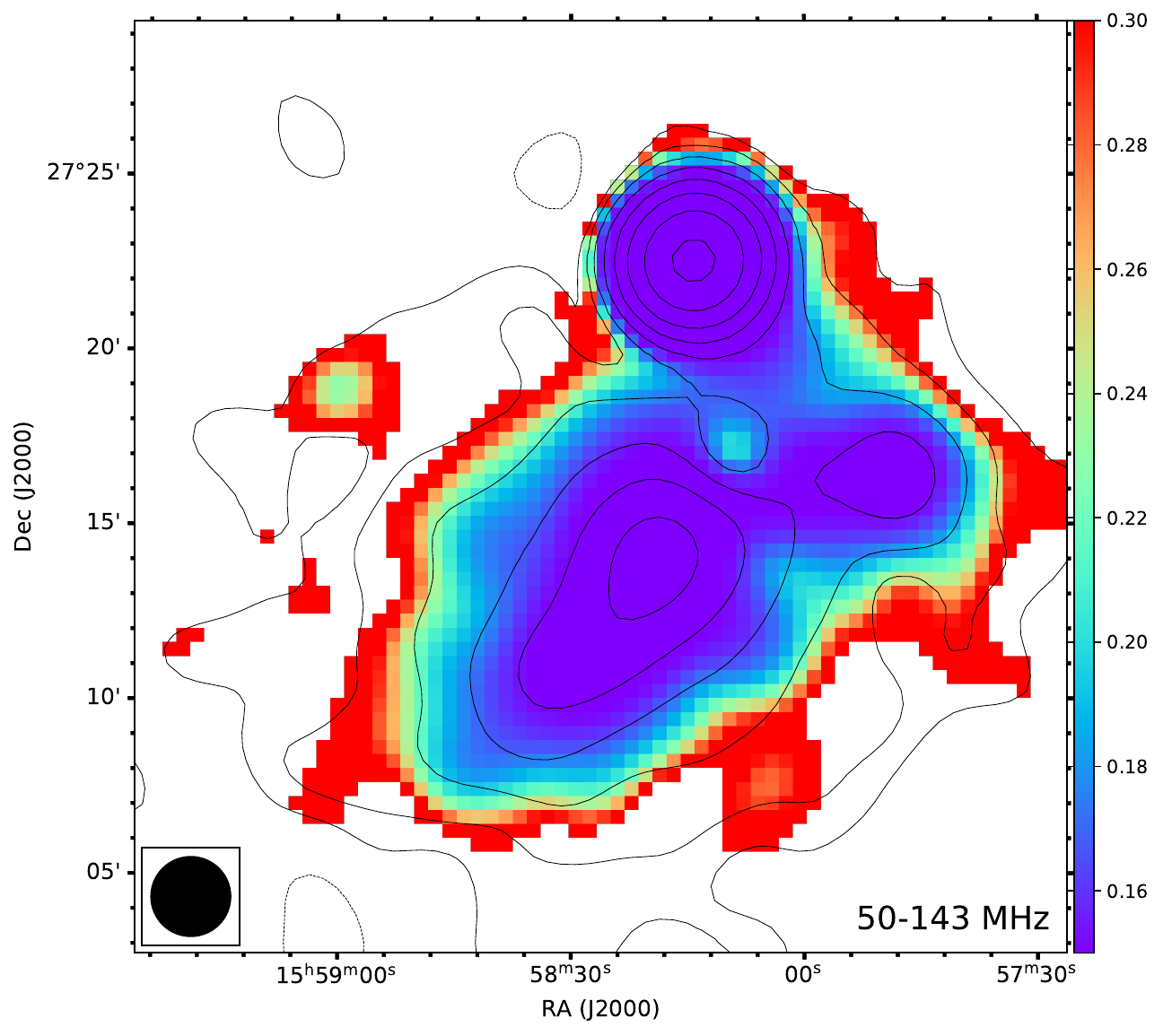}

	\smallskip
	
	\caption{Spectral index error maps between 50 and 143 MHz corresponding to the maps reported in Fig. \ref{SPIX_lofar}.}
	\label{errSPIX_lofar}
\end{figure*}

In Figs. \ref{errSPIX_lofar-gmrt}, \ref{errSPIX_lofar} we report the error maps associated with the spectral index maps shown in Figs. \ref{SPIX_lofar-gmrt}, \ref{SPIX_lofar}. Errors are obtained as 
\begin{equation}
\Delta \alpha =  \left\lvert \frac{1}{\ln{ \left( \frac{\nu_{\rm 1}}{\nu_{\rm 2} } \right) }}\right\lvert \sqrt{ \left( \frac{\Delta S_{\rm 1}}{S_{\rm 1}}\right)^2 + \left( \frac{\Delta S_{\rm 2}}{S_{\rm 2}}\right)^2 } \; \; \; ,
\label{eq:spectralindexerrorformula}
\end{equation}
where $\Delta S$ are computed as in Eq. \ref{erroronflux}.

\section{Thermodynamical maps}
\label{appendix:Thermodynamical maps}
%\FloatBarrier

\begin{figure*}
	\centering
	\includegraphics[width=0.85\textwidth]{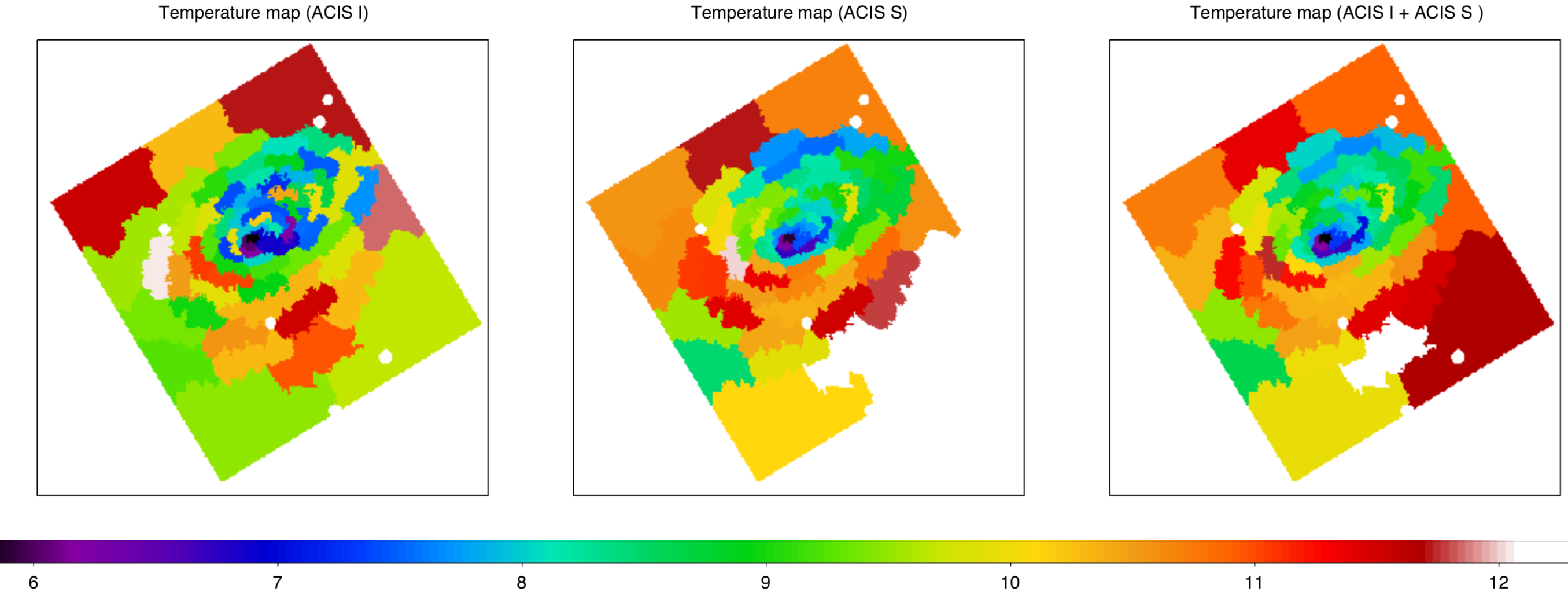} 
	\includegraphics[width=0.85\textwidth]{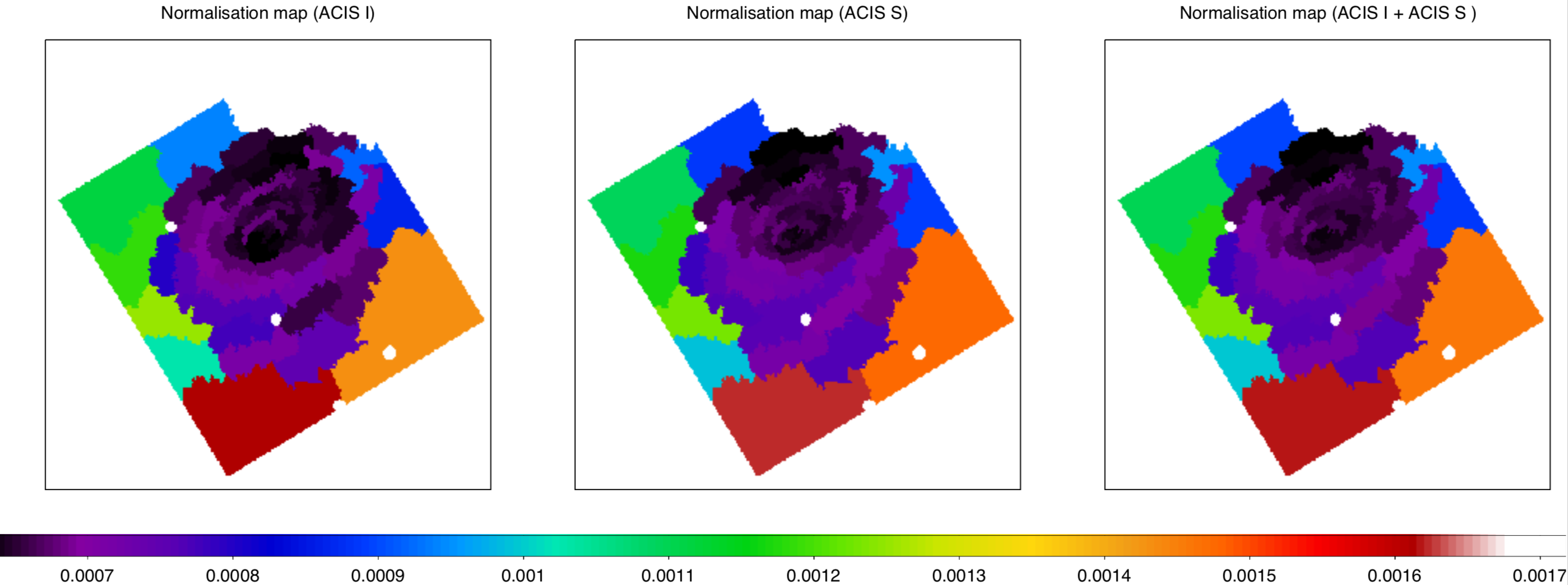}
	\includegraphics[width=0.85\textwidth]{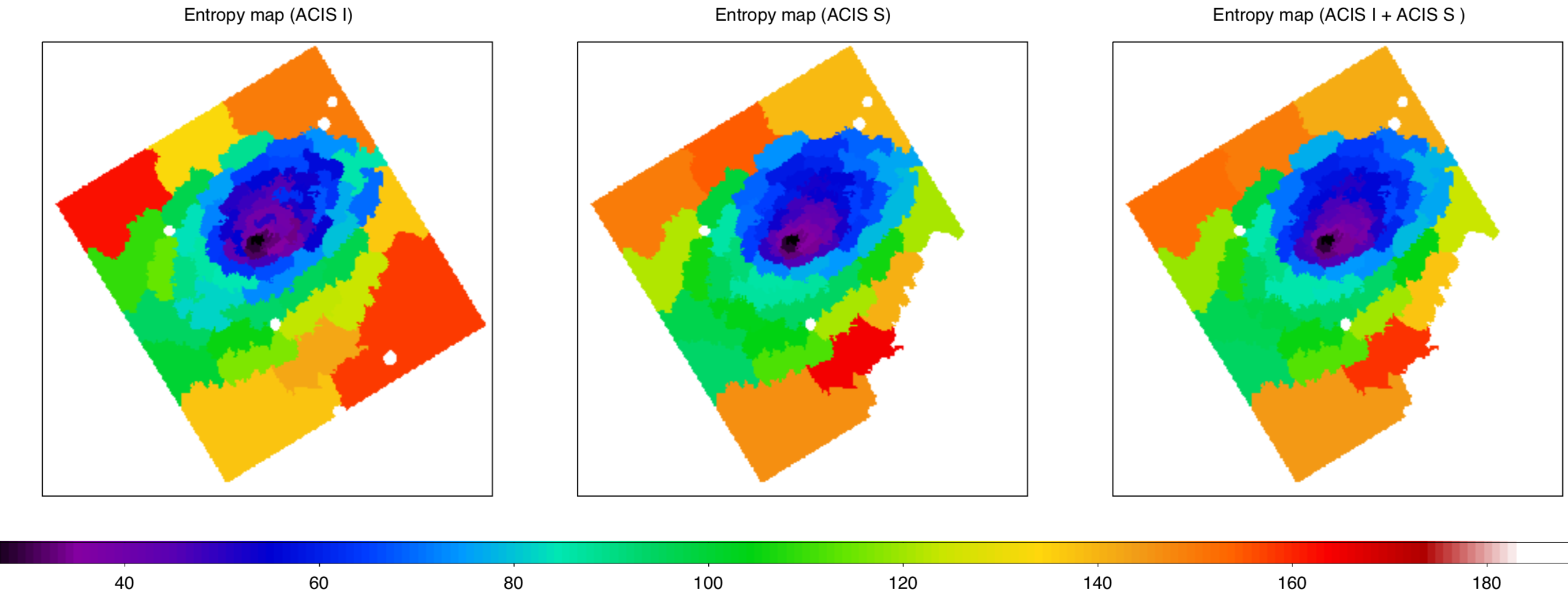}

	\caption{Projected thermodynamical maps of A2142 obtained with Chandra ACIS-I only (left columns), ACIS-S only (middle columns), and combined ACIS-I plus ACIS-S (right columns): temperature (top panels, in units of keV), {\tt apec} normalisation (middle panels, in units of $\rm cm^{-5}$), and entropy (bottom panels, in units of ${\rm keV \; cm^{5/3} \; arcmin^{-2/3}}$).}
	\label{thermomaps}%
\end{figure*}   

\begin{figure*}
	\centering
	\includegraphics[width=0.35\textwidth]{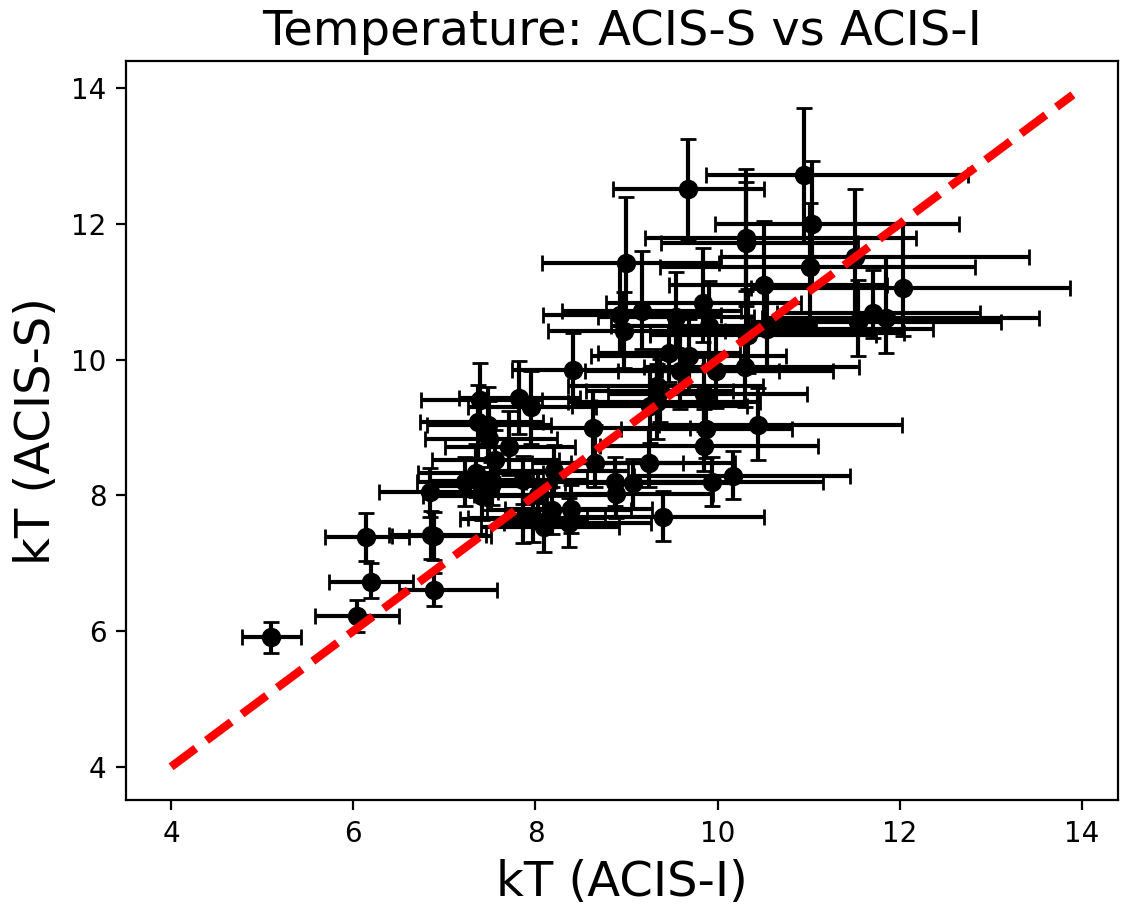} 
	\includegraphics[width=0.37\textwidth]{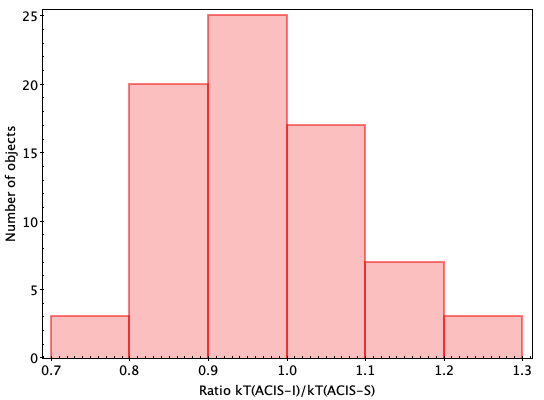}
	\includegraphics[width=0.35\textwidth]{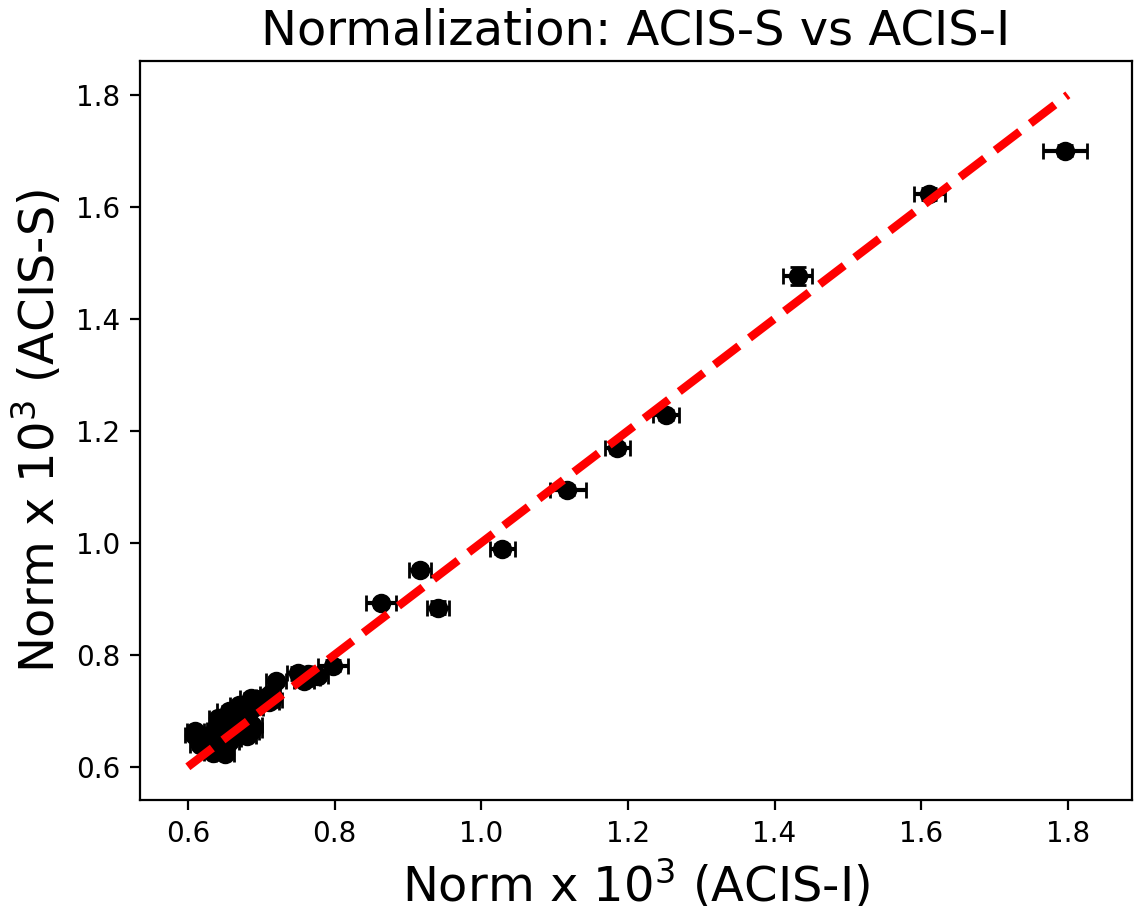}
	\includegraphics[width=0.37\textwidth]{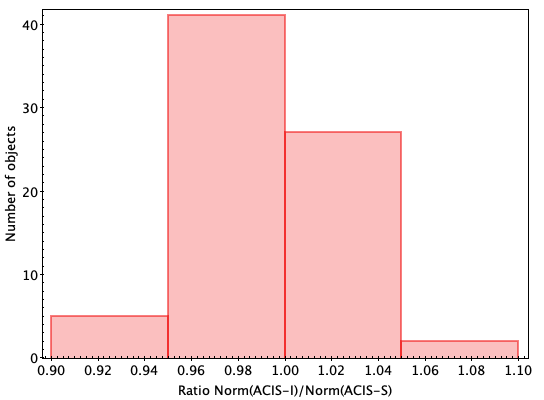}
	\caption{Comparison of fitted temperatures (upper panels) and normalisations (lower panels) derived from ACIS-I and ACIS-S separately. Left panels report the values, and the bisector line is plotted in red. Right panels report the distribution of the ratios of ACIS-I to ACIS-S values.}
	\label{ACIScomparison}%
\end{figure*}  

In the following, we describe the procedure adopted to derive the entropy map reported in Fig. \ref{entropy}, and associated systematic uncertainties. 

The fit of each spectrum provides a value of temperature $kT$ (in units of keV) and normalisation $\mathcal{N}$, where the latter is a proxy for the density. Indeed, the normalisation of the {\tt apec} model is defined as the integral of the squared numerical density ($n_{\rm e}\times n_{\rm H}$) over the volume: 
\begin{equation}
\mathcal{N}=\frac{10^{-14}}{4\pi D_{\rm A}^2 (1+z)^2}\int n_{\rm e} n_{\rm H} {\rm d}V \; \; \; \; \;  {[\rm cm^{-5}]} \; \; \; ,
\label{normAPEC}
\end{equation}
where $D_{\rm A}$ is the angular distance at the cluster redshift. We define the entropy as:
\begin{equation}
s=kT \left( \frac{\mathcal{N}}{A} \right)^{-\frac{1}{3}}
 \; \; \; \; \;  {\rm [keV \; cm^{5/3} \; arcmin^{-2/3}]} \; \; \; ,
\label{s}
\end{equation}
where $A$ is the area of each extracting region (in units of arcmin$^2$). As both the temperature and density are projected (or `pseudo') quantities, the entropy has to be considered as a projected quantity as well. 

As shown in Fig. \ref{thermomaps}, we produced (pseudo-) temperature, normalisation, and entropy maps for both single and combined ACIS-I and ACIS-S chips to check for possible biases. The associated errors maps (not shown) are computed directly from the fitting errors for $kT$ and $\mathcal{N}$, whereas uncertainties on $s$ are derived by assuming the standard error propagation formula as:
\begin{equation}
\Delta s= s \sqrt{\left(\frac{k\Delta T}{k T}\right)^2  + \frac{1}{9} \left(\frac{\Delta \mathcal{N}}{\mathcal{N}}\right)^2}  \; \; \; .
\label{s_err}
\end{equation}

In Fig. \ref{ACIScomparison} we compare the values of the fitted temperature (upper panels) and normalisation (lower panels) obtained separately from ACIS-I and ACIS-S. The fitted normalisations are well in agreement within the $\sim 5\%$; on the other hand, the fitted temperatures show a larger scatter, with typical differences of $\sim10-20\%$, up to $\sim30\%$. Some intrinsic scatter is expected due to the different levels of degradation of the Chandra chips as a function of time \citep{grant14}. In principle, spectra from the earlier ACIS-I observation are expected to provide more accurate fitted temperatures than spectra from the more recent ACIS-S observations; nevertheless, the ACIS-S observations are deeper than the ACIS-I observation by a factor 3.5. Therefore, when the spectra of all the chips are fitted simultaneously, the temperatures are primarily driven by the ACIS-S observations, and statistical errors reduce. Despite this bias unavoidably affects the entropy, we recover consistent trends for both single and combined cases in the inner regions of interest, where $s\lesssim120$ ${\rm keV \; cm^{5/3} \; arcmin^{-2/3}}$; fitting errors for the combined case are $\Delta s\lesssim 5$ ${\rm keV \; cm^{5/3} \; arcmin^{-2/3}}$ and $\Delta s\lesssim 10$ ${\rm keV \; cm^{5/3} \; arcmin^{-2/3}}$ in the inner and outer regions, respectively. Therefore, we can genuinely rely on our results, which moreover are supported by the agreement with those of \cite{rossetti13} derived from XMM-Newton data.

\end{appendix}

\end{document}